 \documentclass[%
  aip,jcp,
 amsmath,amssymb,
  reprint,%
 ]{revtex4-1}

\usepackage{graphicx}
\usepackage{dcolumn}
\usepackage{bm}

\usepackage{mathtools}
\usepackage{multirow}
\usepackage{makecell}

\begin{document}

\preprint{APS/123-QED}

\title{In search of a precursor for crystal nucleation of hard and charged colloids}

\author{Marjolein de Jager}
\email{m.e.dejager@uu.nl}
\affiliation{Soft Condensed Matter, Debye Institute of Nanomaterials Science, Utrecht University, Utrecht, Netherlands}
\author{Frank Smallenburg}
\affiliation{Universit\'e Paris-Saclay, CNRS, Laboratoire de Physique des Solides, 91405 Orsay, France}
\author{Laura Filion}
\affiliation{Soft Condensed Matter, Debye Institute of Nanomaterials Science, Utrecht University, Utrecht, Netherlands}


\date{\today}

\begin{abstract}
The interplay between crystal nucleation and the structure of the metastable fluid has been a topic of significant debate over recent years. In particular, it has been suggested that even in simple model systems such as hard or charged colloids, crystal nucleation might be foreshadowed by significant fluctuations in local structure around the location where the first nucleus arises. We investigate this using computer simulations of spontaneous nucleation events in both hard and charged colloidal particles. To detect local structural variations, we use both standard and unsupervised machine learning methods capable of finding hidden structures in the metastable fluid phase. We track numerous nucleation events for the face-centered cubic and body-centered cubic crystal on a local level, and demonstrate that all signs of crystallinity emerge simultaneously from the very start of the nucleation process. We thus conclude that there is no precursor for the nucleation of charged colloids.
\end{abstract}

\maketitle


\newcommand{\figwidthB}{0.40\linewidth}
\newcommand{\figwidthC}{0.45\linewidth}


\section{Introduction}
Crystal nucleation plays an important role in fields ranging from colloidal self-assembly, to protein crystallization, and even polymorph selection in pharmaceuticals \cite{karthika2016review, sosso2016crystal}.   However, despite its importance in a  number of essential fields, the detailed mechanism of forming a crystal nucleus still remains a topic of continuous debate.

The simplest theory which addresses crystal nucleation is classical nucleation theory (CNT). In CNT, the metastable fluid is continuously undergoing thermal fluctuations, where small, solid clusters form and dissolve until one appears which is large enough (critically large) to grow out into a macroscopic crystal. The size of such a critical cluster is given simply by balancing the bulk free-energy gain associated with transitioning into the more stable solid phase, with the surface free-energy cost of having a finite crystal cluster immersed in the fluid.  This picture, however, becomes significantly more complicated when one considers the possibility of multiple competing crystal structures, typically referred to as polymorphs. In systems with crystal polymorphs, the crystalline phase that first nucleates in the metastable fluid is not necessarily the stable phase. Theories to address such situations, such as the Ostwald step rule \cite{ostwald1897studien}, and the Alexander-McTague theory \cite{alexander1978should} have proven unreliable in explaining polymorph selection (see e.g. Refs. \onlinecite{russo2012selection,taffs2016role,ouyang2016polymorph}).

One complication when studying such questions is the close interplay between local structural motifs that occur naturally in the fluid, and the ones that might emerge when the crystal forms. It has been suggested that motifs hiding in the fluid are predictive of, or even responsible for, the location or polymorph of the nucleus that forms \cite{russo2012microscopic, gispen2023crystal}. To investigate this possibility, one avenue forward could be to explore just how much information the metastable liquid is hiding regarding the nucleation process.  Over the last two decades a plethora of studies have appeared presenting contradictory observations \cite{ten1999homogeneous,schilling2010precursor, russo2016crystal, tan2014visualizing, li2020revealing,russo2012microscopic, hu2022revealing, lu2015experimental,berryman2016early,tanaka2019revealing,lechner2011role}.  In particular, in simple systems such as charged colloids which nucleate into either the face-centered cubic (FCC) or body-centered cubic (BCC) crystal, some studies have argued that local structural order develops before the local density increases \cite{russo2012microscopic, hu2022revealing, lu2015experimental}, while other authors found evidence that the two processes happen simultaneously \cite{berryman2016early}.

To address this issue, some recent, intriguing studies have explored how modifying (via biasing) the structure of the fluid -- either enhancing or suppressing specific local motifs -- affects the nucleation process \cite{hu2022revealing,taffs2016role}.  In principle, such studies might be able to give one direct evidence that a specific local structure either enhances or suppresses the nucleation process.  Unfortunately, however, biasing the structure of the fluid modifies not only its local structure but also its thermodynamics, meaning that comparisons with the unbiased case are inconclusive.

The more direct route to trying to explore how various kinds of local ordering interplay in crystal nucleation is simply to simulate the nucleation event, and follow the various structural and density features as nucleation happens. At first glance this would appear to be a straightforward approach.  However, the challenge in this case lies in the difficulty in creating local order parameters that are unbiased. For example, order parameters that are tuned to recognize the crystalline regions from fluid might struggle at the boundary between the fluid and crystal  -- a highly important aspect at the beginning of nucleation. Similar issues exist for other order parameters, making it very difficult to pinpoint the start of the nucleation process and hence to determine whether structural order emerges before, during, or after densification. Hence, in some cases instead of capturing accurately whether local structure exists in the highly fluctuating metastable fluid, one ends up examining the properties of the order parameter instead of the properties of the fluid. 

While this problem is never fully avoidable, one option to try and avoid accidental biases is to exploit multiple different measures for local order -- for instance measures associated with symmetries like bond-order parameters and order associated with the topological connections between neighboring particles -- such as topological cluster classification (TCC).  Interestingly, new unsupervised machine learning (UML) algorithms also give new avenues to probe structure (see e.g. Refs. \onlinecite{becker2022unsupervised,reinhart2017machine,boattini2019unsupervised,boattini2020autonomously,coli2021artificial,van2020classifying,gardin2021classifying,coslovich2022dimensionality,paret2020assessing,adorf2019analysis})
. Recent studies have even demonstrated that simple, UML-based approaches are able to extract variations in disorder in the structure of supercooled fluids from e.g. a simple vector of bond order parameters \cite{boattini2020autonomously,paret2020assessing,coslovich2022dimensionality}.  Intriguingly, this  includes identifying variations in local structure that are not easily extracted by looking at each element of the vector individually.

In this paper, we attempt to take the utmost care in identifying local signatures of the fluid and and revisit the question: are there hidden local structures present in metastable fluid that foreshadow the location of the imminent formation of a crystal nucleus? 
Specifically, we  apply both classical and UML-based methods to the nucleation of hard and charged colloids in both the regime of strong screening and weak screening, for which respectively the FCC and BCC crystals nucleate. To this end, we simulate numerous spontaneous nucleation events, and closely follow all nucleation events as a function of time. In particular, similar to Ref. \onlinecite{berryman2016early} we zoom in on the regions where the nuclei are born and analyze the local fluctuations in density and structure of the metastable fluid. By doing this we can locally track whether there is a delay between the increase in local structural ordering and local density prior to the start of nucleation. Such a delay would indicate the presence of a precursor. However, within the limits of this study, we find no evidence of such a precursor in the systems we studied.

\section{Model}
We consider a system of $N$ like-charged hard spheres of diameter $\sigma$ suspended in a solvent containing salt. The effective interaction potential between these colloids is given by the repulsive hard-core Yukawa potential
\begin{equation}
	\label{eq:pothcyuk}
	\beta\phi(r) = \begin{dcases}
		\beta\epsilon \; \frac{e^{-\kappa\sigma (r/\sigma -1)} }{r/\sigma}  & \quad \text{for } r\geq\sigma,\\
		\infty       & \quad \text{for } r<\sigma,
	\end{dcases}
\end{equation}
with contact value $\beta\epsilon = Z^2\lambda_B/\sigma(1+\kappa\sigma/2)^2$, where $Z$ is the charge of the colloids in electron charge, $\lambda_B$ is the Bjerrum length, $\kappa$ is the inverse Debye screening length, and $\beta=1/k_BT$, with $k_B$ the Boltzmann constant and $T$ the temperature. Note that in the limit of zero charge ($Z\to0$) or infinite screening ($\kappa\sigma\to\infty$), this potential reduces to the hard-sphere potential.
The interaction potential was truncated and shifted such that the shift was never more than $10^{-5}k_BT$.

Nucleation of both the BCC and FCC phases in this system has been studied in the past (see e.g. Refs. \onlinecite{auer2005numerical, desgranges2007polymorph, browning2008nucleation, gispen2022kinetic, jager2022crystal}). In a previous study \cite{jager2022crystal}, we used umbrella sampling to calculate the nucleation barriers and rates of highly screened charged particles. In this paper, we will study the nucleation of some of these (nearly-)hard systems, as well as the nucleation of weakly screened charged particles. 
To be able to compare the nucleation processes of different systems, we select state points with approximately equal barrier heights.
In particular, we will simulate brute-force nucleation events of systems with barrier heights around 15-18$k_BT$. 
Information on the nucleation barriers of the systems studied is given in Tab. \ref{tab:info}. 
Note that systems with a Debye screening length of $1/\kappa\sigma=0.01$ were found to behave essentially as ``hard'' spheres when mapped with an effective hard-sphere diameter \cite{jager2022crystal}.
A brief explanation of the methods used for computing the nucleation barriers as well as some additional information on these systems can be found in the Supplemental Materials (SM).

\begin{table}[t!]
\caption{\label{tab:info} For each system studied, the packing fraction of the supersaturated fluid $\eta^*$ at which the brute force nucleation is performed together with the corresponding supersaturation $\beta|\Delta\mu|$. The last columns give the critical nucleus size $n^*$ and barrier height $\beta\Delta G^*$ obtained using umbrella sampling. The error in $\beta\Delta G^*$ is no more than 1. 
}
\begin{ruledtabular}
\begin{tabular}{ccccccc}
 & $\beta\epsilon$ & $1/\kappa\sigma$ & $\eta^*$ & $\beta |\Delta\mu|$ & $n^*$ & $\beta\Delta G^*$ \\ \hline
\multirow{3}{*}{FCC} & \multicolumn{2}{c}{hard spheres}  & 0.5385   & 0.585   & 75   & 16.5  \\
 & 81 & 0.01     & 0.4681    & 0.584   & 84   & 16.3  \\ 
 & 8  & 0.04     & 0.4400    & 0.541   & 69   & 14.8  \\ \hline
 BCC & 81 & 0.40     & 0.1305    & 0.321   & 122  & 18.0  \\
\end{tabular}
\end{ruledtabular}
\end{table}


\section{Methods}
To explain the methods we use for studying the nucleation events, we need to discuss two things: i) how we identify local structure, and ii) how we track nucleation events locally.

\subsection{Identifying local structure}
We use three different methods to classify the local structure. The first method considers just the averaged bond-orientational order parameters (BOPs) of Lechner and Dellago \cite{lechner2008accurate}. For this, we first calculate for each particle $i$ the complex quantities
\begin{equation}\label{eq:qlm}
    q_{lm}(i) = \frac{1}{N_b(i)} \sum_{j\in\mathcal{N}_b(i)} Y_l^m(\theta_{ij}, \phi_{ij}),
\end{equation}
where $\mathcal{N}_b(i)$ is the set of the $N_b(i)$ nearest neighbors of particle $i$, $\,Y_{lm}\left( \theta,\phi \right)$ are the spherical harmonics with $m\in[-l,l]$, and $\theta_{ij}$ and $\phi_{ij}$ are the polar and azimuthal angles of the vector $\mathbf{r}_{ij}=\mathbf{r}(j)-\mathbf{r}(i)$ connecting particles $i$ and $j$. We use the SANN algorithm \cite{van2012parameter} to determine the nearest neighbors.
Next, we average these complex quantities over the set of nearest neighbors as well as the particle itself
\begin{equation}\label{eq:qlmbar}
    \bar{q}_{lm}(i) = \frac{1}{N_b(i)+1} \sum_{j\in\{i,\mathcal{N}_b(i)\}} q_{lm}(j).
\end{equation}
Finally, we compute the rotationally invariant averaged BOPs
\begin{equation}\label{eq:ql}
    \bar{q}_l(i) = \sqrt{\frac{4\pi}{2l+1} \sum_{m=-l}^l |\bar{q}_{lm}(i)|^2},
\end{equation}
and
\begin{equation}\label{eq:wl}
    \bar{w}_l(i) = \frac{w_l(i)}{ \left( \sum_{m=-l}^l |\bar{q}_{lm}(i)|^2 \right)^{3/2} },
\end{equation}
with
\begin{equation}
    w_l(i) =\!\!\!\!\!\! \sum_{\begin{matrix}
    \scriptstyle m_1,m_2,m_3\\ 
    \scriptstyle m_1+m_2+m_3=0
    \end{matrix}} \!\!\!\!\!\!
    \begin{pmatrix}
    l & l & l\\
    m_1 & m_2 & m_3
    \end{pmatrix}
    q_{lm_1}(i) q_{lm_2}(i) q_{lm_3}(i),
\end{equation}
where the term in brackets is the Wigner $3j$ symbol, which is only non-zero when $m_1+m_2+m_3=0$. Note that $w_l=0$ when $l$ is odd. Depending on the choice of $l$, these BOPs are sensitive to different (crystal) symmetries. For example, $\bar{q}_6$ is very helpful in distinguishing more fluid-like environments from more solid-like environments such as FCC and BCC\cite{lechner2008accurate}.

While the BOPs are extremely useful for detecting specific symmetries in the local structure of the fluid, they are not necessarily optimal for detecting the most important structural variations in disordered systems such as the metastable fluid. Recent work has shown that BOPs in combination with unsupervised machine learning algorithms is highly effective at autonomously detecting variations that might be difficult to see by studying the individual BOPs \cite{boattini2019unsupervised,boattini2020autonomously}. Hence, for the second method, we use an unsupervised algorithm to autonomously detect local structural fluctuations in the metastable fluid. In Ref. \onlinecite{boattini2019unsupervised}, Boattini \textit{et al.} showed that a neural-network-based autoencoder, which is given $\bar{q}_l$ with $l\in[1,8]$ as input, does a good job in distinguishing a whole range of different local structures. Similarly, in Ref. \onlinecite{van2020classifying} van Damme \textit{et al.} showed that using principal component analysis (PCA) as dimensionality reduction method also does a good job in distinguishing the sizable assortment of crystal structures formed by rounded tetrahedra. For our system we found that PCA and an autoencoder preformed equally well in distinguishing order, and hence chose to use the simpler PCA algorithm in our analysis. To specifically focus on finding local fluctuations or signatures in the metastable fluid, we train the PCA model on $\bar{q}_l$ with $l\in[1,8]$ of configurations containing only fluid particles and no (significant) solid nuclei. We then use this trained PCA model to analyze the entire nucleation trajectory. Note that in this way, the first principal component corresponds to the largest BOPs-related structural variation in the metastable fluid.

For the third and last method, we use an altogether different approach for classifying local structure. In particular we use the topological cluster classification (TCC) algorithm developed by Malins \textit{et al.} \cite{malins2013identification} to detect any local motifs that the BOPs might have overlooked. More specifically, we use TCC to calculate the population of certain types of clusters, as well as the number of clusters of a certain type a particle is involved in. 

Note that for computing the nucleation barriers, we additionally need a binary classification method which labels a particle as either fluid or solid. For this we use the 6-fold Ten Wolde bonds \cite{tenwolde1996simulation}
\begin{equation}\label{eq:d6}
    d_6(i,j) = \frac{ \sum_{m=-6}^{m=6} q_{6m}(i) q_{6m}^*(j) }{ \sqrt{ \left(\sum_{m=-6}^{m=6} |q_{6m}(i)|^2 \right) \left(\sum_{m=-6}^{m=6} |q_{6m}(j)|^2 \right) } },
\end{equation}
where $^*$ indicates the complex conjugate and $q_{6m}$ are the bond-orientational order parameters given by Eq. \eqref{eq:qlm}. Particle $i$ is classified as solid if it has 6 or more neighboring particles $j$ with which it has a solid-like bond, i.e. $d_6(i,j)>0.7$. 
We also use this fluid-solid classification to initially locate and follow the nucleation event. However, we want to point out that, although this provides a general overview of the nucleation event, it is not an ideal order parameter to study the onset of nucleation. Its binary nature with the thresholds for $d_6(i,j)$ and the number of solid-like bonds causes it to overlook subtle increases in the local structural ordering of the fluid and hence reacts more slowly to the nucleation than other continuous order parameters.

\subsection{Simulating and tracking nucleation events}
To obtain the spontaneous nucleation events we use brute force MC, KMC, and MD simulations in the $NVT$-ensemble, where we simulate $N=10976$ and $N=11664$ particles for the systems forming the FCC and BCC phase, respectively. The difference between the MC and KMC simulations is the acceptance ratio of the trial particle moves. For the MC simulations this acceptance ratio is around 30\%, whilst for the KMC simulations it is around 85\% resulting in dynamics that mimic Brownian motion \cite{sanz2010dynamic}. The MD simulations are performed using LAMMPS with a Nose-Hoover thermostat \cite{plimpton1995fast}. We only perform MD simulations for the soft hard-core Yukawa system, i.e. the one with $1/\kappa\sigma=0.40$. As this system does not feel its hard core, it can simply be ignored.
Once we have obtained the numerous nucleation events, we want to track the local density and structure to determine if there is a difference between the increase in local structural ordering and local density at the start of nucleation. To this end, for each nucleation event we find the position $\mathbf{r}_0$ that best captures the center of the nucleus at the start of nucleation. For this we use the average center-of-mass of the precritical nucleus as a starting point and, if needed, by eye adjust it to best capture the birthplace of the crystal nucleus. 
Next, for each snapshot of the nucleation trajectory, starting well before the start of nucleation, we determine all particles inside a sphere of radius $R$ around $\mathbf{r}_0$, and take the average of the local properties of these particles. This is similar to what Berryman \textit{et al.} did in Ref. \onlinecite{berryman2016early}. The local structural properties that we consider are explained in the previous subsection.  Additionally, we define for each particle a local packing fraction measured via the volume of its Voronoi cell. The volumes of the Voronoi cells were obtained using \texttt{voro++} \cite{rycroft2009voro}.
As we are searching for local precursors, we choose $R$ such that the selected region contains around 30-40 particles. This size provides a good balance between being large enough to obtain relatively stable averages of the local properties, and being small enough to ensure that the averaged properties still represent the local situation.


\section{Results}

\subsection{Structure of the metastable fluid}

Before we look into the actual crystal nucleation, we first characterize the structural properties of the metastable fluid. 

\begin{figure*}[t!]
\begin{tabular}{lll}
     a) & \hspace{0.5cm} & b)  \\[-0.45cm]
     \includegraphics[width=\figwidthB]{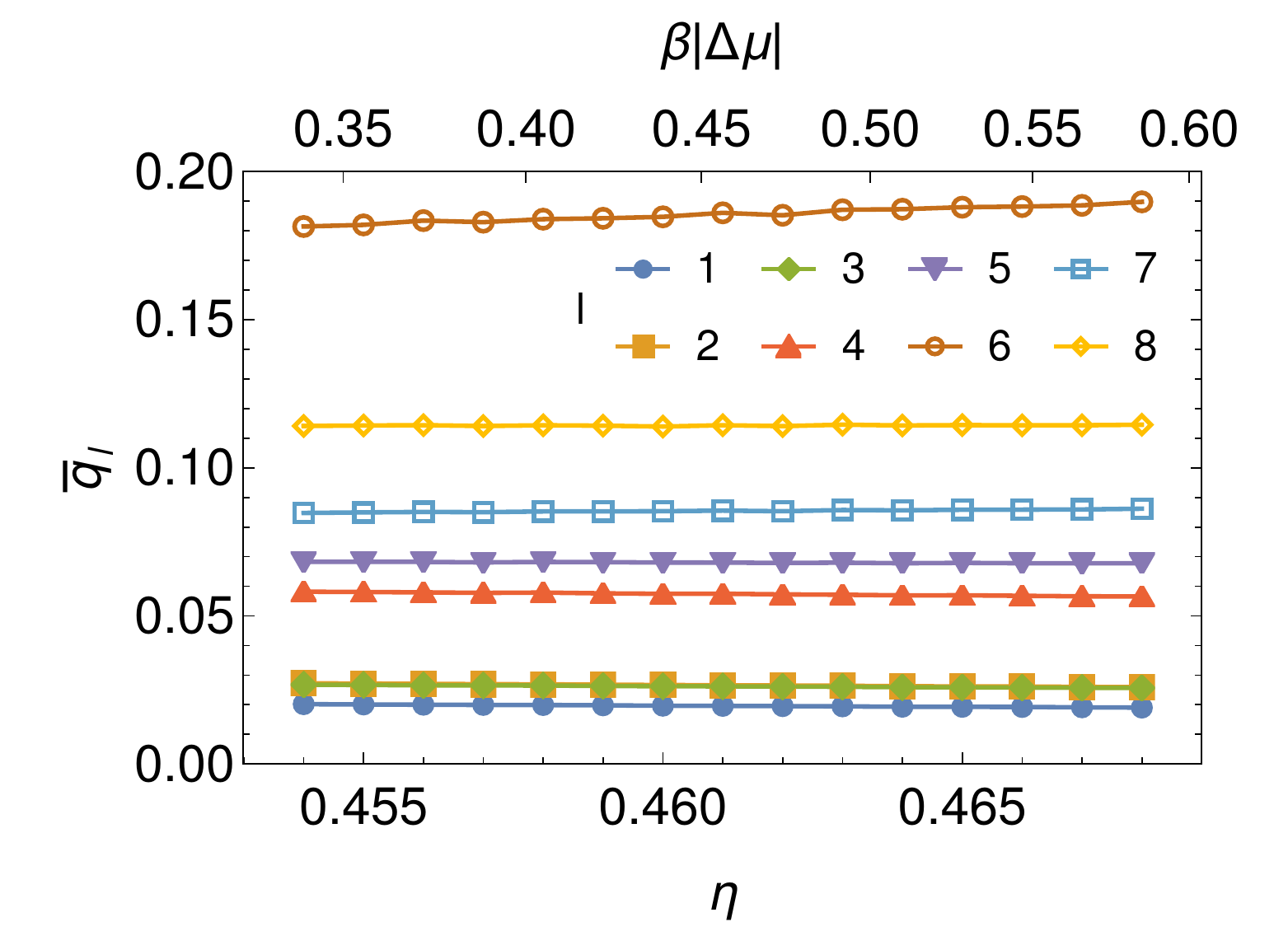} & & \includegraphics[width=\figwidthB]{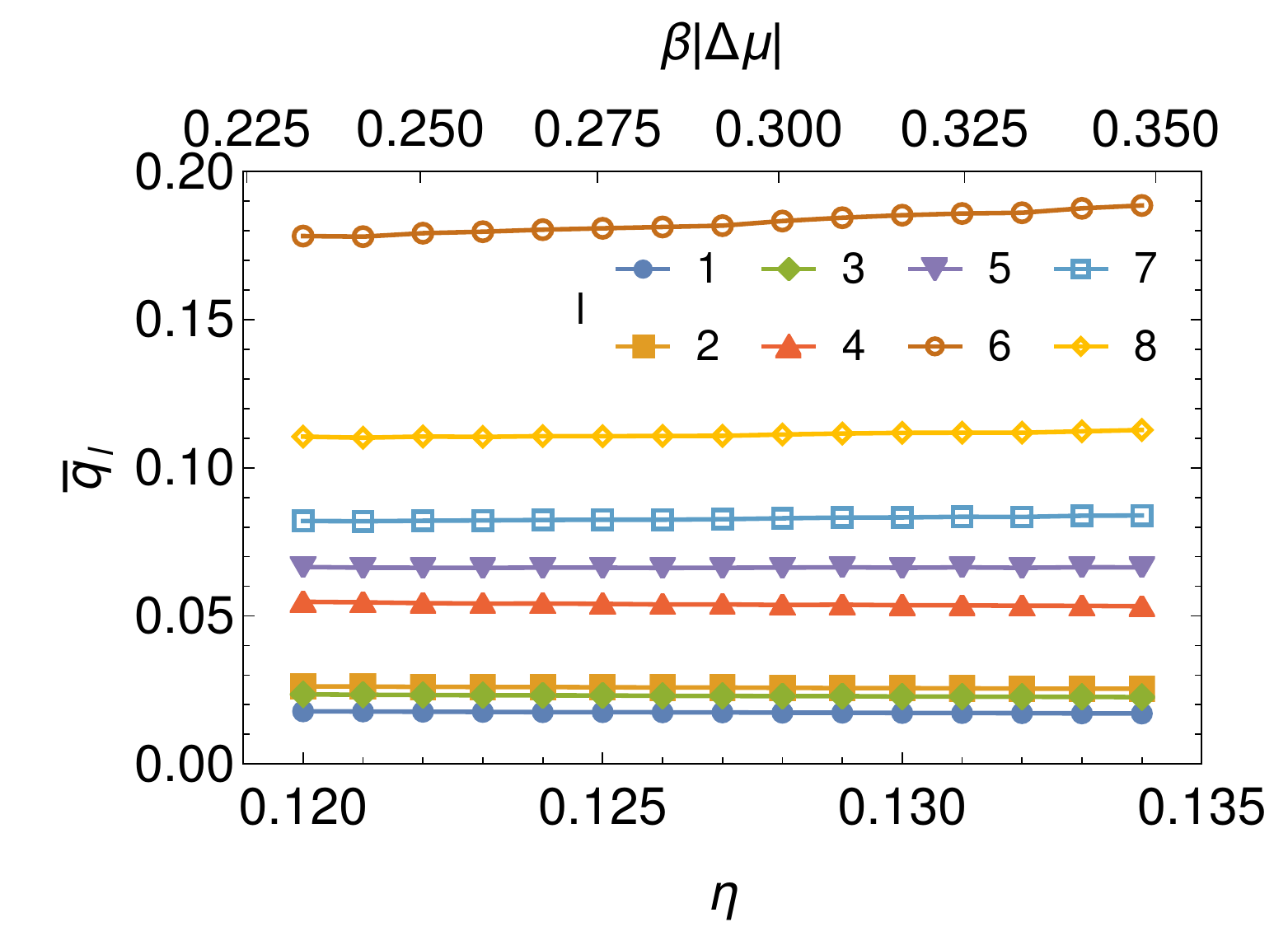} \\
     c) & \hspace{0.5cm} & d)  \\[-0.45cm]
     \includegraphics[width=\figwidthB]{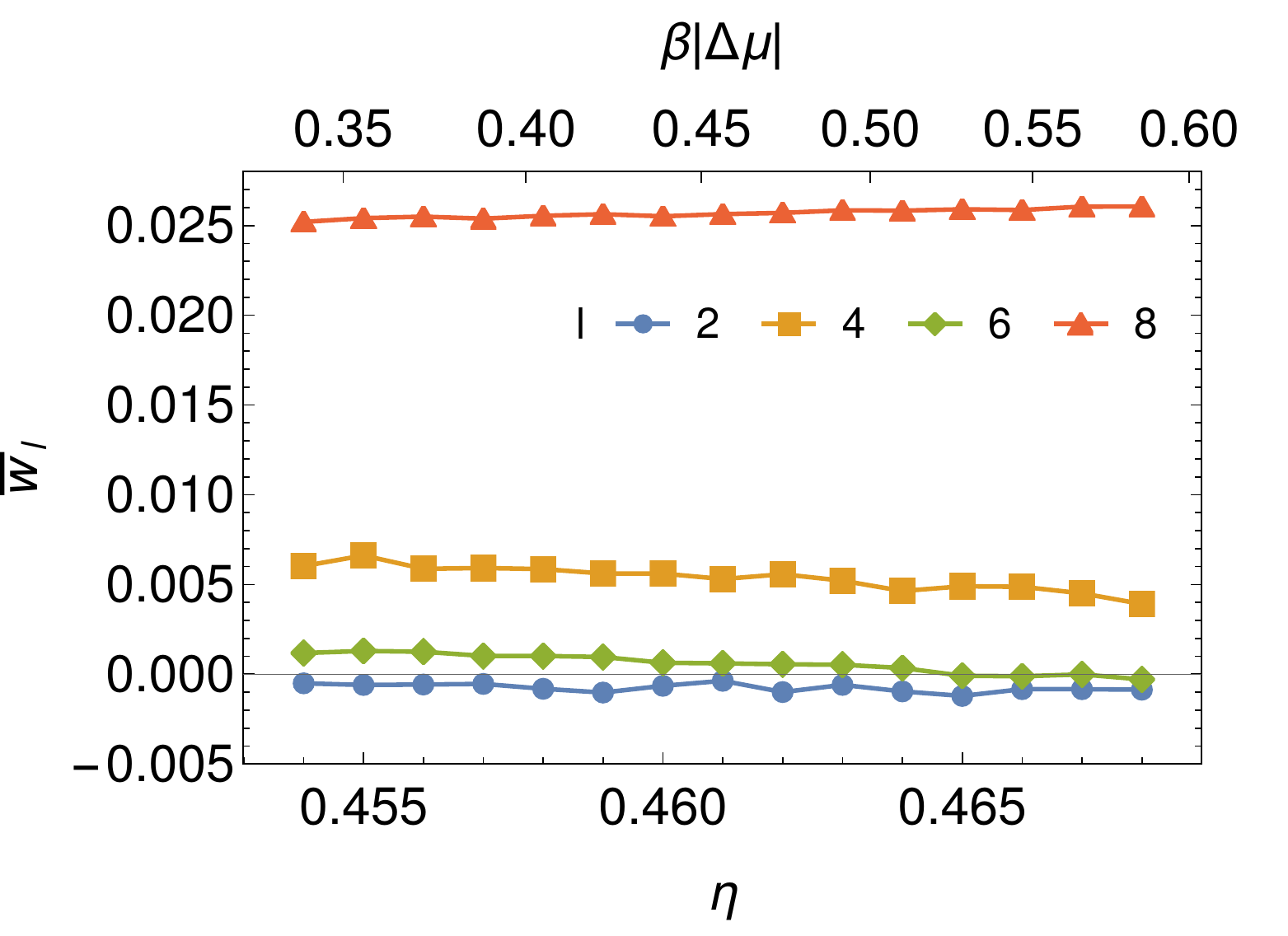} & & \includegraphics[width=\figwidthB]{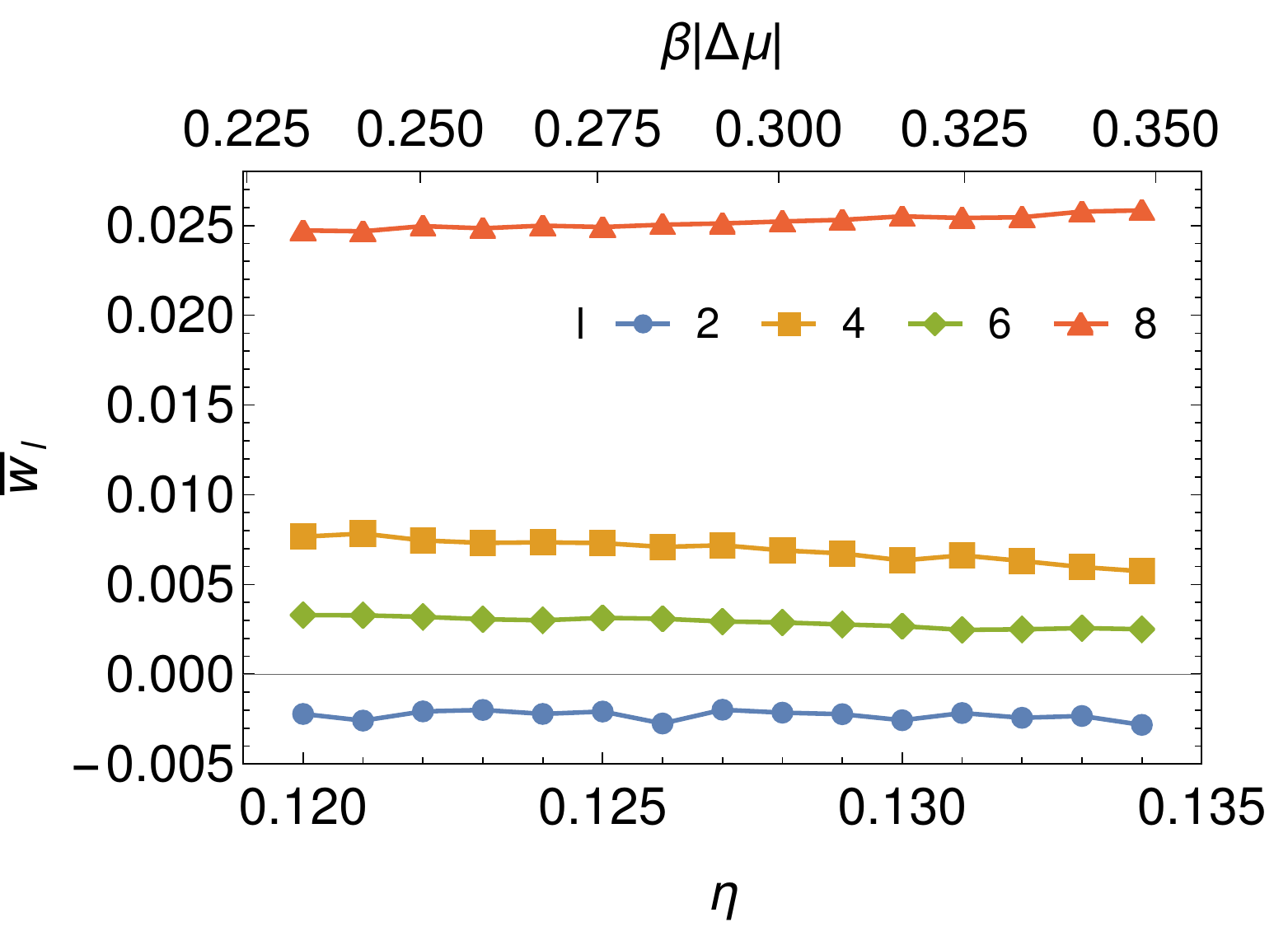}
\end{tabular}
    \caption[width=1\linewidth]{\label{fig:fluidbops} The mean of a,b) the first eight $\bar{q}_l$'s and c,d) the first four even $\bar{w}_l$'s as a function of the supersaturation for the metastable fluids of hard-core Yukawa with a,c) $\beta\epsilon=81$ and $1/\kappa\sigma=0.01$, and b,d) $\beta\epsilon=81$ and $1/\kappa\sigma=0.40$.
    }
\end{figure*}

First, we examine the globally averaged values of the local BOPs, and plot the results in  Fig. \ref{fig:fluidbops} for the metastable fluids of essentially hard spheres and of soft spheres as a function of the supersaturation. We see that $\bar{q}_6$, $\bar{q}_8$, and $\bar{w}_8$ are most prominent in both metastable fluids, and that all BOPs are only marginally affected by the increase in supersaturation. Furthermore, notice that the values in both systems are surprisingly similar, even though the metastable fluid of essentially hard spheres later forms an FCC crystal, whereas the fluid of the soft spheres will form a BCC crystal. The most prevalent difference between the two systems can be found in $\bar{w}_6$, which is smaller for the nearly-hard spheres than for soft spheres, and for high supersaturation even becomes on average negative for the nearly-hard spheres whereas it stays positive for the soft spheres. See the SM for more analysis on the $\bar{w}_l$'s.
We, thus, conclude that the fluid's ``knowledge'' about which crystal phase it should nucleate into is difficult to distinguish from the global values of the BOPs.

\begin{figure*}[t!]
    \includegraphics[width=0.75\linewidth]{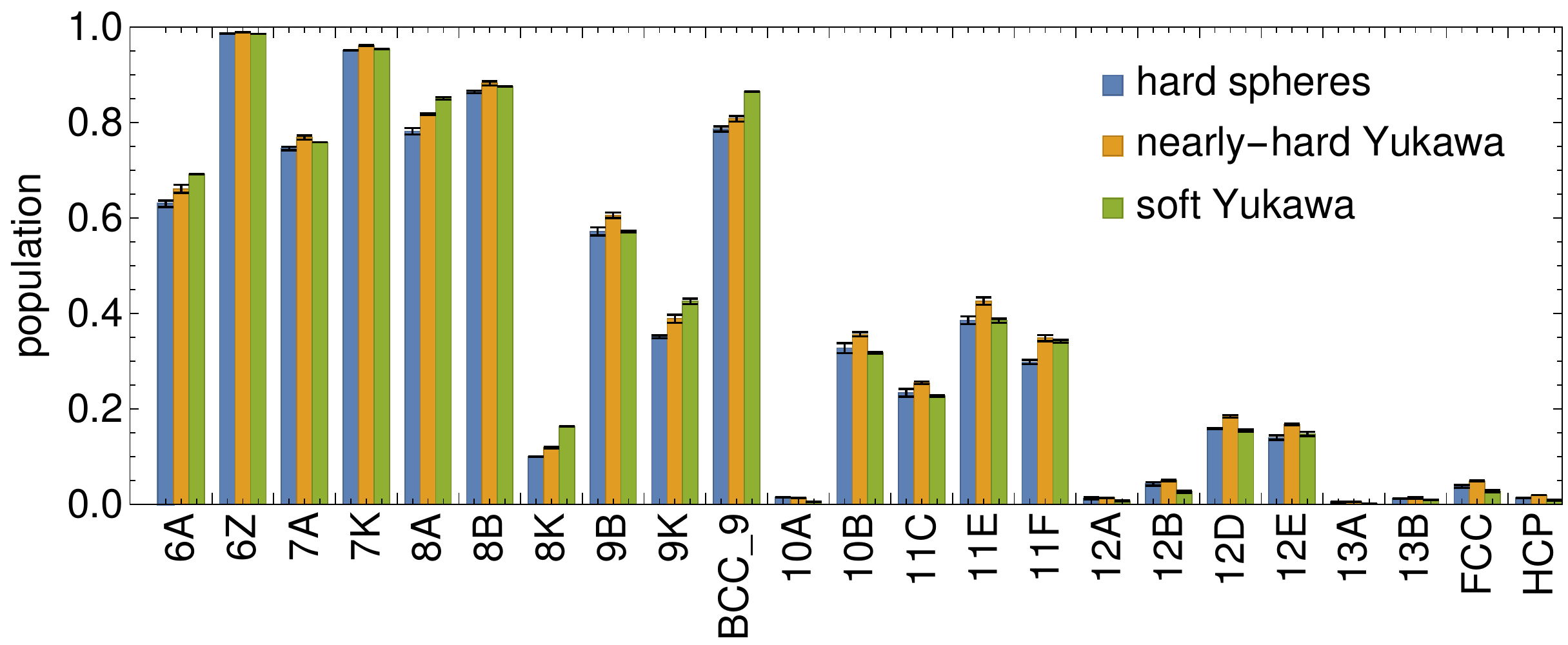}
    \caption[width=1\linewidth]{\label{fig:fluidtcc} The population of various TCC clusters in the metastable fluids of hard spheres ($\eta=0.5385$), nearly-hard hard-core Yukawa particles ($\beta\epsilon=8$, $1/\kappa\sigma=0.04$, $\eta=0.4400$), and soft hard-core Yukawa particles ($\beta\epsilon=81$, $1/\kappa\sigma=0.40$, $\eta=0.1305$).
    }
\end{figure*}

In addition to the BOPs, we take a look at the presence of the different TCC clusters in the  metastable fluids. Figure \ref{fig:fluidtcc} shows the population of various TCC clusters in the metastable fluid of hard spheres, and soft spheres. Even though we see some small deviations in the populations of the different metastable fluids -- e.g. clusters 6A, 8A, 8K, 9K, and BCC\_9 have a slightly higher population in the fluid of soft spheres and 9B, 10B, 11C, 11E, and 12D have a slightly higher population in the fluid of nearly-hard spheres -- the values are again surprisingly similar. This indicates once more that it is difficult to determine which crystal phase will nucleate from the metastable fluid for the systems studied here.

\begin{figure*}[t]
    \includegraphics[width=0.98\linewidth]{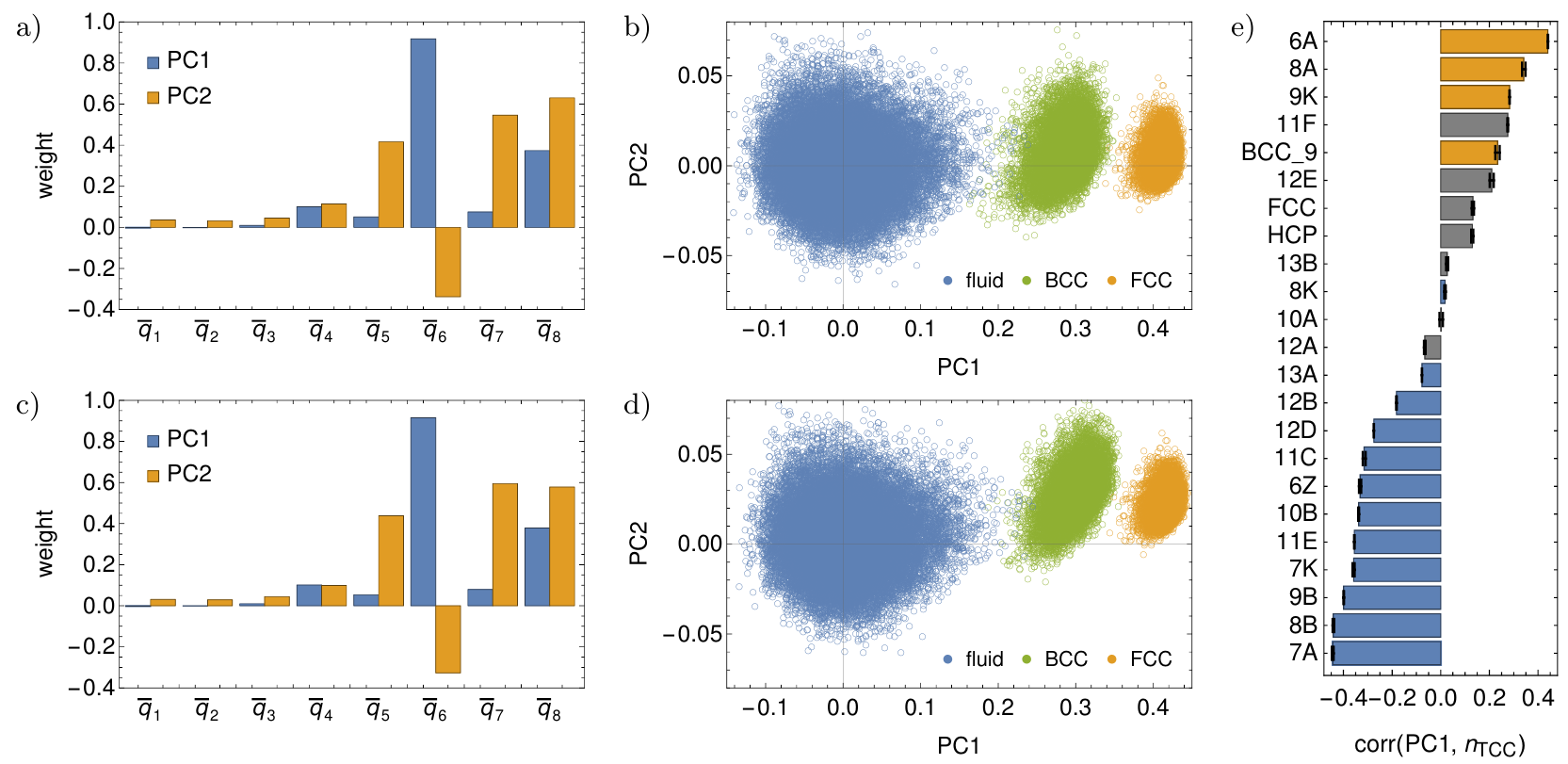}
    \caption[width=1\linewidth]{\label{fig:fluidpca} PCA on the metastable fluids of a,b) hard spheres ($\eta=0.5385$) and c,d) soft hard-core Yukawa particles ($\beta\epsilon=81$, $1/\kappa\sigma=0.40$, $\eta=0.1305$). a,c) give the weight of each BOP in the first and second principal component. b,d) show the distribution of the fluid particles in the PC1-PC2 plane, with the addition of the corresponding bulk FCC phase of hard spheres ($\eta=0.5981$) and bulk BCC phase of soft hard-core Yukawa ($\beta\epsilon=81$, $1/\kappa\sigma=0.40$, $\eta=0.1311$). e) Pearson correlation between PC1 and the number of TCC clusters a particle is involved in for the hard spheres. The color of the bars indicate clusters that consist of one or more tetrahedral subclusters (blue), one or more square pyramidal subclusters (yellow), or both/neither (gray).
    }
\end{figure*}

\newcommand{\figwidthS}{0.23\linewidth}
\newcommand{\figwidthL}{0.22\linewidth}
\begin{figure*}[t]
\begin{tabular}{llll}
	a) \hspace{0.08\linewidth} $\eta$ & b) \hspace{0.07\linewidth} PC1 & c) \hspace{0.07\linewidth} 9B & d) \hspace{0.07\linewidth} 11F \\
	\includegraphics[width=\figwidthS]{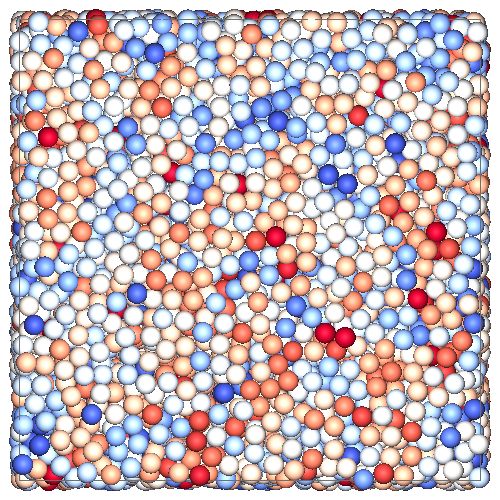}
	& \includegraphics[width=\figwidthS]{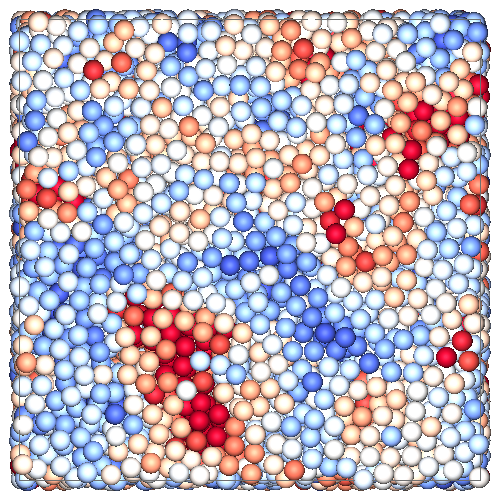}
	& \includegraphics[width=\figwidthS]{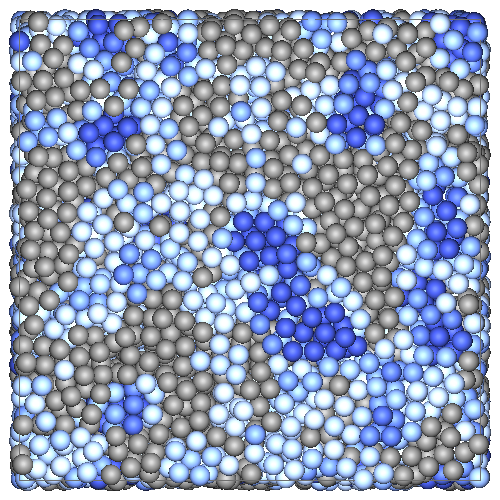}
	& \includegraphics[width=\figwidthS]{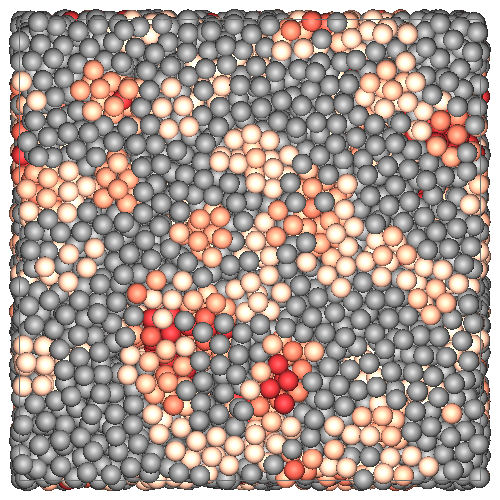}\\[-0.2cm]
	\includegraphics[width=\figwidthL]{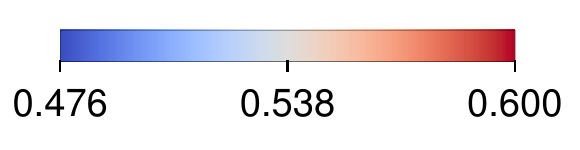}
	& \includegraphics[width=\figwidthL]{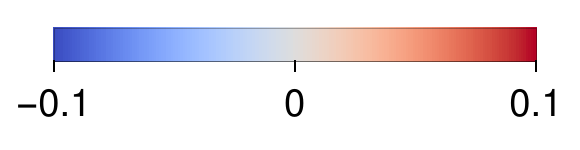}
	& \includegraphics[width=\figwidthL]{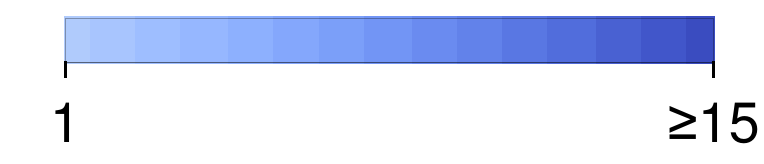}
	& \includegraphics[width=\figwidthL]{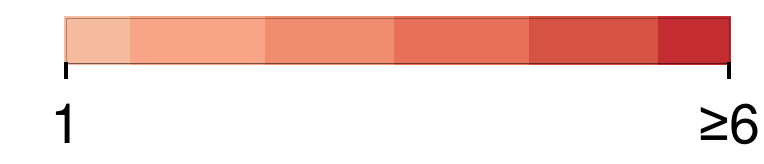} 
	\end{tabular}
	\caption[width=1\linewidth]{\label{fig:fluidsnap} Snapshot of the metastable fluid of hard spheres ($\eta=0.5385$) colored by a) the local packing fraction, b) the first principal component, c) the number of 9B clusters per particle, and d) the number of 11F clusters per particle. 
	}
\end{figure*}

Next, we characterize the local ordering of the metastable fluid on the  single-particle level. As explained in the Methods, we train a PCA model using only configurations of the metastable fluid. In all cases the first principal component (PC1) explains around 70\% of the total variance of the input. To illustrate what kind of fluctuations PCA picks up in the fluid, Figs. \ref{fig:fluidpca}a,c) show for the metastable fluids of hard spheres and of soft spheres the weight of each BOP in the first and second principal component. We see that PC1 is mostly made up of $\bar{q}_6$ and $\bar{q}_8$. Furthermore, Figs. \ref{fig:fluidpca}b,d) show the distribution of these metastable fluid particles in the PC1-PC2 plane, as well as the distributions of the corresponding FCC and BCC phase. (Recall that the data for the crystalline particles was not used in training the PCA models.) In this scatter plot we can neatly see that the crystal phases lie in the region of large PC1. 
To get a better understanding of the real-space distribution of these particles with above or below average PC1, we take a look at a single snapshot of the metastable fluid of hard spheres and color the particles according to their local packing fraction, PC1, and the number of 9B and 11F clusters a particles is involved in, see Fig. \ref{fig:fluidsnap}. 
Even though the spatial correlations in the local packing fraction are not clearly visible, we can clearly distinguish by eye large spatial regions of above or below average PC1. The autocorrelation functions of these spatial correlations can be found in the SM. 
Notice that the regions with above average PC1 correspond to an absence of 9B clusters and a high presence of 11F clusters, while regions with below average PC1 correspond to a high presence of 9B clusters and an absence of 11F clusters. 
Thus, there is a negative correlation between PC1 and 9B clusters and positive correlation between PC1 and 11F. The precise correlations of these two and other TCC clusters with PC1 are shown in Fig. \ref{fig:fluidpca}e). Analogous to what was found in Ref. \onlinecite{boattini2020autonomously}, the TCC clusters can be roughly divided into two groups: those with a negative correlation, which essentially are all clusters consisting of one or more tetrahedral subclusters, and those with a positive correlation, which contain the clusters consisting of one or more square pyramidal subclusters.

\begin{figure*}[t!]
\begin{tabular}{lll}
     a) & \hspace{0.5cm} & b)  \\[-0.4cm]
     \includegraphics[width=\figwidthB]{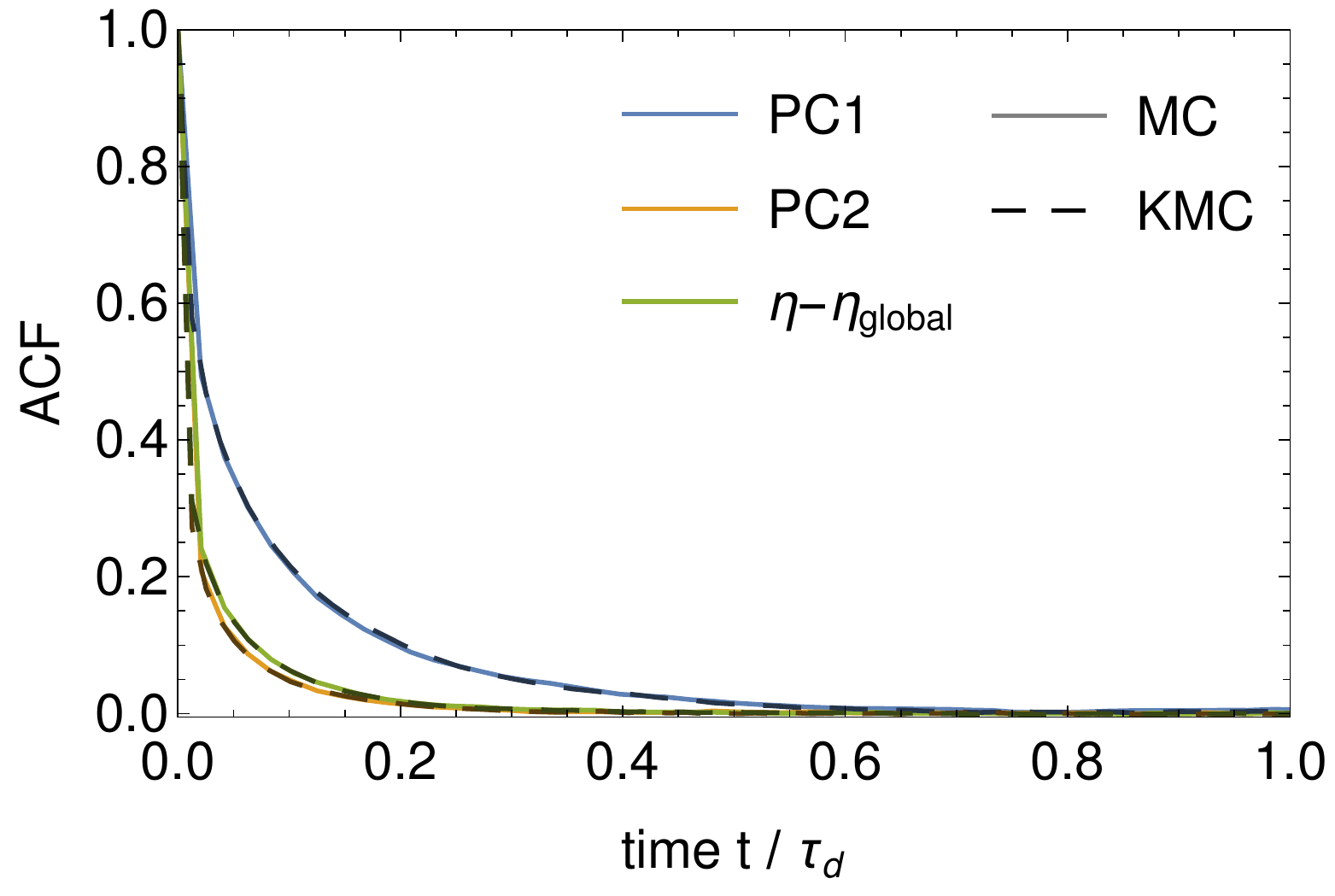} & & \includegraphics[width=\figwidthB]{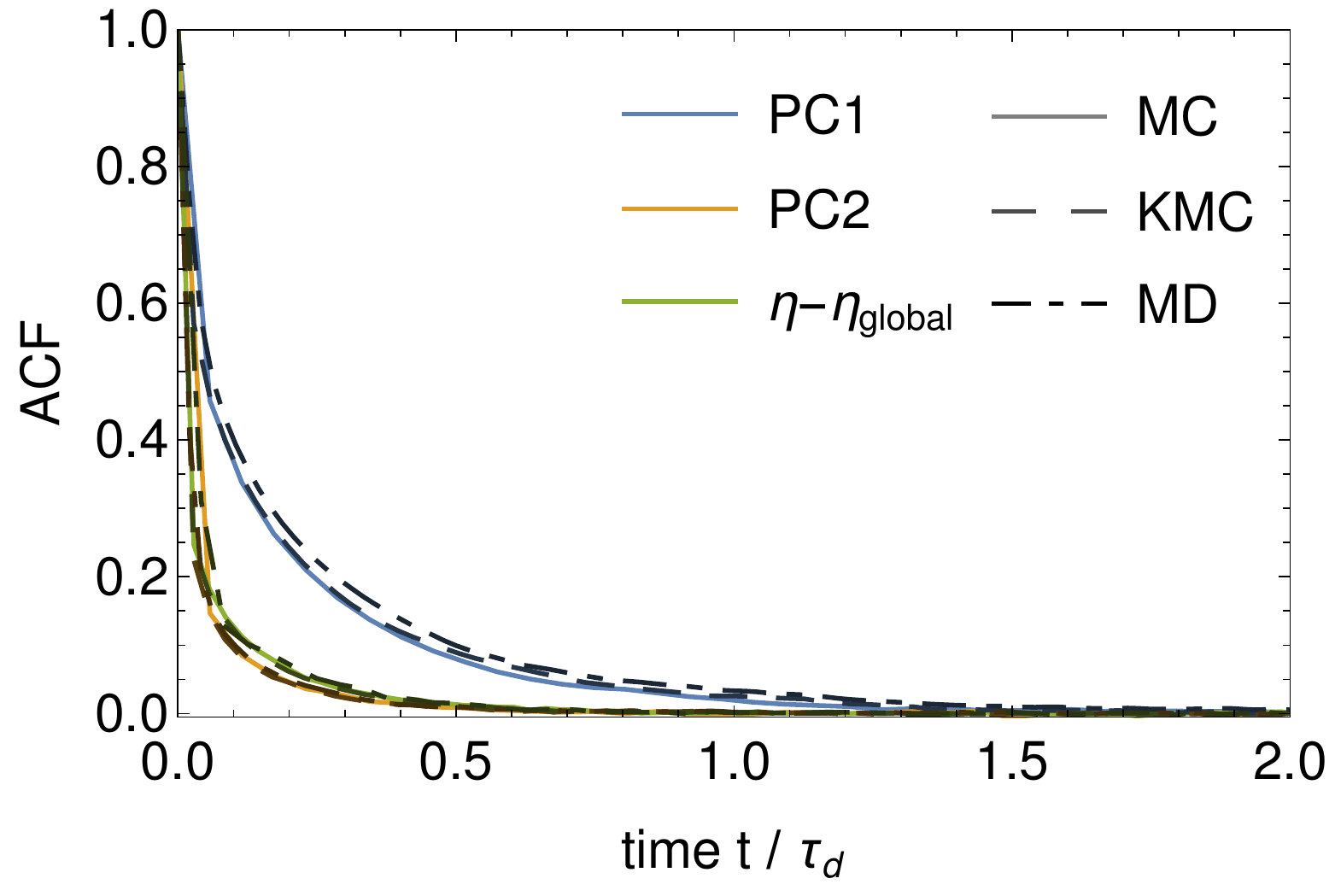}
\end{tabular}
    \caption[width=1\linewidth]{\label{fig:acfpca} Autocorrelation function of the first and second principal components and the local packing fraction in the metastable fluids of a) hard spheres ($\eta=0.5385$) and b) soft hard-core Yukawa particles ($\beta\epsilon=81$, $1/\kappa\sigma=0.40$, $\eta=0.1305$). The different dashing and darkness of the color indicate the simulation method, and the time is in terms of the long-time diffusion time $\tau_d$. }
\end{figure*}

\newcommand{\figwidthSS}{0.24\linewidth}
\begin{figure*}[t!]
\begin{tabular}{llll}
	a) $\;\quad(t-t_0)/\tau_d=-0.18$ & b) $\;\quad(t-t_0)/\tau_d=0.40$ & c) $\;\quad(t-t_0)/\tau_d=0.96$ & d) $\;\quad(t-t_0)/\tau_d=1.43$ \\
	\includegraphics[width=\figwidthSS]{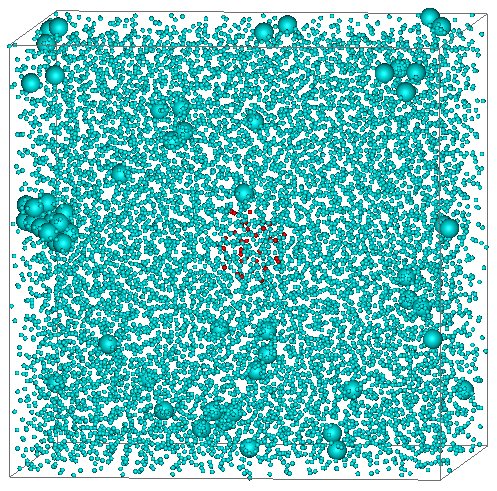}
	& \includegraphics[width=\figwidthSS]{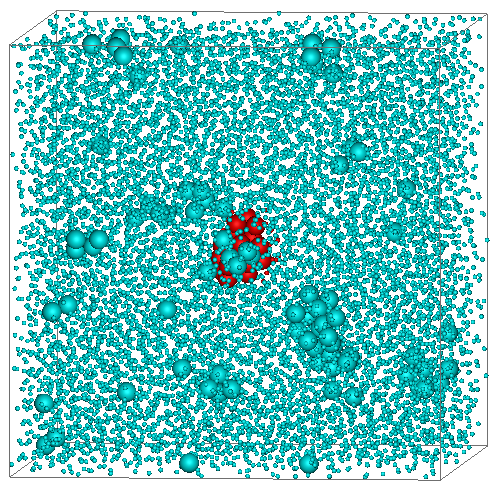}
	& \includegraphics[width=\figwidthSS]{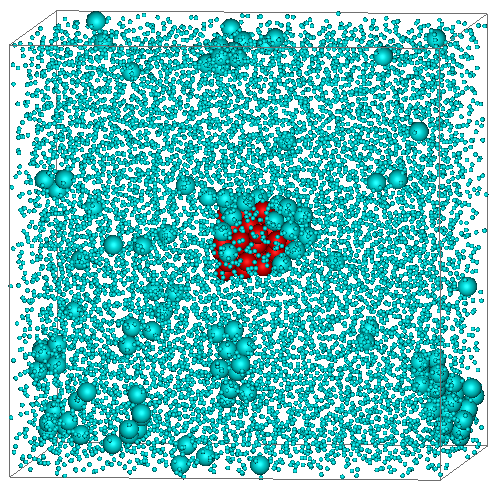}
	& \includegraphics[width=\figwidthSS]{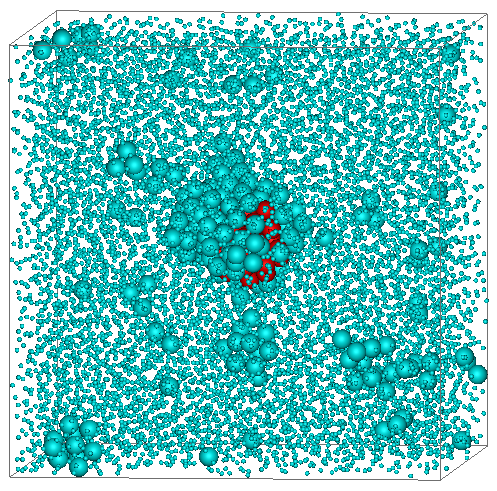}
	\end{tabular}
	\caption[width=1\linewidth]{\label{fig:selectregion} Four snapshots of a typical nucleation event of hard spheres ($\eta=0.5385$). Here $t_0$ indicates the start of nucleation. Fluid particles are displayed at a quarter of their actual size to make the nucleus visible, and red indicates the particles inside the studied region, i.e. those inside the sphere of radius $R$ around the center of nucleation $\mathbf{r}_0$. 
	}
\end{figure*}

Combining the observation of large spatial regions of above average PC1 with the proximity of these particles to the crystal phases in the PC1-PC2 scatter plot (Fig. \ref{fig:fluidpca}), we can conclude that regions of above average PC1 form a good candidate for harboring a precursor for crystal nucleation. 
In the next section, we will investigate these regions whilst tracking the nucleation events. However, before we turn our attention to that, we need to determine the temporal correlations of the local structure such that we know the time window before the start of nucleation during which we can search for a precursor.
Figure \ref{fig:acfpca} shows, for multiple simulation methods, the autocorrelation functions (ACFs) of the first two principal components and the local packing fraction in the metastable fluids of hard spheres and of soft spheres.
Here, we give the time in terms of the long-time diffusion time $\tau_d=\sigma^2/6D_l$, where $D_l$ is the long-time diffusion coefficient obtained from the mean-squared displacement. Notice that the ACFs are essentially independent of the simulation method, which confirms that the dynamics are also independent of the choice of simulation method. We see in both systems that PC2 and the local packing fraction decay extremely fast in time. Although PC1 decays more slowly, i.e. within half a diffusion time for  the hard spheres and one diffusion time for the soft spheres, this is still relatively fast. The decay time of PC1 provides a good estimate for the time window before the start of nucleation in which we can search for a precursor.

\begin{figure*}[t!]
\begin{tabular}{lll}
     a) & \hspace{0.5cm} & b)  \\[-0.6cm]
     \includegraphics[width=\figwidthC]{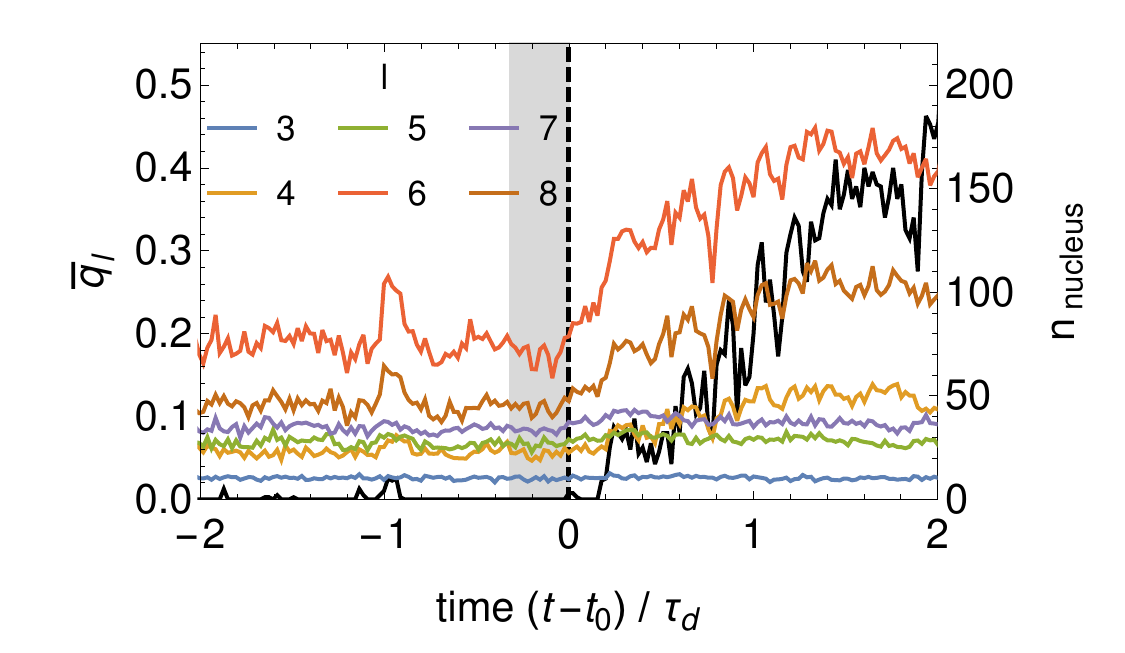} & & \includegraphics[width=\figwidthC]{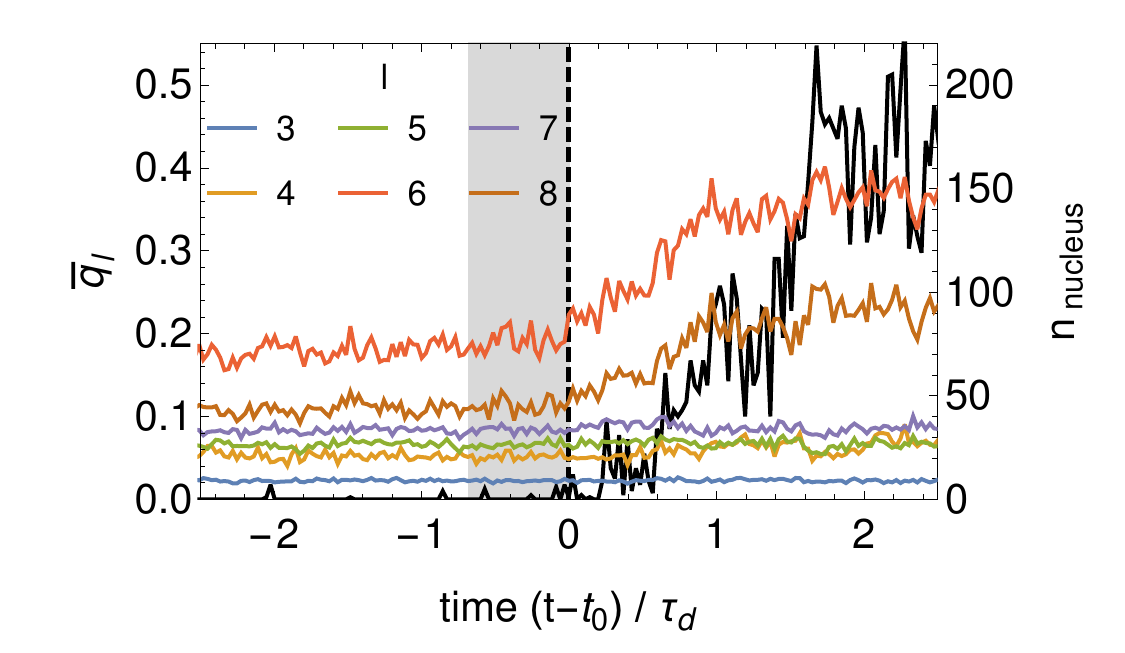} \\
     c) & \hspace{0.5cm} & d)  \\[-0.6cm]
     \includegraphics[width=\figwidthC]{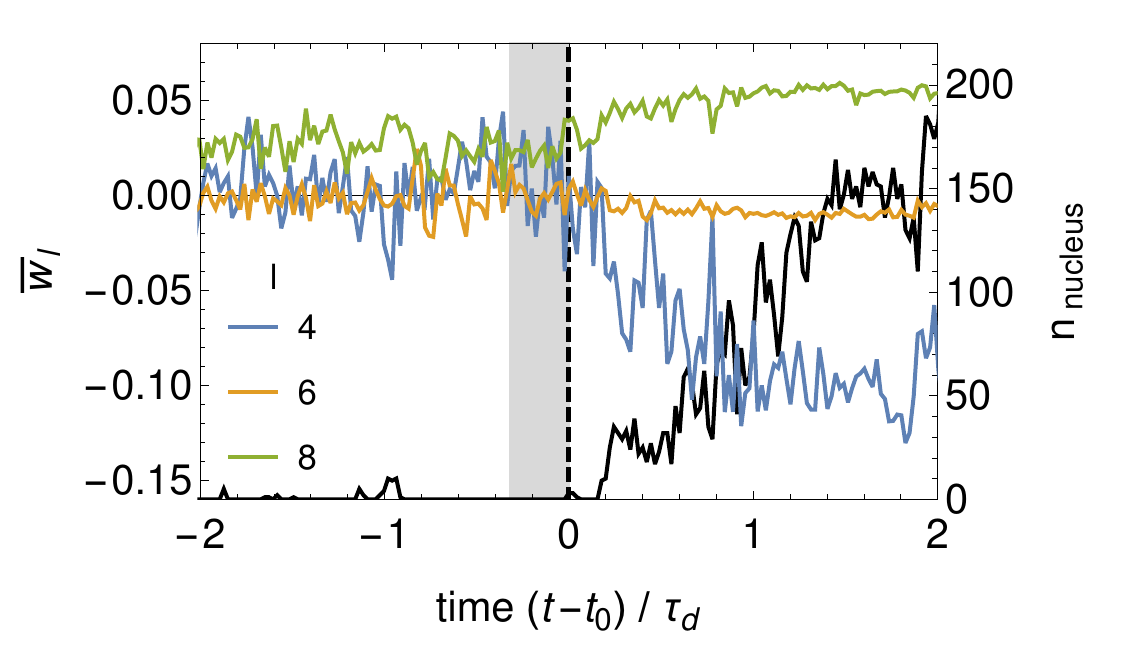} & & \includegraphics[width=\figwidthC]{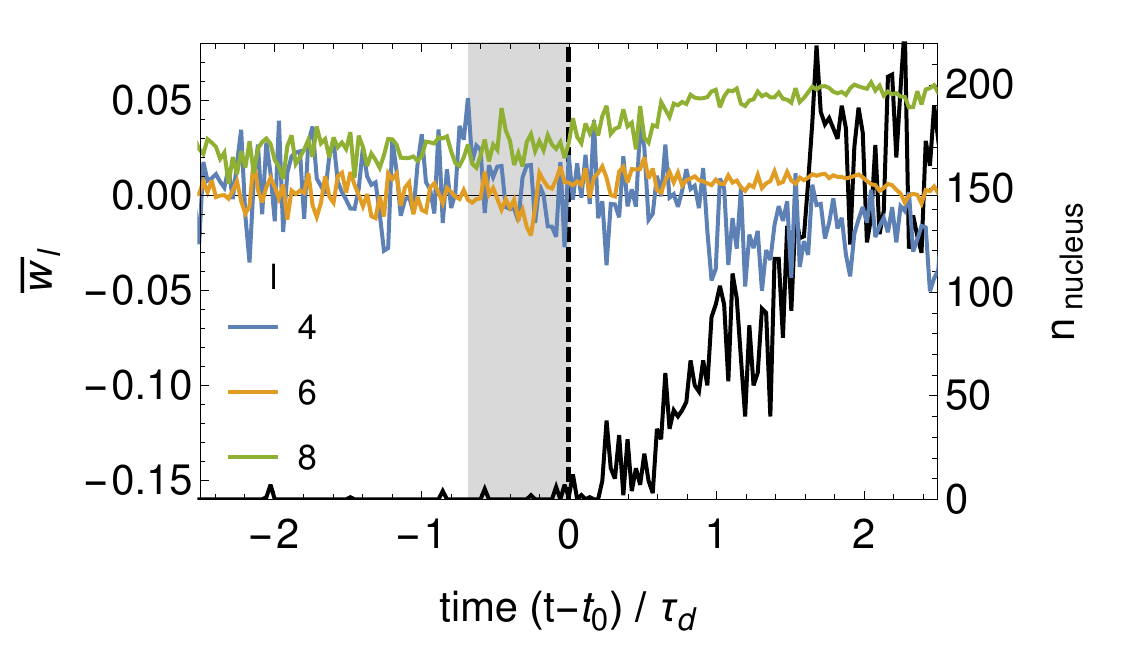} \\
     e) & \hspace{0.5cm} & f)  \\[-0.6cm]
     \includegraphics[width=\figwidthC]{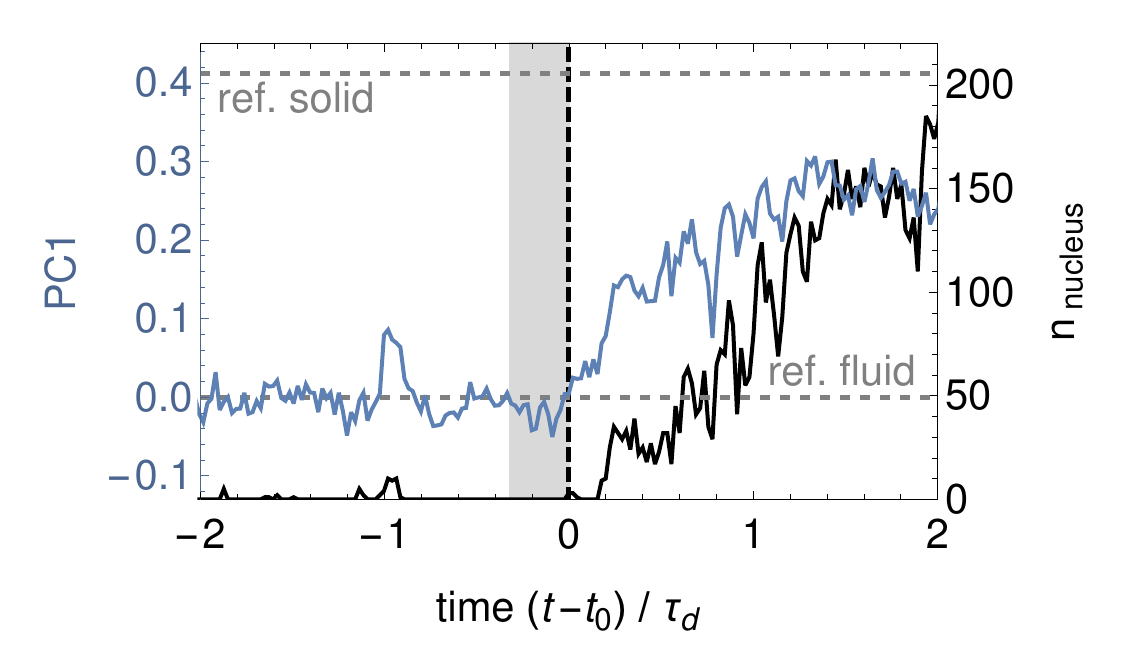} & & \includegraphics[width=\figwidthC]{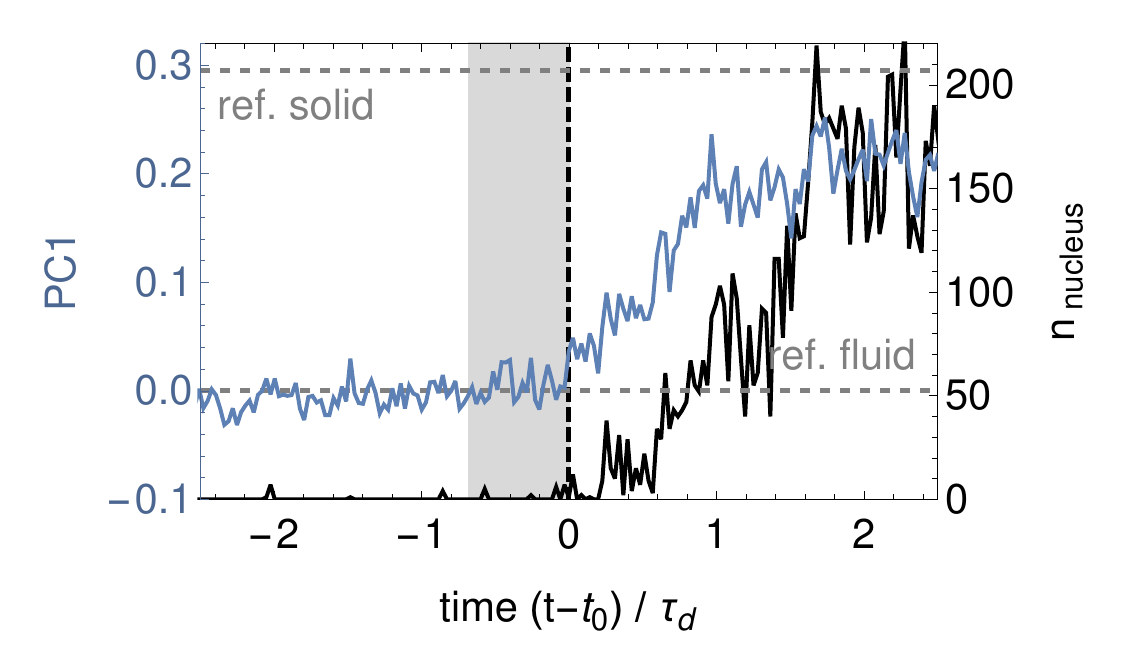} \\
     g) & \hspace{0.5cm} & h)  \\[-0.6cm]
     \includegraphics[width=\figwidthC]{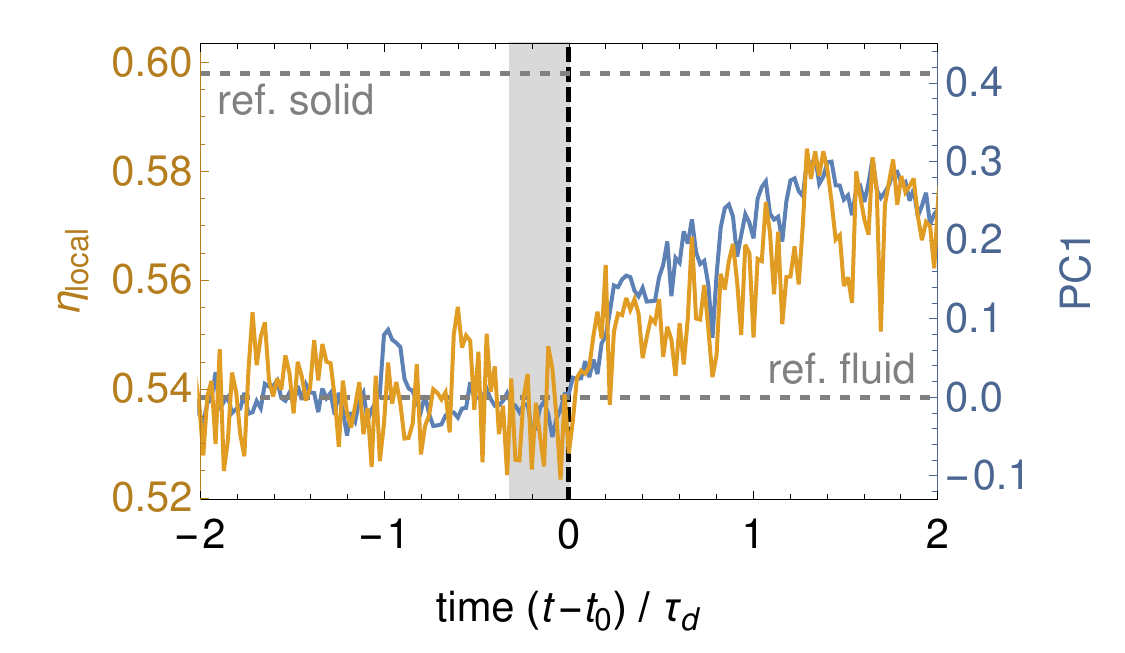} & & \includegraphics[width=\figwidthC]{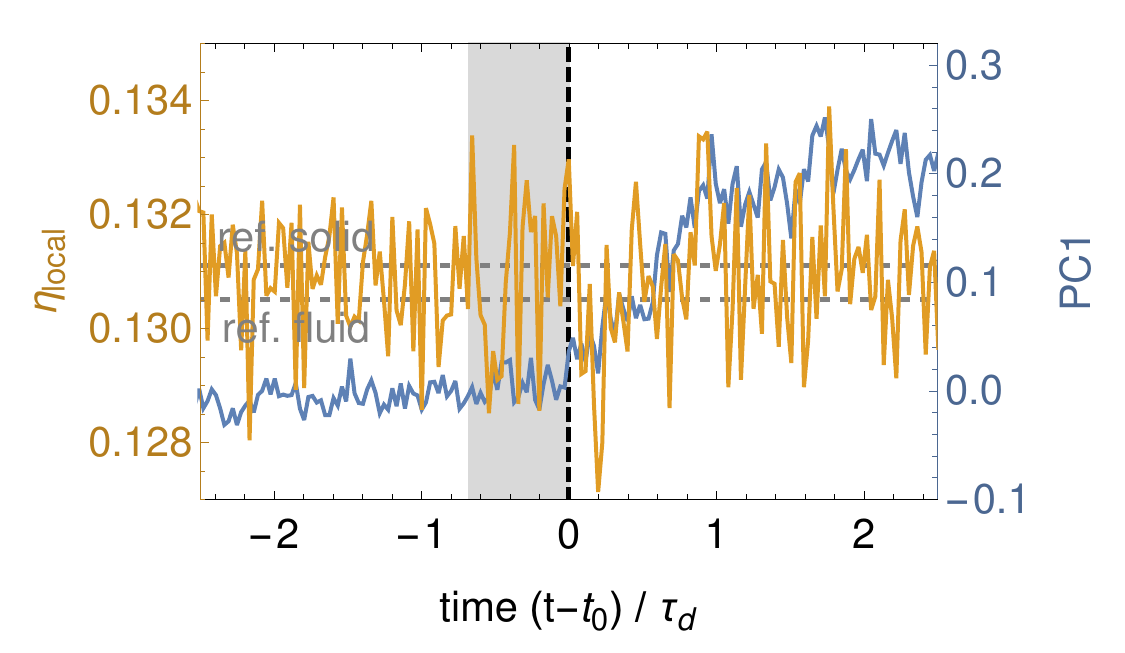}
\end{tabular}
    \caption[width=1\linewidth]{\label{fig:mainevents} Left: typical nucleation event of hard spheres ($\eta=0.5385$), same as in Fig. \ref{fig:selectregion}. Right: typical nucleation event of soft hard-core Yukawa particles ($\beta\epsilon=81$, $1/\kappa\sigma=0.40$, $\eta=0.1305$). Both events were obtained using MC simulations. The vertical dashed line in each figure indicates the start of nucleation $t_0$, and the shaded area indicates the time window before $t_0$ for which the ACF of PC1 $>0.05$ (see Fig. \ref{fig:acfpca}). In a-f) the black line (right axis) gives the size of the biggest nucleus present in the studied region. The other lines give the average value of a-b) $\bar{q}_l$ for $l\in[3,8]$, c-d) $\bar{w}_l$ for $l\in[4,6,8]$, e-f) PC1. In g-h) the blue line (right axis) gives PC1, while the yellow line (left axis) gives the local packing fraction $\eta_\text{local}$. Note that in e-g) the horizontal dashed lines give the reference value of PC1 in the fluid and solid phase. In g) the right axis is scaled in such a way that the reference values in the fluid and solid of PC1 and $\eta_\text{local}$ lie on top of each other. In h) the horizontal dashed lines give the reference value of $\eta_\text{local}$ in the fluid and solid phase. }
\end{figure*}

\begin{figure*}[t!]
\begin{tabular}{lll}
     a) & \hspace{0.5cm} & b)  \\[-0.6cm]
     \includegraphics[width=\figwidthC]{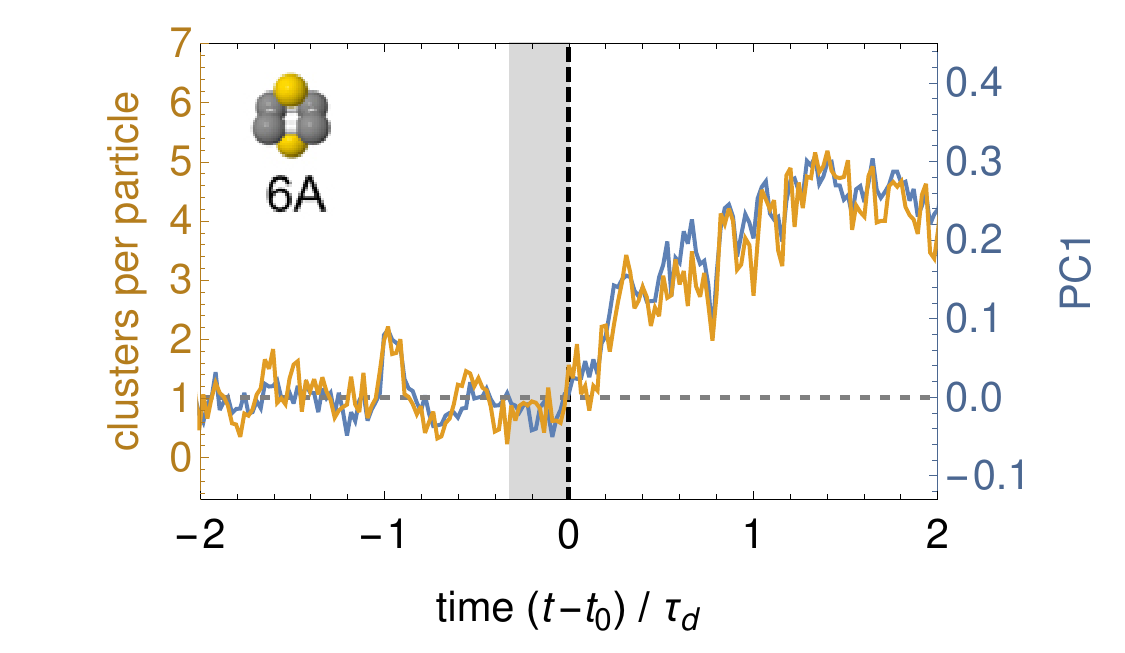} & & \includegraphics[width=\figwidthC]{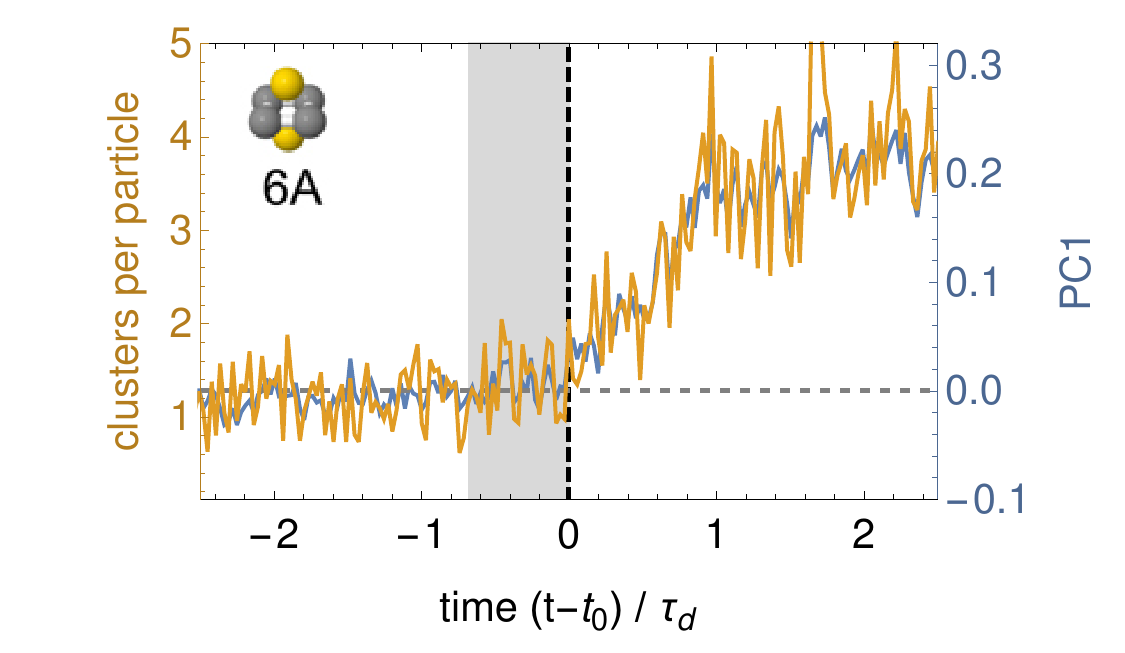} \\
     c) & \hspace{0.5cm} & d)  \\[-0.6cm]
     \includegraphics[width=\figwidthC]{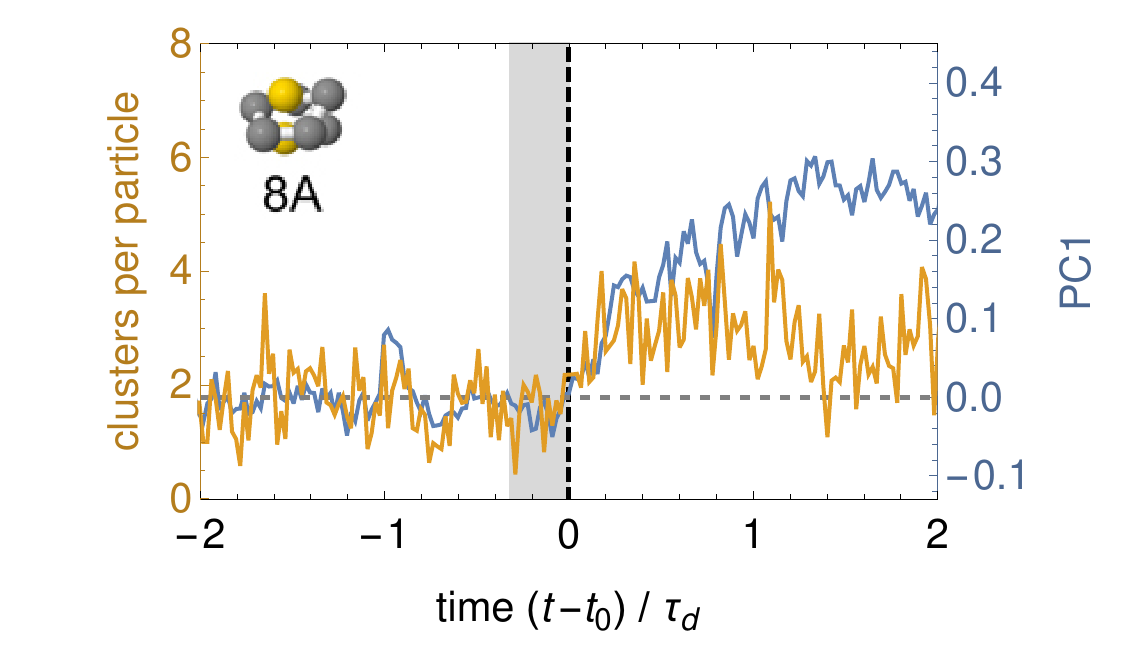} & & \includegraphics[width=\figwidthC]{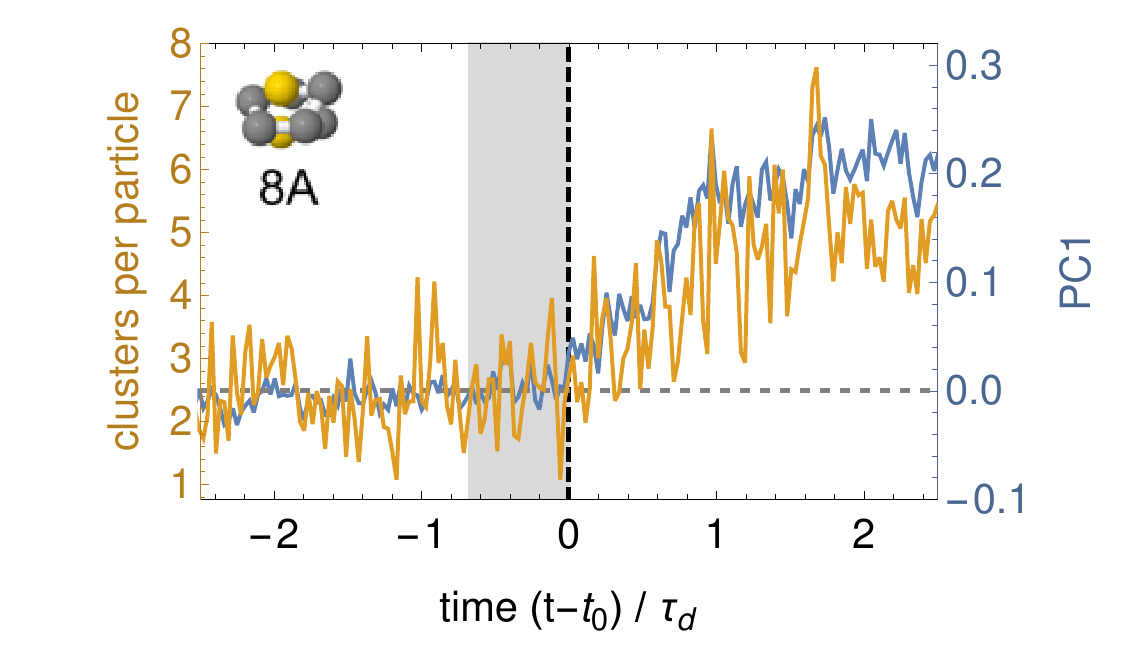} \\
     e) & \hspace{0.5cm} & f)  \\[-0.6cm]
     \includegraphics[width=\figwidthC]{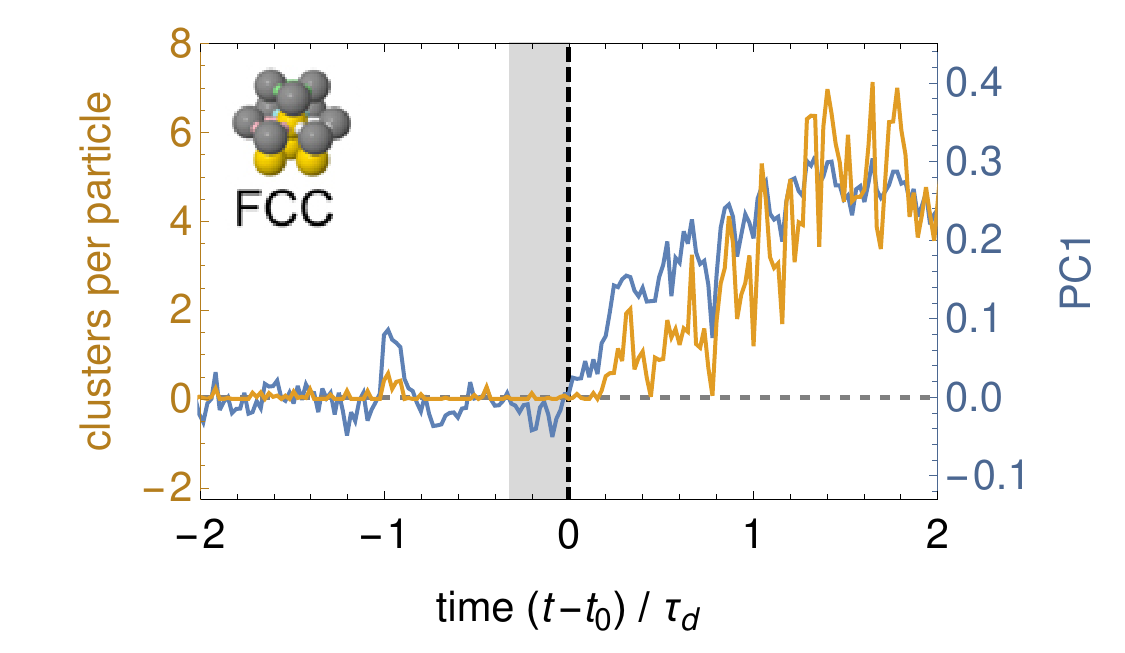} & & \includegraphics[width=\figwidthC]{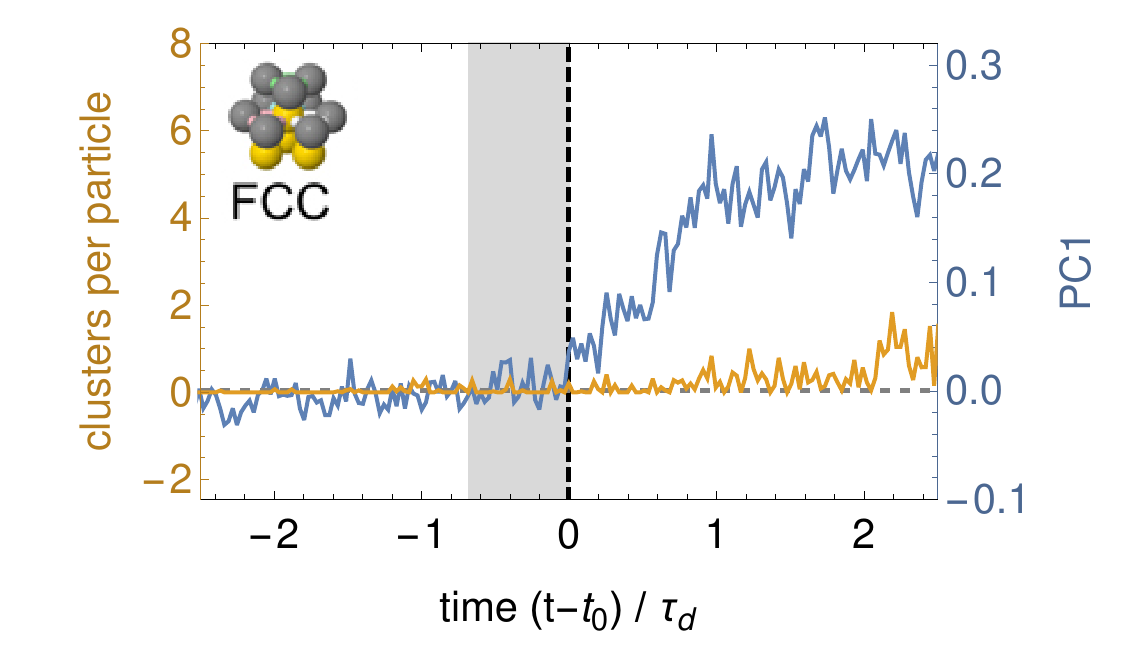} \\
     g) & \hspace{0.5cm} & h)  \\[-0.6cm]
     \includegraphics[width=\figwidthC]{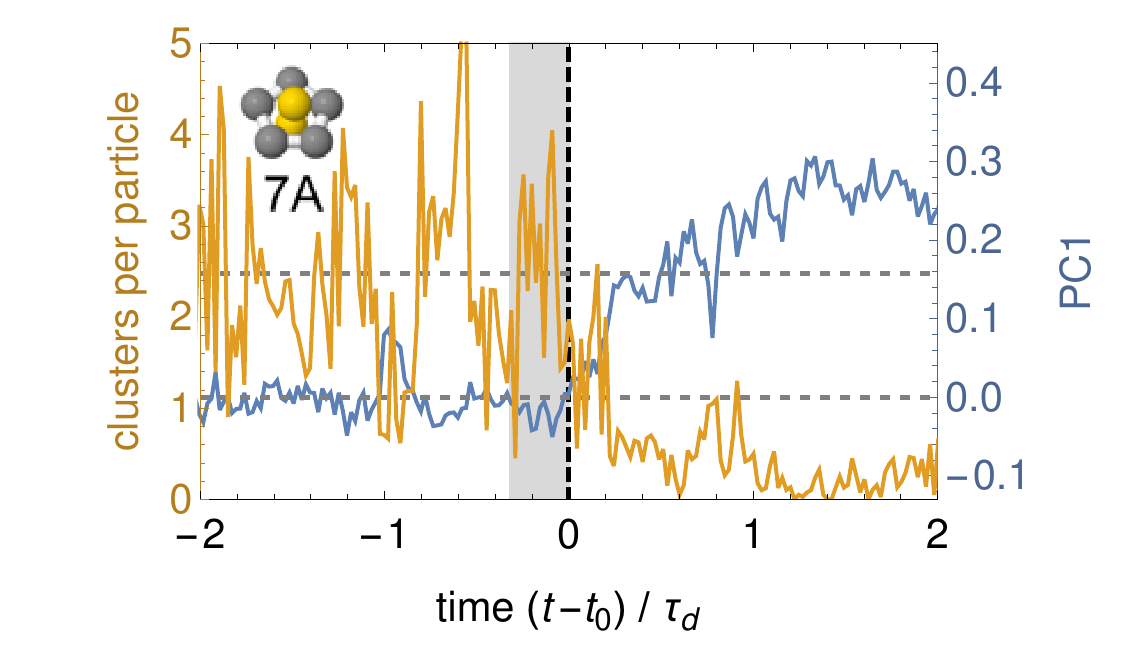} & & \includegraphics[width=\figwidthC]{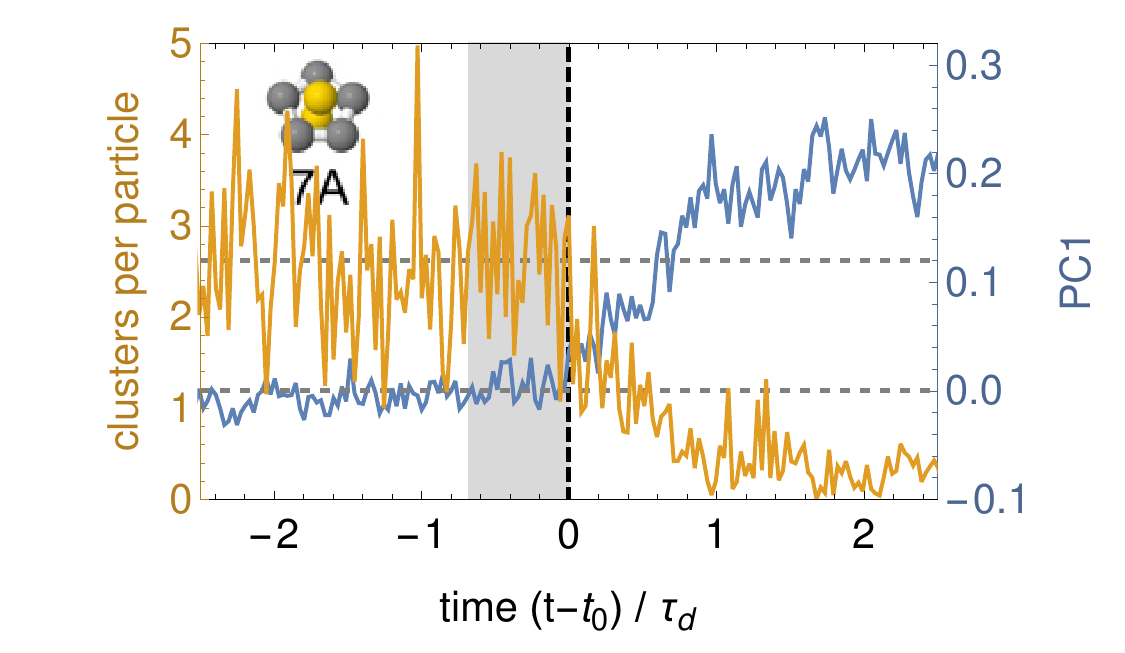}
\end{tabular}
    \caption[width=1\linewidth]{\label{fig:maineventstcc} For the same events as in Fig. \ref{fig:mainevents}, i.e. hard spheres (left) and soft hard-core Yukawa particles (right), the average number of clusters per particle (left axis, yellow) for a couple of TCC clusters together with the average value of PC1 (right axis, blue). The horizontal dashed lines indicate the reference values of the number of clusters per particle and PC1 in the fluid. In a-f) the left axis is scaled in such a way that these lines lie on top of each other. In g-h) this was not possible without inverting one of the axes. 
    }
\end{figure*}

\subsection{Nucleation study}

We now turn our attention to crystal nucleation. As explained in the methods, we simulate numerous spontaneous nucleation events using (K)MC and MD simulations, and track for all these events the local properties of the region where nucleation starts. Here, we discuss our observations using two typical nucleation events: one of the hard-spheres system and one of soft spheres ($\beta\epsilon=81$, $1/\kappa\sigma=0.40$). Both these nucleation events were obtained using MC simulations. More nucleation events, where we either used other simulation methods or studied the other systems mentioned in Tab. \ref{tab:info}, can be found in the SM. 
To better illustrate which region we study while tracking a nucleation event, Fig. \ref{fig:selectregion} shows a couple of snapshots of the nucleation event of hard spheres where the particles inside the studied region are colored red. Figure \ref{fig:mainevents} shows for this event and the nucleation event of soft spheres the average properties of the particles in this studied region. 
Before we discuss what we see, let us again point out that the nucleus size (black line) is not an ideal order parameter for tracking nucleation since its binary nature causes it to overlook subtle increases in the local structural ordering at the onset of nucleation. It does, however, provide a general overview of the nucleation event, such as when the nucleus reaches its critical size (see Tab. \ref{tab:info}). 
That being said, let us first discuss the BOPs of the studied region. We observe no notable change in the behavior of the BOPs before the start of nucleation, but, as soon as nucleation starts, we see a sharp increase in the values of $\bar{q}_6$ and $\bar{q}_8$ for both systems. Furthermore, for the hard spheres, we see that, once nucleation starts, $\bar{q}_4$ increases, $\bar{w}_6$ stays negative, and $\bar{w}_4$ decreases. This all indicates that indeed the FCC phase nucleates. On the other hand, for the soft spheres, we see that as nucleation starts $\bar{q}_4$ barely increases, $\bar{w}_6$ stays positive, and $\bar{w}_4$ keeps fluctuating around zero, which all indicates that indeed the BCC phase nucleates.
Similar to the behavior of the BOPs, we observe no notable change in the behavior of PC1 before the start of nucleation, but see a sharp increase in its value once nucleation starts. Note that this increase in PC1 is visible before the number of particles classified as crystalline starts to rise (black line), demonstrating that PC1 is a better order parameter for tracking the start of nucleation than the nucleus size according to our definition. 
Lastly, for the hard spheres, we see that the local packing fraction increases simultaneously with PC1 as soon as nucleation starts, and that no notable behavior can be observed before the start of nucleation. This strongly indicates that increase in structural ordering and local density go together and, thus, that there is no apparent precursor.
Unfortunately, as the difference between the packing fraction of the fluid and solid phases is extremely small for the soft spheres, i.e. less than 0.001, it is not possible to observe any increase in the local packing fraction on top of the normal fluctuations. Hence, we cannot draw any conclusions on the local packing fraction of the soft spheres.

Next, to show that we have missed no subtle changes in the local structure and thus confirm that there is no precursor for nucleation, we further examine the local structure using TCC. In Fig. \ref{fig:maineventstcc} we show for four of the most relevant TCC clusters the average number of clusters a particle is involved in and compare it with PC1. 
These four clusters are: i) 6A, which has the strongest positive correlation with PC1 and is present in both bulk FCC and bulk BCC, ii) 8A, which has the second strongest positive correlation with PC1 and is present in bulk BCC but not bulk FCC, iii) FCC, which is present in bulk FCC but not bulk BCC, and iv) 7A, which has the strongest negative correlation with PC1 and can neither be found in bulk FCC nor bulk BCC.
Similar figures of other TCC clusters can be found in the SM.
For all clusters we see that there is no significant change prior to the start of nucleation. Furthermore, we see that the trends of the 6A cluster coincide almost perfectly with those of PC1. 
Similarly, we see that the trends of the 8A cluster closely follow the trends of PC1. 
However, for hard spheres the initial increase in 8A clusters is followed by a decrease. Since 8A is a cluster that is usually found in bulk BCC and not in bulk FCC, this initial increase might be surprising. This can be explained via the observation that 8A clusters are found in high concentrations near the surface of growing nuclei\cite{gispen2023crystal}. As a result, the number of these clusters decreases once the nucleus grows beyond our averaging radius.
For the FCC cluster, we observe a sharp increase during the nucleation of hard spheres. Notice, however, that this increase starts slightly later than the increase in PC1. This is not surprising as the FCC cluster is a relatively large cluster, i.e. it contains 13 particles, and consequently is not present in the first stages of nucleation. For the soft spheres there is no significant increase in FCC clusters, as expected. 
Lastly, we take a look at the 7A cluster. In contrast to the other three clusters, this five-fold symmetric cluster has a strong negative correlation with PC1. Moreover, it is strongly present in the metastable fluid phases, whereas its presence in the FCC and BCC phases is negligible.  It is, therefore, not surprising that we observe an immediate and sharp decrease in 7A clusters as soon as nucleation starts.


\section{Conclusions}
To conclude, we have characterized the local structure of various metastable fluids of charged colloids using multiple methods: bond-orientational order parameters (BOPs), principal component analysis (PCA) on the BOPs, and topological cluster classification (TCC). In doing this have attempted to avoid artefacts due to biases in our chosen order parameters. For all systems we have found that any local structural ordering has a relatively short lifetime, resulting in a short time window prior to the start of nucleation in which a precursor could exist. By tracking the local structure of the spatial region coinciding with the birthplace of the crystal nucleus, we show that inside this time window no atypical behavior in the local structural order is observed using any of our structural order parameters. Furthermore, we demonstrate that all structural characteristics that differ significantly between the fluid and crystal phase start changing simultaneously as soon as nucleation starts. Specifically in the case of FCC, this includes the local density, which starts growing immediately as soon as structural order emerges.
We, thus, conclude that we find no evidence for a precursor for the crystal nucleation of hard and charged colloids.


\section{Acknowledgements}
L.F. and M.d.J. acknowledge funding from the Vidi research program with project number VI.VIDI.192.102 which is financed by the Dutch Research Council (NWO).



\bibliography{paper}

\begin{thebibliography}{41}%
\makeatletter
\providecommand \@ifxundefined [1]{%
 \@ifx{#1\undefined}
}%
\providecommand \@ifnum [1]{%
 \ifnum #1\expandafter \@firstoftwo
 \else \expandafter \@secondoftwo
 \fi
}%
\providecommand \@ifx [1]{%
 \ifx #1\expandafter \@firstoftwo
 \else \expandafter \@secondoftwo
 \fi
}%
\providecommand \natexlab [1]{#1}%
\providecommand \enquote  [1]{``#1''}%
\providecommand \bibnamefont  [1]{#1}%
\providecommand \bibfnamefont [1]{#1}%
\providecommand \citenamefont [1]{#1}%
\providecommand \href@noop [0]{\@secondoftwo}%
\providecommand \href [0]{\begingroup \@sanitize@url \@href}%
\providecommand \@href[1]{\@@startlink{#1}\@@href}%
\providecommand \@@href[1]{\endgroup#1\@@endlink}%
\providecommand \@sanitize@url [0]{\catcode `\\12\catcode `\$12\catcode
  `\&12\catcode `\#12\catcode `\^12\catcode `\_12\catcode `\%12\relax}%
\providecommand \@@startlink[1]{}%
\providecommand \@@endlink[0]{}%
\providecommand \url  [0]{\begingroup\@sanitize@url \@url }%
\providecommand \@url [1]{\endgroup\@href {#1}{\urlprefix }}%
\providecommand \urlprefix  [0]{URL }%
\providecommand \Eprint [0]{\href }%
\providecommand \doibase [0]{http://dx.doi.org/}%
\providecommand \selectlanguage [0]{\@gobble}%
\providecommand \bibinfo  [0]{\@secondoftwo}%
\providecommand \bibfield  [0]{\@secondoftwo}%
\providecommand \translation [1]{[#1]}%
\providecommand \BibitemOpen [0]{}%
\providecommand \bibitemStop [0]{}%
\providecommand \bibitemNoStop [0]{.\EOS\space}%
\providecommand \EOS [0]{\spacefactor3000\relax}%
\providecommand \BibitemShut  [1]{\csname bibitem#1\endcsname}%
\let\auto@bib@innerbib\@empty
\bibitem [{\citenamefont {Karthika}, \citenamefont {Radhakrishnan},\ and\
  \citenamefont {Kalaichelvi}(2016)}]{karthika2016review}%
  \BibitemOpen
  \bibfield  {author} {\bibinfo {author} {\bibfnamefont {S.}~\bibnamefont
  {Karthika}}, \bibinfo {author} {\bibfnamefont {T.}~\bibnamefont
  {Radhakrishnan}}, \ and\ \bibinfo {author} {\bibfnamefont {P.}~\bibnamefont
  {Kalaichelvi}},\ }\href@noop {} {\bibfield  {journal} {\bibinfo  {journal}
  {Crystal Growth \& Design}\ }\textbf {\bibinfo {volume} {16}},\ \bibinfo
  {pages} {6663} (\bibinfo {year} {2016})}\BibitemShut {NoStop}%
\bibitem [{\citenamefont {Sosso}\ \emph {et~al.}(2016)\citenamefont {Sosso},
  \citenamefont {Chen}, \citenamefont {Cox}, \citenamefont {Fitzner},
  \citenamefont {Pedevilla}, \citenamefont {Zen},\ and\ \citenamefont
  {Michaelides}}]{sosso2016crystal}%
  \BibitemOpen
  \bibfield  {author} {\bibinfo {author} {\bibfnamefont {G.~C.}\ \bibnamefont
  {Sosso}}, \bibinfo {author} {\bibfnamefont {J.}~\bibnamefont {Chen}},
  \bibinfo {author} {\bibfnamefont {S.~J.}\ \bibnamefont {Cox}}, \bibinfo
  {author} {\bibfnamefont {M.}~\bibnamefont {Fitzner}}, \bibinfo {author}
  {\bibfnamefont {P.}~\bibnamefont {Pedevilla}}, \bibinfo {author}
  {\bibfnamefont {A.}~\bibnamefont {Zen}}, \ and\ \bibinfo {author}
  {\bibfnamefont {A.}~\bibnamefont {Michaelides}},\ }\href@noop {} {\bibfield
  {journal} {\bibinfo  {journal} {Chemical reviews}\ }\textbf {\bibinfo
  {volume} {116}},\ \bibinfo {pages} {7078} (\bibinfo {year}
  {2016})}\BibitemShut {NoStop}%
\bibitem [{\citenamefont {Ostwald}(1897)}]{ostwald1897studien}%
  \BibitemOpen
  \bibfield  {author} {\bibinfo {author} {\bibfnamefont {W.}~\bibnamefont
  {Ostwald}},\ }\href@noop {} {\bibfield  {journal} {\bibinfo  {journal}
  {Zeitschrift f{\"u}r physikalische Chemie}\ }\textbf {\bibinfo {volume}
  {22}},\ \bibinfo {pages} {289} (\bibinfo {year} {1897})}\BibitemShut
  {NoStop}%
\bibitem [{\citenamefont {Alexander}\ and\ \citenamefont
  {McTague}(1978)}]{alexander1978should}%
  \BibitemOpen
  \bibfield  {author} {\bibinfo {author} {\bibfnamefont {S.}~\bibnamefont
  {Alexander}}\ and\ \bibinfo {author} {\bibfnamefont {J.}~\bibnamefont
  {McTague}},\ }\href@noop {} {\bibfield  {journal} {\bibinfo  {journal} {Phys.
  Rev. Lett.}\ }\textbf {\bibinfo {volume} {41}},\ \bibinfo {pages} {702}
  (\bibinfo {year} {1978})}\BibitemShut {NoStop}%
\bibitem [{\citenamefont {Russo}\ and\ \citenamefont
  {Tanaka}(2012{\natexlab{a}})}]{russo2012selection}%
  \BibitemOpen
  \bibfield  {author} {\bibinfo {author} {\bibfnamefont {J.}~\bibnamefont
  {Russo}}\ and\ \bibinfo {author} {\bibfnamefont {H.}~\bibnamefont {Tanaka}},\
  }\href@noop {} {\bibfield  {journal} {\bibinfo  {journal} {Soft Matter}\
  }\textbf {\bibinfo {volume} {8}},\ \bibinfo {pages} {4206} (\bibinfo {year}
  {2012}{\natexlab{a}})}\BibitemShut {NoStop}%
\bibitem [{\citenamefont {Taffs}\ and\ \citenamefont
  {Patrick~Royall}(2016)}]{taffs2016role}%
  \BibitemOpen
  \bibfield  {author} {\bibinfo {author} {\bibfnamefont {J.}~\bibnamefont
  {Taffs}}\ and\ \bibinfo {author} {\bibfnamefont {C.}~\bibnamefont
  {Patrick~Royall}},\ }\href@noop {} {\bibfield  {journal} {\bibinfo  {journal}
  {Nat. Commun.}\ }\textbf {\bibinfo {volume} {7}},\ \bibinfo {pages} {13225}
  (\bibinfo {year} {2016})}\BibitemShut {NoStop}%
\bibitem [{\citenamefont {Ouyang}\ \emph {et~al.}(2016)\citenamefont {Ouyang},
  \citenamefont {Fu}, \citenamefont {Sun},\ and\ \citenamefont
  {Xu}}]{ouyang2016polymorph}%
  \BibitemOpen
  \bibfield  {author} {\bibinfo {author} {\bibfnamefont {W.}~\bibnamefont
  {Ouyang}}, \bibinfo {author} {\bibfnamefont {C.}~\bibnamefont {Fu}}, \bibinfo
  {author} {\bibfnamefont {Z.}~\bibnamefont {Sun}}, \ and\ \bibinfo {author}
  {\bibfnamefont {S.}~\bibnamefont {Xu}},\ }\href@noop {} {\bibfield  {journal}
  {\bibinfo  {journal} {Phys. Rev. E}\ }\textbf {\bibinfo {volume} {94}},\
  \bibinfo {pages} {042805} (\bibinfo {year} {2016})}\BibitemShut {NoStop}%
\bibitem [{\citenamefont {Russo}\ and\ \citenamefont
  {Tanaka}(2012{\natexlab{b}})}]{russo2012microscopic}%
  \BibitemOpen
  \bibfield  {author} {\bibinfo {author} {\bibfnamefont {J.}~\bibnamefont
  {Russo}}\ and\ \bibinfo {author} {\bibfnamefont {H.}~\bibnamefont {Tanaka}},\
  }\href@noop {} {\bibfield  {journal} {\bibinfo  {journal} {Sci. Rep.}\
  }\textbf {\bibinfo {volume} {2}},\ \bibinfo {pages} {1} (\bibinfo {year}
  {2012}{\natexlab{b}})}\BibitemShut {NoStop}%
\bibitem [{\citenamefont {Gispen}\ \emph {et~al.}(2023)\citenamefont {Gispen},
  \citenamefont {Coli}, \citenamefont {van Damme}, \citenamefont {Royall},\
  and\ \citenamefont {Dijkstra}}]{gispen2023crystal}%
  \BibitemOpen
  \bibfield  {author} {\bibinfo {author} {\bibfnamefont {W.}~\bibnamefont
  {Gispen}}, \bibinfo {author} {\bibfnamefont {G.~M.}\ \bibnamefont {Coli}},
  \bibinfo {author} {\bibfnamefont {R.}~\bibnamefont {van Damme}}, \bibinfo
  {author} {\bibfnamefont {C.~P.}\ \bibnamefont {Royall}}, \ and\ \bibinfo
  {author} {\bibfnamefont {M.}~\bibnamefont {Dijkstra}},\ }\href@noop {}
  {\bibfield  {journal} {\bibinfo  {journal} {ACS Nano}\ }\textbf {\bibinfo
  {volume} {17}},\ \bibinfo {pages} {8807–8814} (\bibinfo {year}
  {2023})}\BibitemShut {NoStop}%
\bibitem [{\citenamefont {ten Wolde}\ and\ \citenamefont
  {Frenkel}(1999)}]{ten1999homogeneous}%
  \BibitemOpen
  \bibfield  {author} {\bibinfo {author} {\bibfnamefont {P.~R.}\ \bibnamefont
  {ten Wolde}}\ and\ \bibinfo {author} {\bibfnamefont {D.}~\bibnamefont
  {Frenkel}},\ }\href@noop {} {\bibfield  {journal} {\bibinfo  {journal}
  {Physical Chemistry Chemical Physics}\ }\textbf {\bibinfo {volume} {1}},\
  \bibinfo {pages} {2191} (\bibinfo {year} {1999})}\BibitemShut {NoStop}%
\bibitem [{\citenamefont {Schilling}\ \emph {et~al.}(2010)\citenamefont
  {Schilling}, \citenamefont {Sch{\"o}pe}, \citenamefont {Oettel},
  \citenamefont {Opletal},\ and\ \citenamefont
  {Snook}}]{schilling2010precursor}%
  \BibitemOpen
  \bibfield  {author} {\bibinfo {author} {\bibfnamefont {T.}~\bibnamefont
  {Schilling}}, \bibinfo {author} {\bibfnamefont {H.~J.}\ \bibnamefont
  {Sch{\"o}pe}}, \bibinfo {author} {\bibfnamefont {M.}~\bibnamefont {Oettel}},
  \bibinfo {author} {\bibfnamefont {G.}~\bibnamefont {Opletal}}, \ and\
  \bibinfo {author} {\bibfnamefont {I.}~\bibnamefont {Snook}},\ }\href@noop {}
  {\bibfield  {journal} {\bibinfo  {journal} {Phys. Rev. Lett.}\ }\textbf
  {\bibinfo {volume} {105}},\ \bibinfo {pages} {025701} (\bibinfo {year}
  {2010})}\BibitemShut {NoStop}%
\bibitem [{\citenamefont {Russo}\ and\ \citenamefont
  {Tanaka}(2016)}]{russo2016crystal}%
  \BibitemOpen
  \bibfield  {author} {\bibinfo {author} {\bibfnamefont {J.}~\bibnamefont
  {Russo}}\ and\ \bibinfo {author} {\bibfnamefont {H.}~\bibnamefont {Tanaka}},\
  }\href@noop {} {\bibfield  {journal} {\bibinfo  {journal} {J. Chem. Phys.}\
  }\textbf {\bibinfo {volume} {145}},\ \bibinfo {pages} {211801} (\bibinfo
  {year} {2016})}\BibitemShut {NoStop}%
\bibitem [{\citenamefont {Tan}, \citenamefont {Xu},\ and\ \citenamefont
  {Xu}(2014)}]{tan2014visualizing}%
  \BibitemOpen
  \bibfield  {author} {\bibinfo {author} {\bibfnamefont {P.}~\bibnamefont
  {Tan}}, \bibinfo {author} {\bibfnamefont {N.}~\bibnamefont {Xu}}, \ and\
  \bibinfo {author} {\bibfnamefont {L.}~\bibnamefont {Xu}},\ }\href@noop {}
  {\bibfield  {journal} {\bibinfo  {journal} {Nat. Phys.}\ }\textbf {\bibinfo
  {volume} {10}},\ \bibinfo {pages} {73} (\bibinfo {year} {2014})}\BibitemShut
  {NoStop}%
\bibitem [{\citenamefont {Li}\ \emph {et~al.}(2020)\citenamefont {Li},
  \citenamefont {Chen}, \citenamefont {Tanaka},\ and\ \citenamefont
  {Tan}}]{li2020revealing}%
  \BibitemOpen
  \bibfield  {author} {\bibinfo {author} {\bibfnamefont {M.}~\bibnamefont
  {Li}}, \bibinfo {author} {\bibfnamefont {Y.}~\bibnamefont {Chen}}, \bibinfo
  {author} {\bibfnamefont {H.}~\bibnamefont {Tanaka}}, \ and\ \bibinfo {author}
  {\bibfnamefont {P.}~\bibnamefont {Tan}},\ }\href@noop {} {\bibfield
  {journal} {\bibinfo  {journal} {Science advances}\ }\textbf {\bibinfo
  {volume} {6}},\ \bibinfo {pages} {eaaw8938} (\bibinfo {year}
  {2020})}\BibitemShut {NoStop}%
\bibitem [{\citenamefont {Hu}\ and\ \citenamefont
  {Tanaka}(2022)}]{hu2022revealing}%
  \BibitemOpen
  \bibfield  {author} {\bibinfo {author} {\bibfnamefont {Y.-C.}\ \bibnamefont
  {Hu}}\ and\ \bibinfo {author} {\bibfnamefont {H.}~\bibnamefont {Tanaka}},\
  }\href@noop {} {\bibfield  {journal} {\bibinfo  {journal} {Nat. Commun.}\
  }\textbf {\bibinfo {volume} {13}},\ \bibinfo {pages} {4519} (\bibinfo {year}
  {2022})}\BibitemShut {NoStop}%
\bibitem [{\citenamefont {Lu}\ \emph {et~al.}(2015)\citenamefont {Lu},
  \citenamefont {Lu}, \citenamefont {Qin},\ and\ \citenamefont
  {Shen}}]{lu2015experimental}%
  \BibitemOpen
  \bibfield  {author} {\bibinfo {author} {\bibfnamefont {Y.}~\bibnamefont
  {Lu}}, \bibinfo {author} {\bibfnamefont {X.}~\bibnamefont {Lu}}, \bibinfo
  {author} {\bibfnamefont {Z.}~\bibnamefont {Qin}}, \ and\ \bibinfo {author}
  {\bibfnamefont {J.}~\bibnamefont {Shen}},\ }\href@noop {} {\bibfield
  {journal} {\bibinfo  {journal} {Solid State Communications}\ }\textbf
  {\bibinfo {volume} {217}},\ \bibinfo {pages} {13} (\bibinfo {year}
  {2015})}\BibitemShut {NoStop}%
\bibitem [{\citenamefont {Berryman}\ \emph {et~al.}(2016)\citenamefont
  {Berryman}, \citenamefont {Anwar}, \citenamefont {Dorosz},\ and\
  \citenamefont {Schilling}}]{berryman2016early}%
  \BibitemOpen
  \bibfield  {author} {\bibinfo {author} {\bibfnamefont {J.~T.}\ \bibnamefont
  {Berryman}}, \bibinfo {author} {\bibfnamefont {M.}~\bibnamefont {Anwar}},
  \bibinfo {author} {\bibfnamefont {S.}~\bibnamefont {Dorosz}}, \ and\ \bibinfo
  {author} {\bibfnamefont {T.}~\bibnamefont {Schilling}},\ }\href@noop {}
  {\bibfield  {journal} {\bibinfo  {journal} {J. Chem. Phys.}\ }\textbf
  {\bibinfo {volume} {145}},\ \bibinfo {pages} {211901} (\bibinfo {year}
  {2016})}\BibitemShut {NoStop}%
\bibitem [{\citenamefont {Tanaka}\ \emph {et~al.}(2019)\citenamefont {Tanaka},
  \citenamefont {Tong}, \citenamefont {Shi},\ and\ \citenamefont
  {Russo}}]{tanaka2019revealing}%
  \BibitemOpen
  \bibfield  {author} {\bibinfo {author} {\bibfnamefont {H.}~\bibnamefont
  {Tanaka}}, \bibinfo {author} {\bibfnamefont {H.}~\bibnamefont {Tong}},
  \bibinfo {author} {\bibfnamefont {R.}~\bibnamefont {Shi}}, \ and\ \bibinfo
  {author} {\bibfnamefont {J.}~\bibnamefont {Russo}},\ }\href@noop {}
  {\bibfield  {journal} {\bibinfo  {journal} {Nature Reviews Physics}\ }\textbf
  {\bibinfo {volume} {1}},\ \bibinfo {pages} {333} (\bibinfo {year}
  {2019})}\BibitemShut {NoStop}%
\bibitem [{\citenamefont {Lechner}, \citenamefont {Dellago},\ and\
  \citenamefont {Bolhuis}(2011)}]{lechner2011role}%
  \BibitemOpen
  \bibfield  {author} {\bibinfo {author} {\bibfnamefont {W.}~\bibnamefont
  {Lechner}}, \bibinfo {author} {\bibfnamefont {C.}~\bibnamefont {Dellago}}, \
  and\ \bibinfo {author} {\bibfnamefont {P.~G.}\ \bibnamefont {Bolhuis}},\
  }\href@noop {} {\bibfield  {journal} {\bibinfo  {journal} {Phys. Rev. Lett.}\
  }\textbf {\bibinfo {volume} {106}},\ \bibinfo {pages} {085701} (\bibinfo
  {year} {2011})}\BibitemShut {NoStop}%
\bibitem [{\citenamefont {Becker}\ \emph {et~al.}(2022)\citenamefont {Becker},
  \citenamefont {Devijver}, \citenamefont {Molinier},\ and\ \citenamefont
  {Jakse}}]{becker2022unsupervised}%
  \BibitemOpen
  \bibfield  {author} {\bibinfo {author} {\bibfnamefont {S.}~\bibnamefont
  {Becker}}, \bibinfo {author} {\bibfnamefont {E.}~\bibnamefont {Devijver}},
  \bibinfo {author} {\bibfnamefont {R.}~\bibnamefont {Molinier}}, \ and\
  \bibinfo {author} {\bibfnamefont {N.}~\bibnamefont {Jakse}},\ }\href@noop {}
  {\bibfield  {journal} {\bibinfo  {journal} {Phys. Rev. E}\ }\textbf {\bibinfo
  {volume} {105}},\ \bibinfo {pages} {045304} (\bibinfo {year}
  {2022})}\BibitemShut {NoStop}%
\bibitem [{\citenamefont {Reinhart}\ \emph {et~al.}(2017)\citenamefont
  {Reinhart}, \citenamefont {Long}, \citenamefont {Howard}, \citenamefont
  {Ferguson},\ and\ \citenamefont {Panagiotopoulos}}]{reinhart2017machine}%
  \BibitemOpen
  \bibfield  {author} {\bibinfo {author} {\bibfnamefont {W.~F.}\ \bibnamefont
  {Reinhart}}, \bibinfo {author} {\bibfnamefont {A.~W.}\ \bibnamefont {Long}},
  \bibinfo {author} {\bibfnamefont {M.~P.}\ \bibnamefont {Howard}}, \bibinfo
  {author} {\bibfnamefont {A.~L.}\ \bibnamefont {Ferguson}}, \ and\ \bibinfo
  {author} {\bibfnamefont {A.~Z.}\ \bibnamefont {Panagiotopoulos}},\
  }\href@noop {} {\bibfield  {journal} {\bibinfo  {journal} {Soft Matter}\
  }\textbf {\bibinfo {volume} {13}},\ \bibinfo {pages} {4733} (\bibinfo {year}
  {2017})}\BibitemShut {NoStop}%
\bibitem [{\citenamefont {Boattini}, \citenamefont {Dijkstra},\ and\
  \citenamefont {Filion}(2019)}]{boattini2019unsupervised}%
  \BibitemOpen
  \bibfield  {author} {\bibinfo {author} {\bibfnamefont {E.}~\bibnamefont
  {Boattini}}, \bibinfo {author} {\bibfnamefont {M.}~\bibnamefont {Dijkstra}},
  \ and\ \bibinfo {author} {\bibfnamefont {L.}~\bibnamefont {Filion}},\
  }\href@noop {} {\bibfield  {journal} {\bibinfo  {journal} {J. Chem. Phys.}\
  }\textbf {\bibinfo {volume} {151}},\ \bibinfo {pages} {154901} (\bibinfo
  {year} {2019})}\BibitemShut {NoStop}%
\bibitem [{\citenamefont {Boattini}\ \emph {et~al.}(2020)\citenamefont
  {Boattini}, \citenamefont {Mar{\'\i}n-Aguilar}, \citenamefont {Mitra},
  \citenamefont {Foffi}, \citenamefont {Smallenburg},\ and\ \citenamefont
  {Filion}}]{boattini2020autonomously}%
  \BibitemOpen
  \bibfield  {author} {\bibinfo {author} {\bibfnamefont {E.}~\bibnamefont
  {Boattini}}, \bibinfo {author} {\bibfnamefont {S.}~\bibnamefont
  {Mar{\'\i}n-Aguilar}}, \bibinfo {author} {\bibfnamefont {S.}~\bibnamefont
  {Mitra}}, \bibinfo {author} {\bibfnamefont {G.}~\bibnamefont {Foffi}},
  \bibinfo {author} {\bibfnamefont {F.}~\bibnamefont {Smallenburg}}, \ and\
  \bibinfo {author} {\bibfnamefont {L.}~\bibnamefont {Filion}},\ }\href@noop {}
  {\bibfield  {journal} {\bibinfo  {journal} {Nat. Commun.}\ }\textbf {\bibinfo
  {volume} {11}},\ \bibinfo {pages} {1} (\bibinfo {year} {2020})}\BibitemShut
  {NoStop}%
\bibitem [{\citenamefont {Coli}\ and\ \citenamefont
  {Dijkstra}(2021)}]{coli2021artificial}%
  \BibitemOpen
  \bibfield  {author} {\bibinfo {author} {\bibfnamefont {G.~M.}\ \bibnamefont
  {Coli}}\ and\ \bibinfo {author} {\bibfnamefont {M.}~\bibnamefont
  {Dijkstra}},\ }\href@noop {} {\bibfield  {journal} {\bibinfo  {journal} {ACS
  nano}\ }\textbf {\bibinfo {volume} {15}},\ \bibinfo {pages} {4335} (\bibinfo
  {year} {2021})}\BibitemShut {NoStop}%
\bibitem [{\citenamefont {van Damme}\ \emph {et~al.}(2020)\citenamefont {van
  Damme}, \citenamefont {Coli}, \citenamefont {van Roij},\ and\ \citenamefont
  {Dijkstra}}]{van2020classifying}%
  \BibitemOpen
  \bibfield  {author} {\bibinfo {author} {\bibfnamefont {R.}~\bibnamefont {van
  Damme}}, \bibinfo {author} {\bibfnamefont {G.~M.}\ \bibnamefont {Coli}},
  \bibinfo {author} {\bibfnamefont {R.}~\bibnamefont {van Roij}}, \ and\
  \bibinfo {author} {\bibfnamefont {M.}~\bibnamefont {Dijkstra}},\ }\href@noop
  {} {\bibfield  {journal} {\bibinfo  {journal} {ACS nano}\ }\textbf {\bibinfo
  {volume} {14}},\ \bibinfo {pages} {15144} (\bibinfo {year}
  {2020})}\BibitemShut {NoStop}%
\bibitem [{\citenamefont {Gardin}\ \emph {et~al.}(2021)\citenamefont {Gardin},
  \citenamefont {Perego}, \citenamefont {Doni},\ and\ \citenamefont
  {Pavan}}]{gardin2021classifying}%
  \BibitemOpen
  \bibfield  {author} {\bibinfo {author} {\bibfnamefont {A.}~\bibnamefont
  {Gardin}}, \bibinfo {author} {\bibfnamefont {C.}~\bibnamefont {Perego}},
  \bibinfo {author} {\bibfnamefont {G.}~\bibnamefont {Doni}}, \ and\ \bibinfo
  {author} {\bibfnamefont {G.~M.}\ \bibnamefont {Pavan}},\ }\href@noop {}
  {\bibfield  {journal} {\bibinfo  {journal} {arXiv}\ } (\bibinfo {year}
  {2021})}\BibitemShut {NoStop}%
\bibitem [{\citenamefont {Coslovich}, \citenamefont {Jack},\ and\ \citenamefont
  {Paret}(2022)}]{coslovich2022dimensionality}%
  \BibitemOpen
  \bibfield  {author} {\bibinfo {author} {\bibfnamefont {D.}~\bibnamefont
  {Coslovich}}, \bibinfo {author} {\bibfnamefont {R.~L.}\ \bibnamefont {Jack}},
  \ and\ \bibinfo {author} {\bibfnamefont {J.}~\bibnamefont {Paret}},\
  }\href@noop {} {\bibfield  {journal} {\bibinfo  {journal} {J. Chem. Phys.}\
  }\textbf {\bibinfo {volume} {157}},\ \bibinfo {pages} {204503} (\bibinfo
  {year} {2022})}\BibitemShut {NoStop}%
\bibitem [{\citenamefont {Paret}, \citenamefont {Jack},\ and\ \citenamefont
  {Coslovich}(2020)}]{paret2020assessing}%
  \BibitemOpen
  \bibfield  {author} {\bibinfo {author} {\bibfnamefont {J.}~\bibnamefont
  {Paret}}, \bibinfo {author} {\bibfnamefont {R.~L.}\ \bibnamefont {Jack}}, \
  and\ \bibinfo {author} {\bibfnamefont {D.}~\bibnamefont {Coslovich}},\
  }\href@noop {} {\bibfield  {journal} {\bibinfo  {journal} {J. Chem. Phys.}\
  }\textbf {\bibinfo {volume} {152}},\ \bibinfo {pages} {144502} (\bibinfo
  {year} {2020})}\BibitemShut {NoStop}%
\bibitem [{\citenamefont {Adorf}\ \emph {et~al.}(2019)\citenamefont {Adorf},
  \citenamefont {Moore}, \citenamefont {Melle},\ and\ \citenamefont
  {Glotzer}}]{adorf2019analysis}%
  \BibitemOpen
  \bibfield  {author} {\bibinfo {author} {\bibfnamefont {C.~S.}\ \bibnamefont
  {Adorf}}, \bibinfo {author} {\bibfnamefont {T.~C.}\ \bibnamefont {Moore}},
  \bibinfo {author} {\bibfnamefont {Y.~J.}\ \bibnamefont {Melle}}, \ and\
  \bibinfo {author} {\bibfnamefont {S.~C.}\ \bibnamefont {Glotzer}},\
  }\href@noop {} {\bibfield  {journal} {\bibinfo  {journal} {J. Phys. Chem. B}\
  }\textbf {\bibinfo {volume} {124}},\ \bibinfo {pages} {69} (\bibinfo {year}
  {2019})}\BibitemShut {NoStop}%
\bibitem [{\citenamefont {Auer}\ and\ \citenamefont
  {Frenkel}(2005)}]{auer2005numerical}%
  \BibitemOpen
  \bibfield  {author} {\bibinfo {author} {\bibfnamefont {S.}~\bibnamefont
  {Auer}}\ and\ \bibinfo {author} {\bibfnamefont {D.}~\bibnamefont {Frenkel}},\
  }\href@noop {} {\bibfield  {journal} {\bibinfo  {journal} {Adv. Comput.
  Simul.}\ }\textbf {\bibinfo {volume} {173}},\ \bibinfo {pages} {149}
  (\bibinfo {year} {2005})}\BibitemShut {NoStop}%
\bibitem [{\citenamefont {Desgranges}\ and\ \citenamefont
  {Delhommelle}(2007)}]{desgranges2007polymorph}%
  \BibitemOpen
  \bibfield  {author} {\bibinfo {author} {\bibfnamefont {C.}~\bibnamefont
  {Desgranges}}\ and\ \bibinfo {author} {\bibfnamefont {J.}~\bibnamefont
  {Delhommelle}},\ }\href@noop {} {\bibfield  {journal} {\bibinfo  {journal}
  {The Journal of chemical physics}\ }\textbf {\bibinfo {volume} {126}},\
  \bibinfo {pages} {054501} (\bibinfo {year} {2007})}\BibitemShut {NoStop}%
\bibitem [{\citenamefont {Browning}, \citenamefont {Doherty},\ and\
  \citenamefont {Fredrickson}(2008)}]{browning2008nucleation}%
  \BibitemOpen
  \bibfield  {author} {\bibinfo {author} {\bibfnamefont {A.~R.}\ \bibnamefont
  {Browning}}, \bibinfo {author} {\bibfnamefont {M.~F.}\ \bibnamefont
  {Doherty}}, \ and\ \bibinfo {author} {\bibfnamefont {G.~H.}\ \bibnamefont
  {Fredrickson}},\ }\href@noop {} {\bibfield  {journal} {\bibinfo  {journal}
  {Physical Review E}\ }\textbf {\bibinfo {volume} {77}},\ \bibinfo {pages}
  {041604} (\bibinfo {year} {2008})}\BibitemShut {NoStop}%
\bibitem [{\citenamefont {Gispen}\ and\ \citenamefont
  {Dijkstra}(2022)}]{gispen2022kinetic}%
  \BibitemOpen
  \bibfield  {author} {\bibinfo {author} {\bibfnamefont {W.}~\bibnamefont
  {Gispen}}\ and\ \bibinfo {author} {\bibfnamefont {M.}~\bibnamefont
  {Dijkstra}},\ }\href@noop {} {\bibfield  {journal} {\bibinfo  {journal}
  {Physical Review Letters}\ }\textbf {\bibinfo {volume} {129}},\ \bibinfo
  {pages} {098002} (\bibinfo {year} {2022})}\BibitemShut {NoStop}%
\bibitem [{\citenamefont {de~Jager}\ and\ \citenamefont
  {Filion}(2022)}]{jager2022crystal}%
  \BibitemOpen
  \bibfield  {author} {\bibinfo {author} {\bibfnamefont {M.}~\bibnamefont
  {de~Jager}}\ and\ \bibinfo {author} {\bibfnamefont {L.}~\bibnamefont
  {Filion}},\ }\href@noop {} {\bibfield  {journal} {\bibinfo  {journal} {J.
  Chem. Phys.}\ }\textbf {\bibinfo {volume} {157}},\ \bibinfo {pages} {154905}
  (\bibinfo {year} {2022})}\BibitemShut {NoStop}%
\bibitem [{\citenamefont {Lechner}\ and\ \citenamefont
  {Dellago}(2008)}]{lechner2008accurate}%
  \BibitemOpen
  \bibfield  {author} {\bibinfo {author} {\bibfnamefont {W.}~\bibnamefont
  {Lechner}}\ and\ \bibinfo {author} {\bibfnamefont {C.}~\bibnamefont
  {Dellago}},\ }\href@noop {} {\bibfield  {journal} {\bibinfo  {journal} {J.
  Chem. Phys.}\ }\textbf {\bibinfo {volume} {129}},\ \bibinfo {pages} {114707}
  (\bibinfo {year} {2008})}\BibitemShut {NoStop}%
\bibitem [{\citenamefont {van Meel}\ \emph {et~al.}(2012)\citenamefont {van
  Meel}, \citenamefont {Filion}, \citenamefont {Valeriani},\ and\ \citenamefont
  {Frenkel}}]{van2012parameter}%
  \BibitemOpen
  \bibfield  {author} {\bibinfo {author} {\bibfnamefont {J.~A.}\ \bibnamefont
  {van Meel}}, \bibinfo {author} {\bibfnamefont {L.}~\bibnamefont {Filion}},
  \bibinfo {author} {\bibfnamefont {C.}~\bibnamefont {Valeriani}}, \ and\
  \bibinfo {author} {\bibfnamefont {D.}~\bibnamefont {Frenkel}},\ }\href@noop
  {} {\bibfield  {journal} {\bibinfo  {journal} {J. Chem. Phys.}\ }\textbf
  {\bibinfo {volume} {136}},\ \bibinfo {pages} {234107} (\bibinfo {year}
  {2012})}\BibitemShut {NoStop}%
\bibitem [{\citenamefont {Malins}\ \emph {et~al.}(2013)\citenamefont {Malins},
  \citenamefont {Williams}, \citenamefont {Eggers},\ and\ \citenamefont
  {Royall}}]{malins2013identification}%
  \BibitemOpen
  \bibfield  {author} {\bibinfo {author} {\bibfnamefont {A.}~\bibnamefont
  {Malins}}, \bibinfo {author} {\bibfnamefont {S.~R.}\ \bibnamefont
  {Williams}}, \bibinfo {author} {\bibfnamefont {J.}~\bibnamefont {Eggers}}, \
  and\ \bibinfo {author} {\bibfnamefont {C.~P.}\ \bibnamefont {Royall}},\
  }\href@noop {} {\bibfield  {journal} {\bibinfo  {journal} {J. Chem. Phys.}\
  }\textbf {\bibinfo {volume} {139}},\ \bibinfo {pages} {234506} (\bibinfo
  {year} {2013})}\BibitemShut {NoStop}%
\bibitem [{\citenamefont {ten Wolde}, \citenamefont {Ruiz-Montero},\ and\
  \citenamefont {Frenkel}(1996)}]{tenwolde1996simulation}%
  \BibitemOpen
  \bibfield  {author} {\bibinfo {author} {\bibfnamefont {P.~R.}\ \bibnamefont
  {ten Wolde}}, \bibinfo {author} {\bibfnamefont {M.~J.}\ \bibnamefont
  {Ruiz-Montero}}, \ and\ \bibinfo {author} {\bibfnamefont {D.}~\bibnamefont
  {Frenkel}},\ }\href@noop {} {\bibfield  {journal} {\bibinfo  {journal}
  {Faraday Discuss.}\ }\textbf {\bibinfo {volume} {104}},\ \bibinfo {pages}
  {93} (\bibinfo {year} {1996})}\BibitemShut {NoStop}%
\bibitem [{\citenamefont {Sanz}\ and\ \citenamefont
  {Marenduzzo}(2010)}]{sanz2010dynamic}%
  \BibitemOpen
  \bibfield  {author} {\bibinfo {author} {\bibfnamefont {E.}~\bibnamefont
  {Sanz}}\ and\ \bibinfo {author} {\bibfnamefont {D.}~\bibnamefont
  {Marenduzzo}},\ }\href@noop {} {\bibfield  {journal} {\bibinfo  {journal} {J.
  Chem. Phys.}\ }\textbf {\bibinfo {volume} {132}},\ \bibinfo {pages} {194102}
  (\bibinfo {year} {2010})}\BibitemShut {NoStop}%
\bibitem [{\citenamefont {Plimpton}(1995)}]{plimpton1995fast}%
  \BibitemOpen
  \bibfield  {author} {\bibinfo {author} {\bibfnamefont {S.}~\bibnamefont
  {Plimpton}},\ }\href@noop {} {\bibfield  {journal} {\bibinfo  {journal} {J.
  Comp. Phys.}\ }\textbf {\bibinfo {volume} {117}},\ \bibinfo {pages} {1}
  (\bibinfo {year} {1995})}\BibitemShut {NoStop}%
\bibitem [{\citenamefont {Rycroft}(2009)}]{rycroft2009voro}%
  \BibitemOpen
  \bibfield  {author} {\bibinfo {author} {\bibfnamefont {C.}~\bibnamefont
  {Rycroft}},\ }\href@noop {} {\bibfield  {journal} {\bibinfo  {journal}
  {Chaos}\ }\textbf {\bibinfo {volume} {19}},\ \bibinfo {pages} {041111}
  (\bibinfo {year} {2009})}\BibitemShut {NoStop}%
\end{thebibliography}%

\end{document}


\preprint{APS/123-QED}

\title{Supplemental Material for ``In search of a precursor for crystal nucleation of hard and charged colloids''}

\author{Marjolein de Jager}
\affiliation{Soft Condensed Matter, Debye Institute of Nanomaterials Science, Utrecht University, Utrecht, Netherlands}
\author{Frank Smallenburg}
\affiliation{Universit\'e Paris-Saclay, CNRS, Laboratoire de Physique des Solides, 91405 Orsay, France}
\author{Laura Filion}
\affiliation{Soft Condensed Matter, Debye Institute of Nanomaterials Science, Utrecht University, Utrecht, Netherlands}


\maketitle

\onecolumngrid  


\renewcommand{\thetable}{S\arabic{table}}
\renewcommand{\thefigure}{S\arabic{figure}}
\renewcommand{\theequation}{S\arabic{equation}}

\newcommand{\figwidthA}{0.45\linewidth}
\newcommand{\figwidthB}{0.40\linewidth}
\newcommand{\figwidthC}{0.45\linewidth}


\section{Nucleation barriers}

\begin{figure}[b!]
     \centering
     \includegraphics[width=\figwidthA]{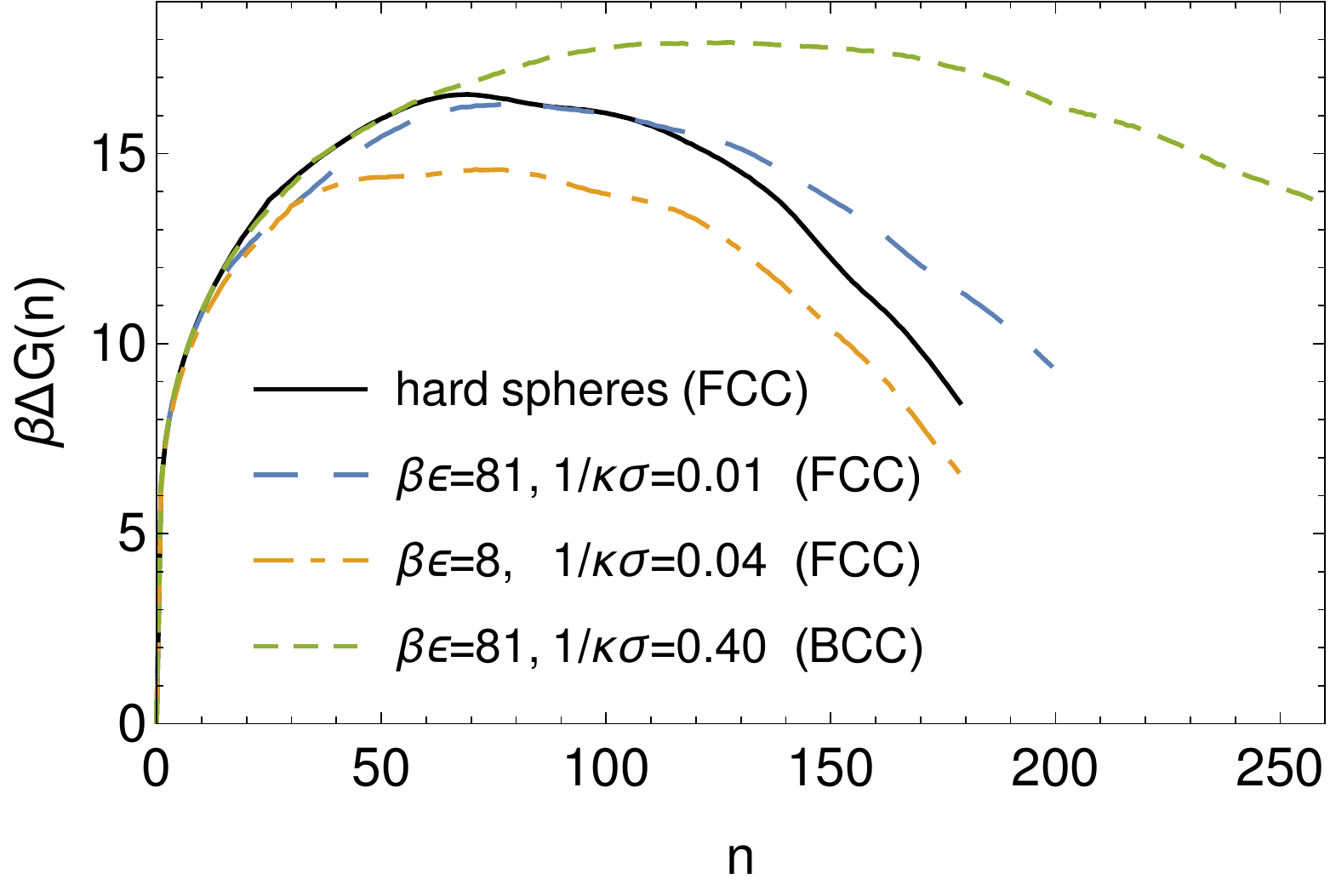}
     \caption{\label{fig:barriers}
     Nucleation barriers of (nearly-)hard spheres forming an FCC crystal and soft spheres forming a BCC crystal. For more information see Tab. \ref{tab:infomore}.}
 \end{figure}

In the main paper, we give the barrier height and critical nucleus size of systems studied for their spontaneous nucleation. Here we provide some more background on these systems and their nucleation barriers. 
To compute the nucleation barriers, we use Monte Carlo (MC) simulations in the $NPT$-ensemble combined with umbrella sampling. We bias the simulations using the number of particles in the nucleus measured via the Ten Wolde order parameter \cite{tenwolde1996simulation} with cutoff values $d_c=0.7$ and $\xi_c=6$. The nearest neighbors for this order parameter are determined using a radial cutoff $r_c$, which is chosen to be approximately the position of the first minimum of the radial distribution function. The full methods are explained in our previous work\cite{jager2022crystal}. For all nucleation barriers, we use a biasing strength of $\lambda=0.02$ and bias around the target nucleus size $n_c$, where we take $n_c$ with an interval of 10. We simulate $N=10976$ and $N=11664$ particles for the systems forming the face-centered cubic (FCC) and body-centered cubic (BCC) phase, respectively. Furthermore, to obtain more accurate nucleation barriers, we perform 2 to 4 independent runs for each window. 
The resulting nucleation barriers are shown in Fig. \ref{fig:barriers} and Tab. \ref{tab:infomore} provides additional information on these barriers, such as the interfacial free energy obtained from fitting the nucleation barrier to classical nucleation theory (see Ref. \onlinecite{jager2022crystal}). 
Note that we use the radial cutoff $r_c$ for determining nearest neighbors in the fluid-solid classification, which is done on the fly, whereas for Lechner and Dellago's bond-orientational order parameters (BOPs)\cite{lechner2008accurate}, which are computed in the post processing, we use the SANN algorithm \cite{van2012parameter} to determine the nearest neighbors.



\begin{table*}[b!]
\caption{\label{tab:infomore} For each system studied, the packing fraction of the supersaturated fluid $\eta^*$ at which the brute force nucleation is performed together with the corresponding packing fraction of the solid phase $\eta_s$, pressure $\beta P\sigma^3$, and supersaturation $\beta|\Delta\mu|$. The last columns give cutoff radius $r_c/\sigma$ as well as the reduced interfacial free energy $\beta\gamma/\rho_s^{2/3}$, critical nucleus size $n^*$, and barrier height $\beta\Delta G^*$ obtained from fitting the nucleation barriers. The error in $\beta\Delta G^*$ is no more than 1.
}
\begin{ruledtabular}
\begin{tabular}{ccccccccccc}
 & $\beta\epsilon$ & $1/\kappa\sigma$ & $\eta^*$ & $\eta_s$ & $\beta P\sigma^3$ & $\beta |\Delta\mu|$ & $r_c/\sigma$ & $\beta\gamma/\rho_s^{2/3}$ & $n^*$ & $\beta\Delta G^*$ \\ \hline
\multirow{3}{*}{FCC} & \multicolumn{2}{c}{hard spheres}   & 0.5385 &  0.5981  & 17.5  & 0.585 & 1.40  & 0.52  & 75   & 16.5  \\
 & 81 & 0.01   & 0.4681 &  0.5190  & 15.4  & 0.584 & 1.46  & 0.49  & 84   & 16.3  \\ 
 & 8  & 0.04   & 0.4400 &  0.4711  & 16.7  & 0.541 & 1.49  & 0.44  & 69   & 14.8  \\ \hline
BCC  & 81 & 0.40   & 0.1305 &  0.1311  & 21.3  & 0.321  & 2.32 & 0.35  & 122  & 18.0  \\
\end{tabular}
\end{ruledtabular}
\end{table*}


\section{Distribution of BOPs in the metastable fluids}
In Fig. 1 of the main text, we show the average values of Lechner and Dellago's BOPs, both $\bar{q_l}$'s and $\bar{w}_l$'s, in the metastable fluids of nearly hard spheres and of soft spheres as a function of the supersaturation. Remember that the metastable fluid of nearly hard spheres will later nucleate the FCC phase, while the one of soft spheres will nucleate the BCC phase.
From this figure we observe that the most prevalent difference between the two systems can be found in $\bar{w}_6$. Here we focus more on this difference by performing a similar analysis to Russo and Tanaka's analysis in Refs. \onlinecite{russo2012microscopic,russo2012selection}. This analysis is based on the observation that $\bar{w}_6$ can be used to distinguish between the BCC crystal and close-packed crystals (FCC and/or HCP), since $\bar{w}_6>0$ for the BCC crystal and $\bar{w}_6<0$ for the FCC and HCP crystals. Furthermore, $\bar{w}_4$ can be used to distinguish FCC from HCP, as $\bar{w}_4<0$ for the FCC crystal and $\bar{w}_4>0$ for the HCP crystal. Thus, by looking at the probability distributions of $\bar{w}_6$ and $\bar{w}_4$ in the metastable fluid for different threshold values of $\bar{q}_6$, Russo and Tanaka demonstrated that the more crystal-like regions in the metastable fluid of hard spheres tend more towards the FCC phase, than the HCP or BCC phase.

\begin{figure*}[b!]
\begin{tabular}{lll}
     a) & \hspace{0.5cm} & b)  \\[-0.5cm]
     \includegraphics[width=\figwidthB]{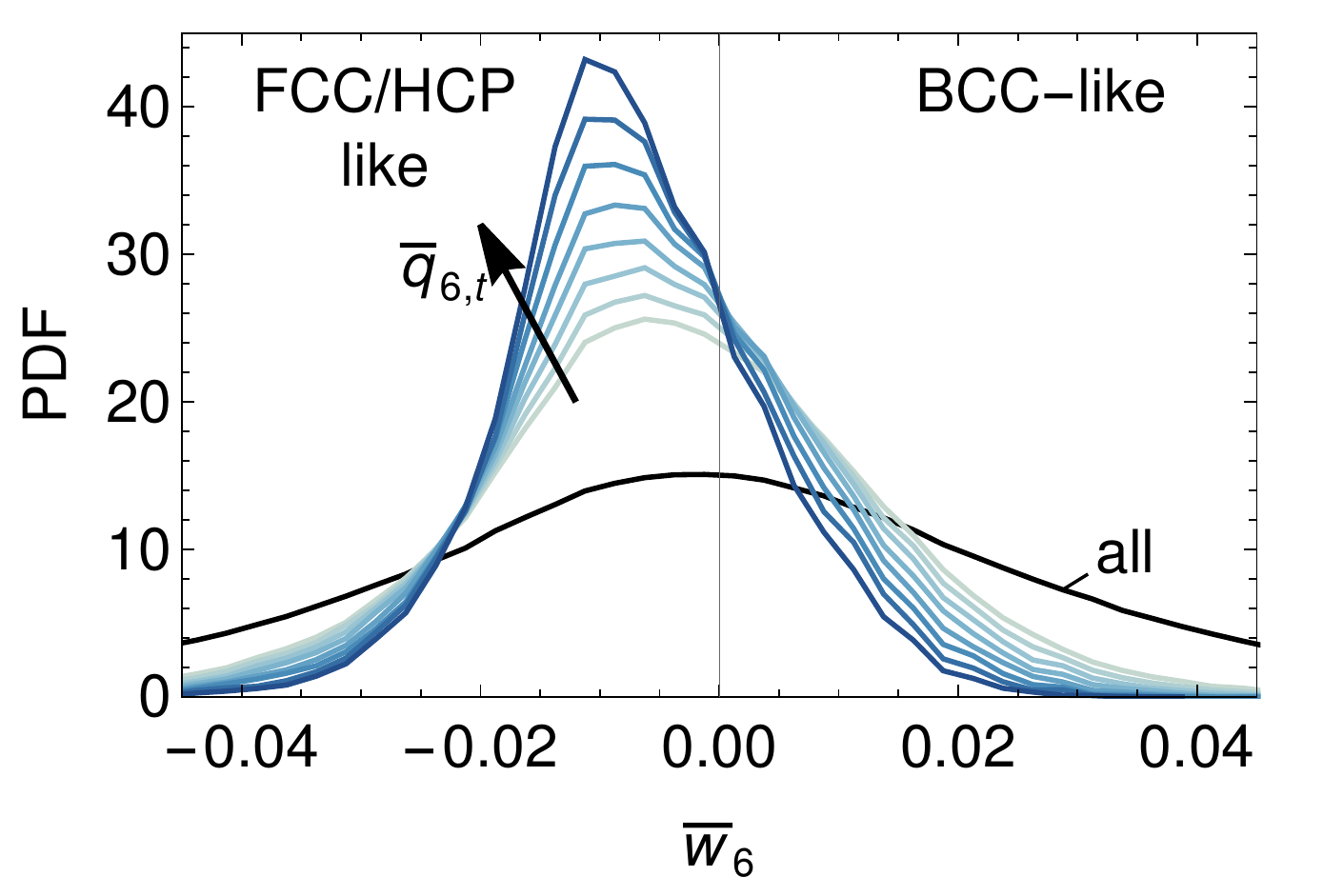} & & \includegraphics[width=\figwidthB]{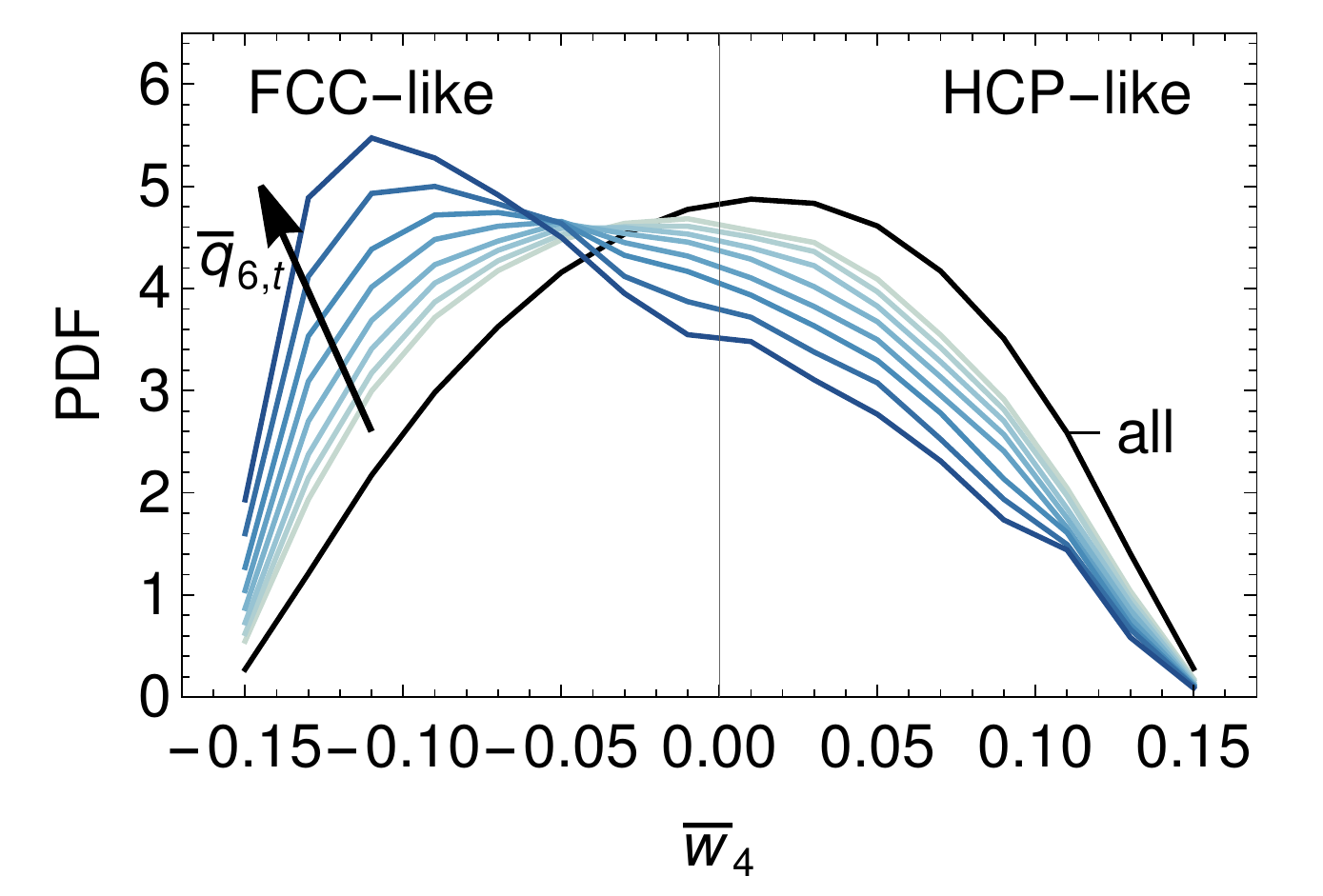} \\
     c) & \hspace{0.5cm} & d)  \\[-0.5cm]
     \includegraphics[width=\figwidthB]{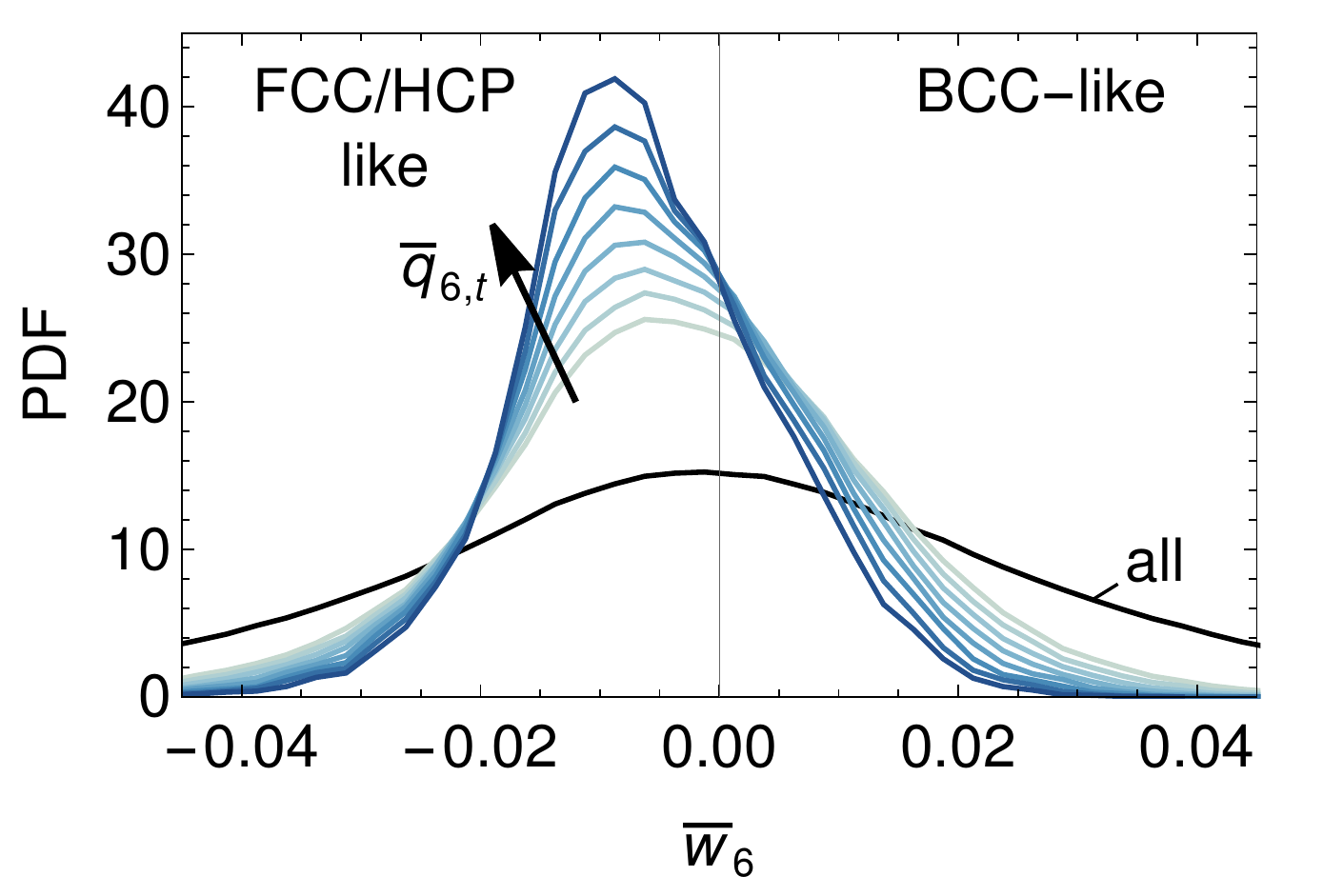} & & \includegraphics[width=\figwidthB]{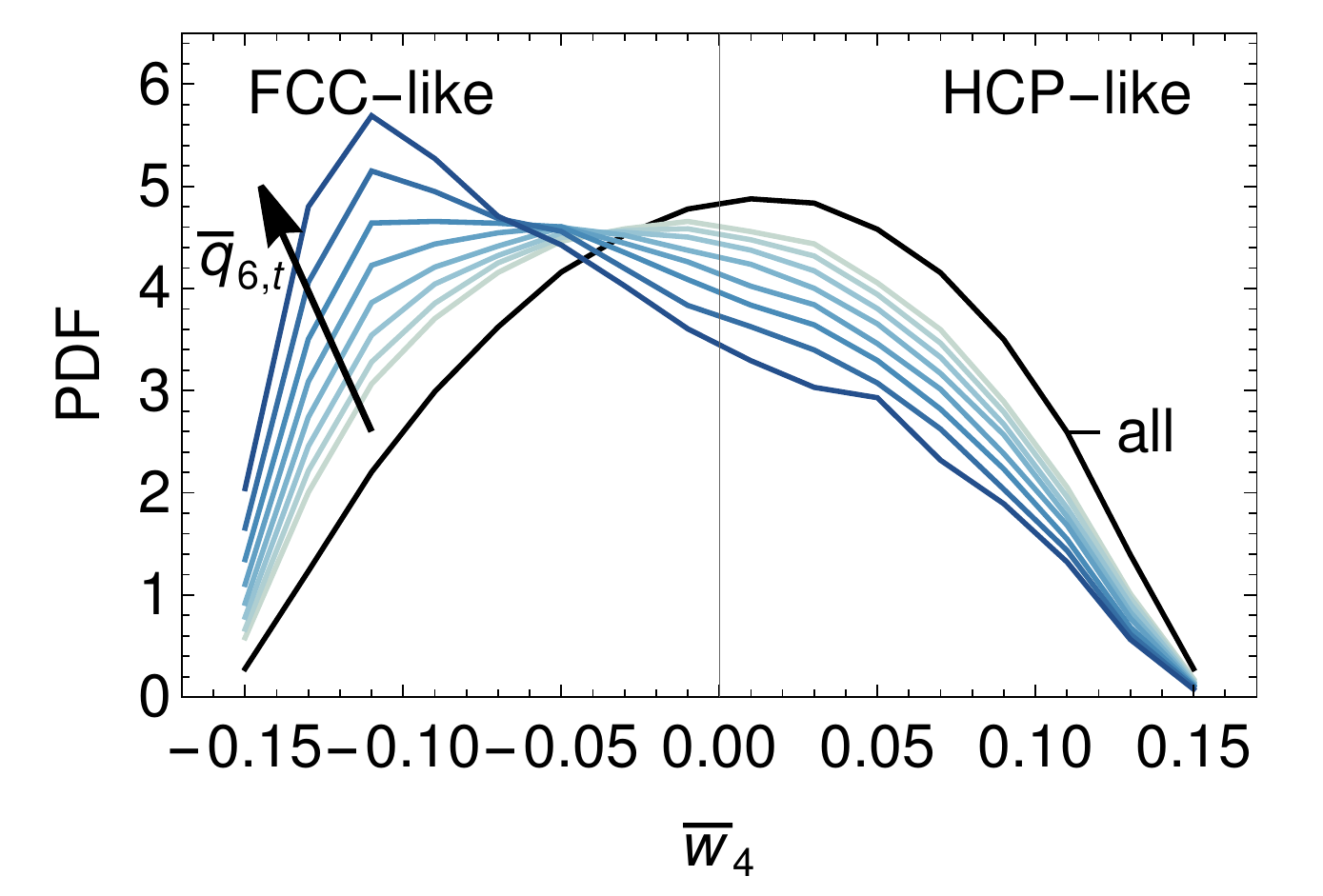} \\
     e) & \hspace{0.5cm} & f)  \\[-0.5cm]
     \includegraphics[width=\figwidthB]{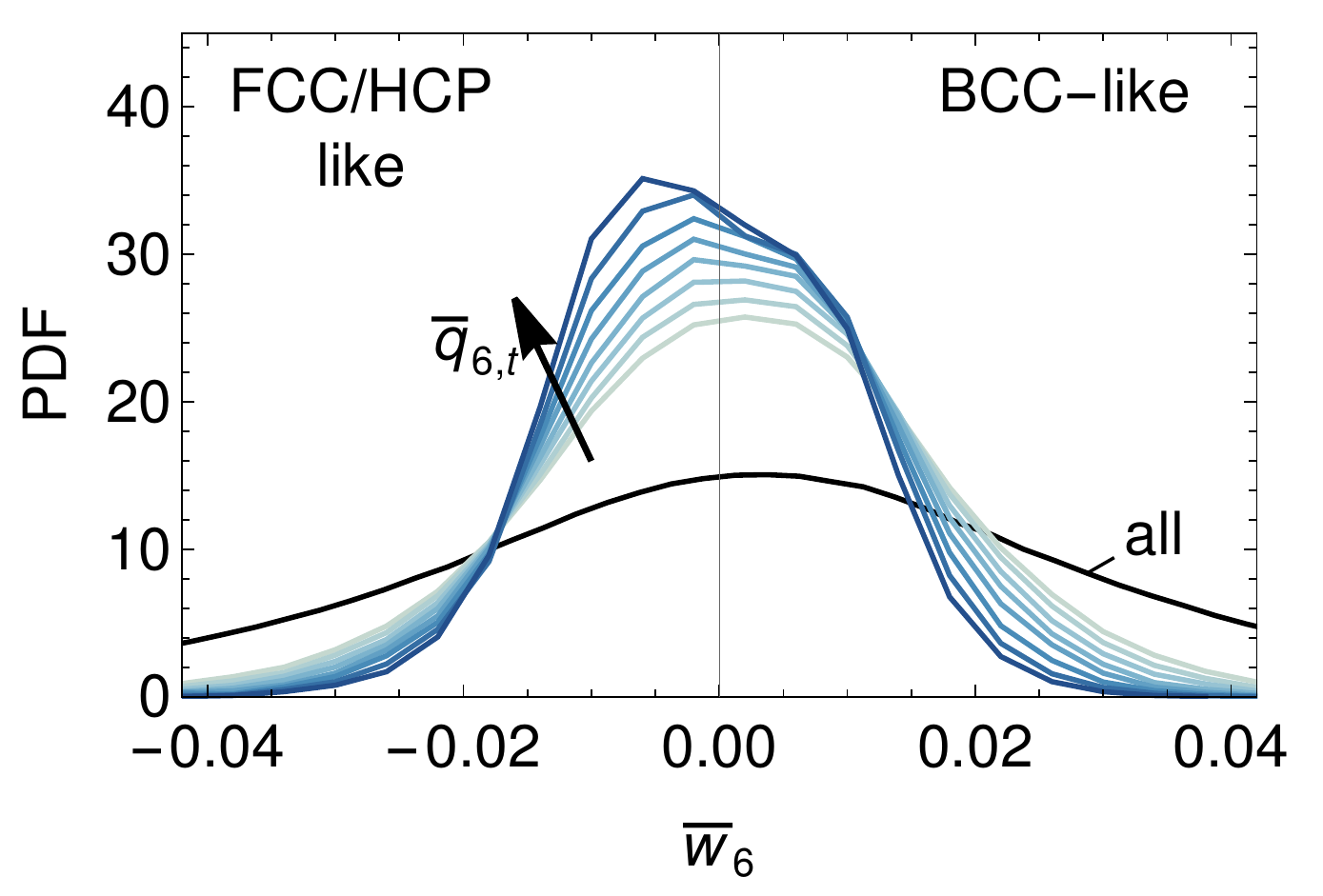} & & \includegraphics[width=\figwidthB]{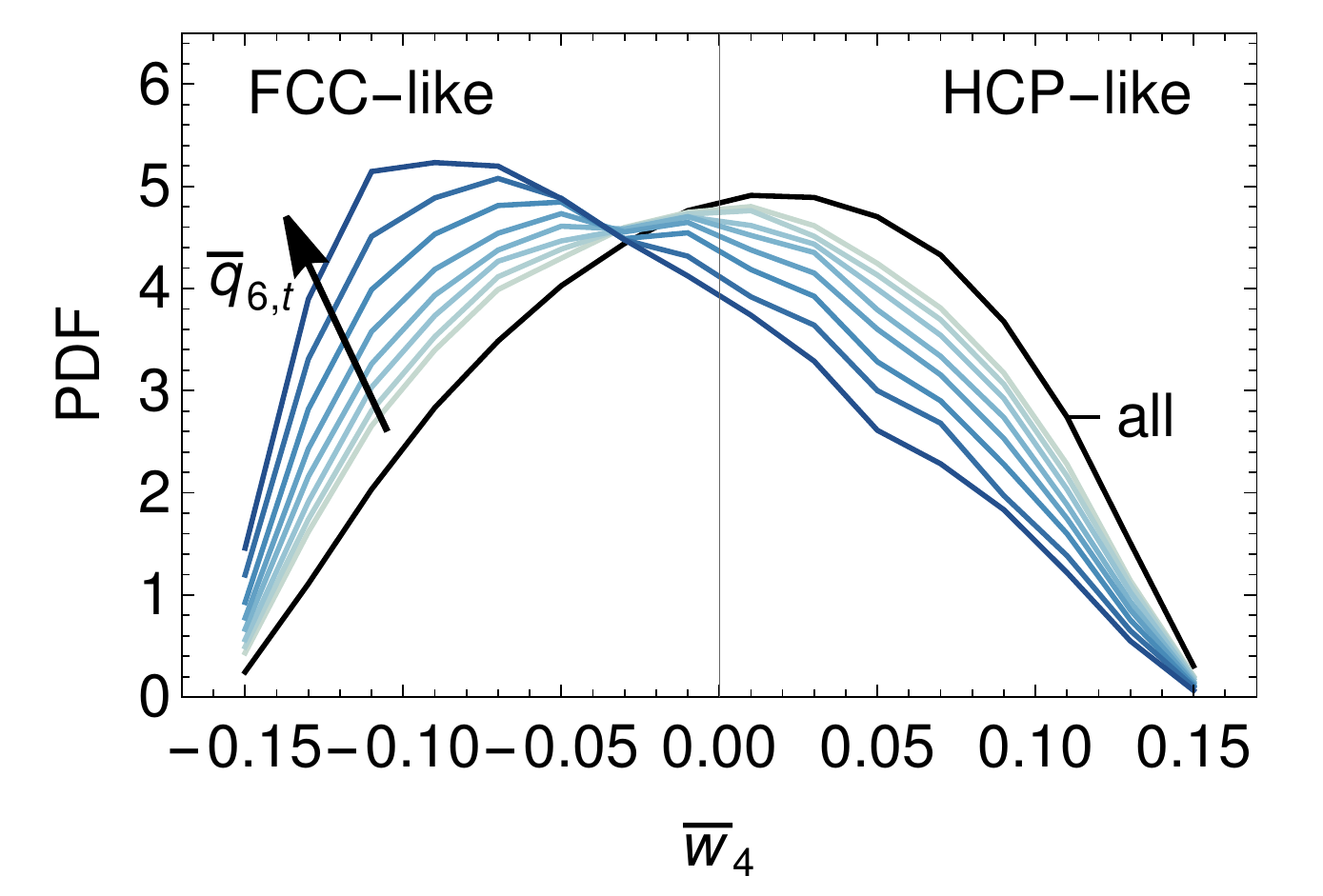} 
\end{tabular}
    \caption[width=1\linewidth]{\label{fig:histws} The probability distribution of a,c,e) $\bar{w}_6$ and b,d,f) $\bar{w}_4$ in the metastable fluids of a,b) hard spheres ($\eta=0.5385$) soft hard-core Yukawa particles with c,d) $\beta\epsilon=8$ and $1/\kappa\sigma=0.04$ ($\eta=0.1305$), and e,f) $\beta\epsilon=81$ and $1/\kappa\sigma=0.40$ ($\eta=0.1305$). In each figure, the black line gives the total distribution and the eight lines with the blue color gradient give the distribution for $\bar{q}_6>\bar{q}_{6,t}$, where the arrow indicates the increase in the threshold value from $\bar{q}_{6,t}=0.25$ to $\bar{q}_{6,t}=0.32$.
    }
\end{figure*}

In Fig. \ref{fig:histws}, we show the probability distribution functions of $\bar{w}_6$ and $\bar{w}_4$ for the metastable fluid of hard spheres, nearly hard spheres ($\beta\epsilon=8$, $1/\kappa\sigma=0.04$), and soft spheres ($\beta\epsilon=81$, $1/\kappa\sigma=0.40$). The black line in each figure shows the probability distribution for the entire system, while the lines with the blue color gradient show the probability distributions for particles with $\bar{q}_6>\bar{q}_{6,t}$ for increasing value of the threshold $\bar{q}_{6,t}$. The probability distributions for hard spheres and nearly hard spheres are practically identical to the ones shown by Russo and Tanaka in Ref. \onlinecite{russo2012microscopic}. From these figures one can draw the conclusion that high $\bar{q}_6$ regions show a predominant preference for negative $\bar{w}_6$ and negative $\bar{w}_4$, and thus correspond to FCC-like symmetries. Furthermore, this preference increases with increasing $\bar{q}_{6,t}$. 
The probability distributions of the soft spheres, on the other hand, show a slight preference for positive $\bar{w}_6$, which corresponds to BCC-like symmetries. However, notice that the distribution shifts towards FCC-like for increasing $\bar{q}_{6,t}$.



\section{Radial correlations}

\begin{figure*}[b!]
\begin{tabular}{lll}
     a) & \hspace{0.5cm} & b)  \\[-0.4cm]
     \includegraphics[width=\figwidthB]{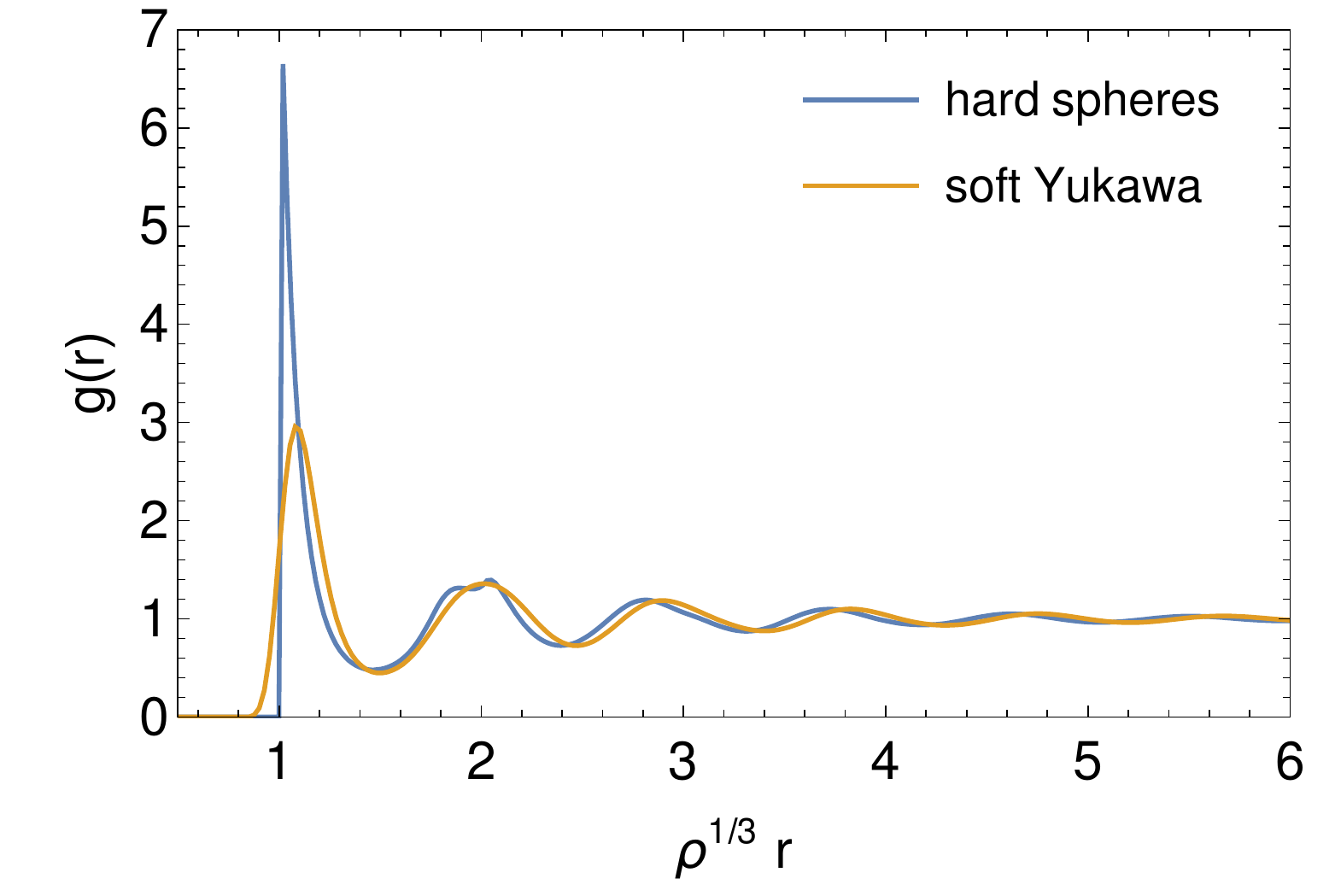} & & \includegraphics[width=\figwidthB]{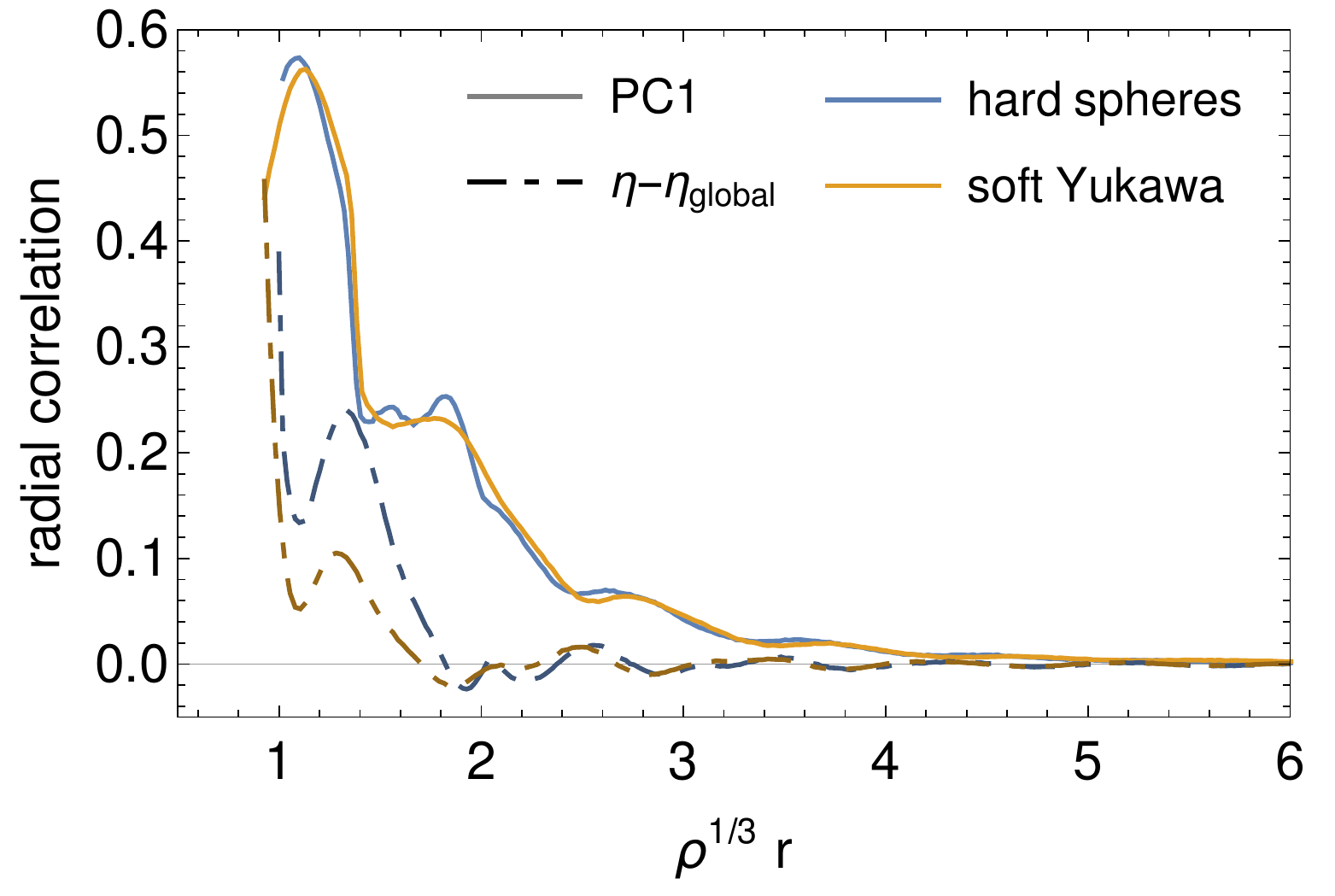} 
\end{tabular}
    \caption[width=1\linewidth]{\label{fig:radialcor} a) radial distribution function and b) radial correlation of the first principal component (solid) and local packing fraction (dashed) in the metastable fluids of hard spheres ($\eta=0.5385$) and soft hard-core Yukawa particles ($\beta\epsilon=81$, $1/\kappa\sigma=0.40$, $\eta=0.1305$). Note that the horizontal axes are scaled by the cubic root of the density such that the first peaks of the $g(r)$ are located a the same position. }
\end{figure*}

In the main text, we qualitative discuss the spatial correlations in the local packing fraction and first principal component (PC1). Here, we quantify the spatial correlations. Figure \ref{fig:radialcor} shows the radial distribution function together with the radial correlations of PC1 and the local packing fraction for the metastable fluids of hard spheres and of soft spheres. We clearly see that PC1 has a measurable correlation even 4 or 5 shells removed, whereas the correlation of the local packing fraction decays to zero within the first two shells. 

\section{More figures and more nucleation events}

In the main paper, we discuss our observations using two typical nucleation events: one of the hard-spheres system and one of the soft spheres ($\beta\epsilon=81$, $1/\kappa\sigma=0.40$). 
Here, we show more tracking results on these two events, as well as the results on some other events.
As in the main paper, all figures show the average values inside the studied region, i.e. the small spatial region containing the birthplace of the crystal nucleus. Moreover, the layout of the figures is the same as in the main paper. The vertical dashed line in each figure indicates the start of nucleation $t_0$, the shaded area indicates the time window before $t_0$ for which the ACF of the first principal component (PC1) is greater than $0.05$, and, if present, the horizontal dashed lines give the reference value of a parameter in the fluid and sometimes in the solid phase as well.

\begin{figure*}[t!]
\begin{tabular}{lll}
     a) & \hspace{0.5cm} & b)  \\[-0.64cm]
     \includegraphics[width=\figwidthC]{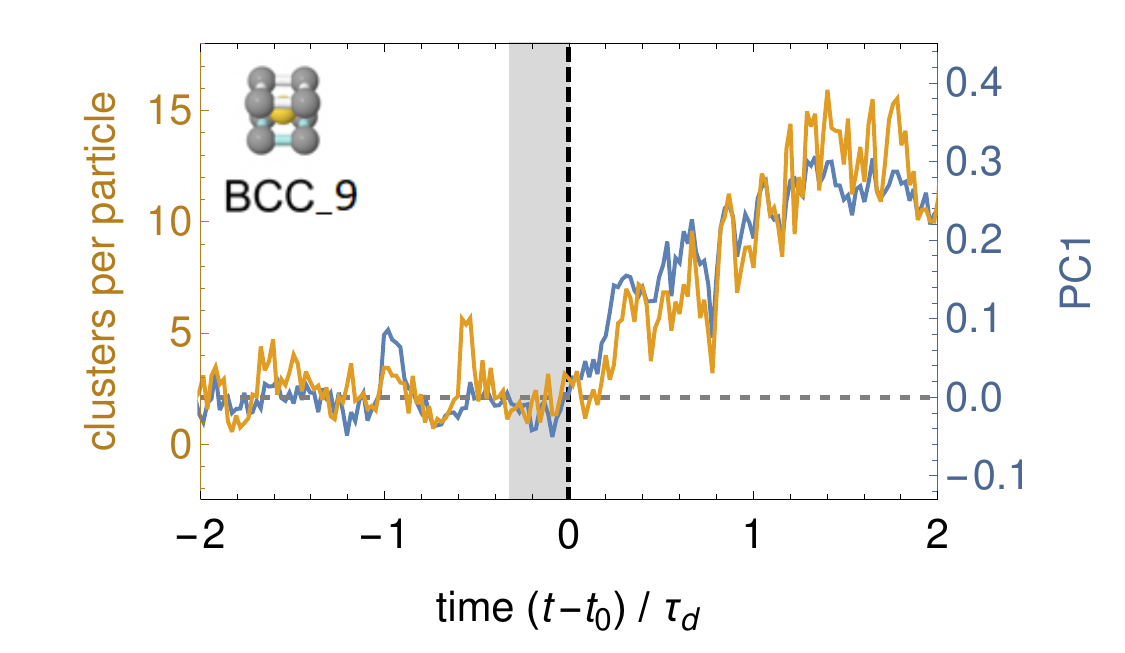} & & \includegraphics[width=\figwidthC]{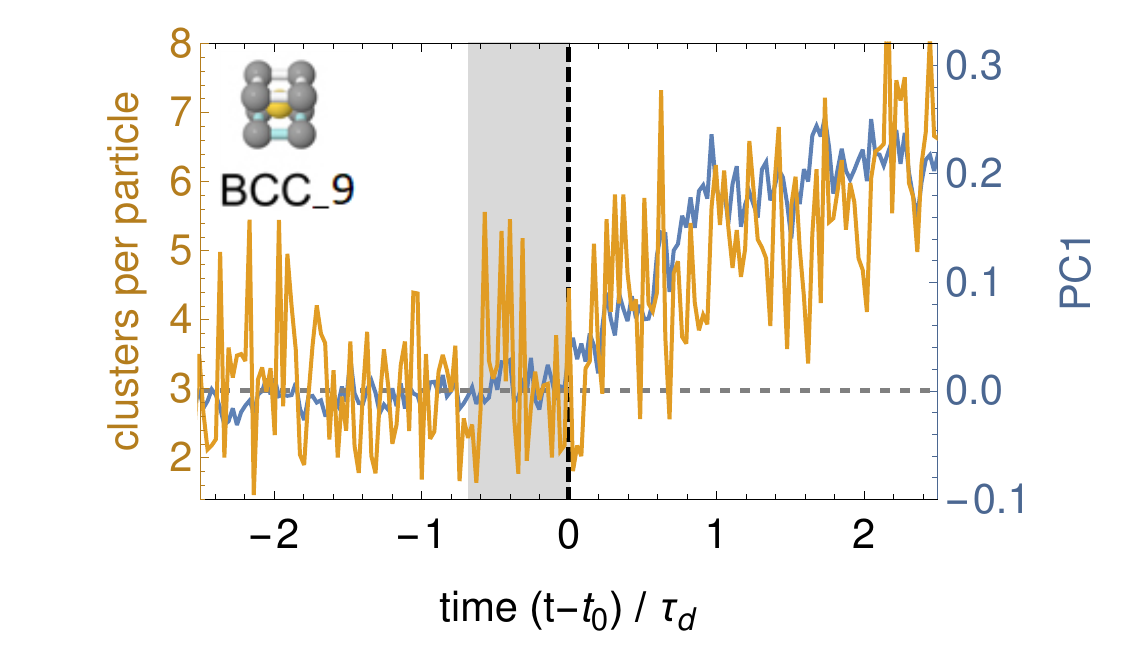} \\
     c) & \hspace{0.5cm} & d)  \\[-0.64cm]
     \includegraphics[width=\figwidthC]{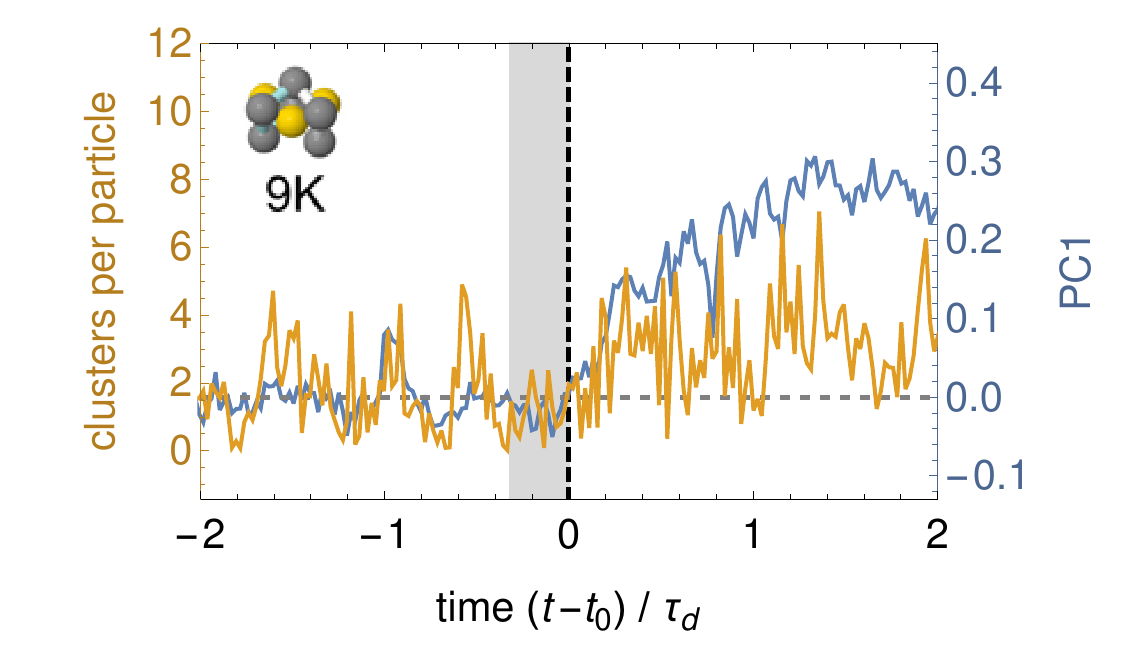} & & \includegraphics[width=\figwidthC]{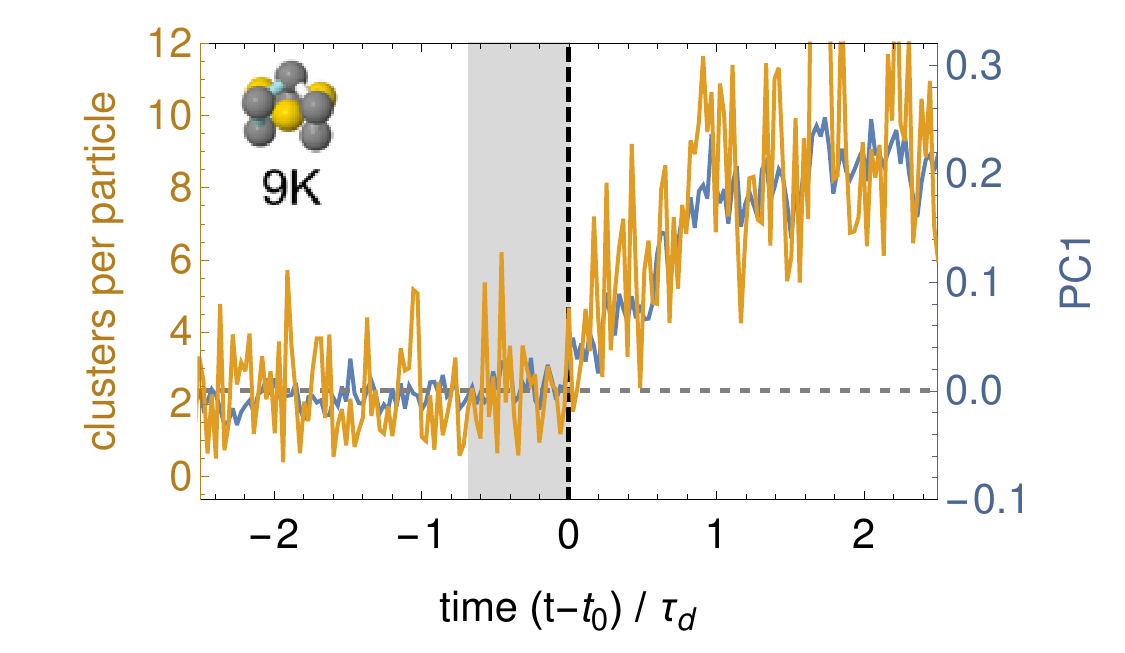} \\
     e) & \hspace{0.5cm} & f)  \\[-0.64cm]
     \includegraphics[width=\figwidthC]{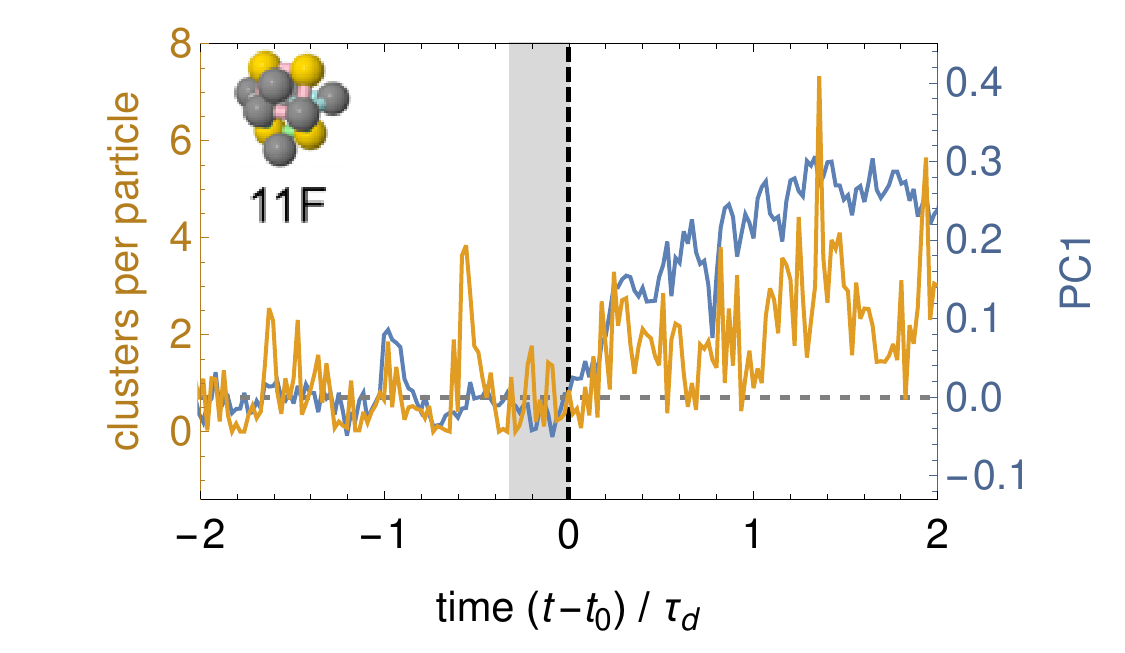} & & \includegraphics[width=\figwidthC]{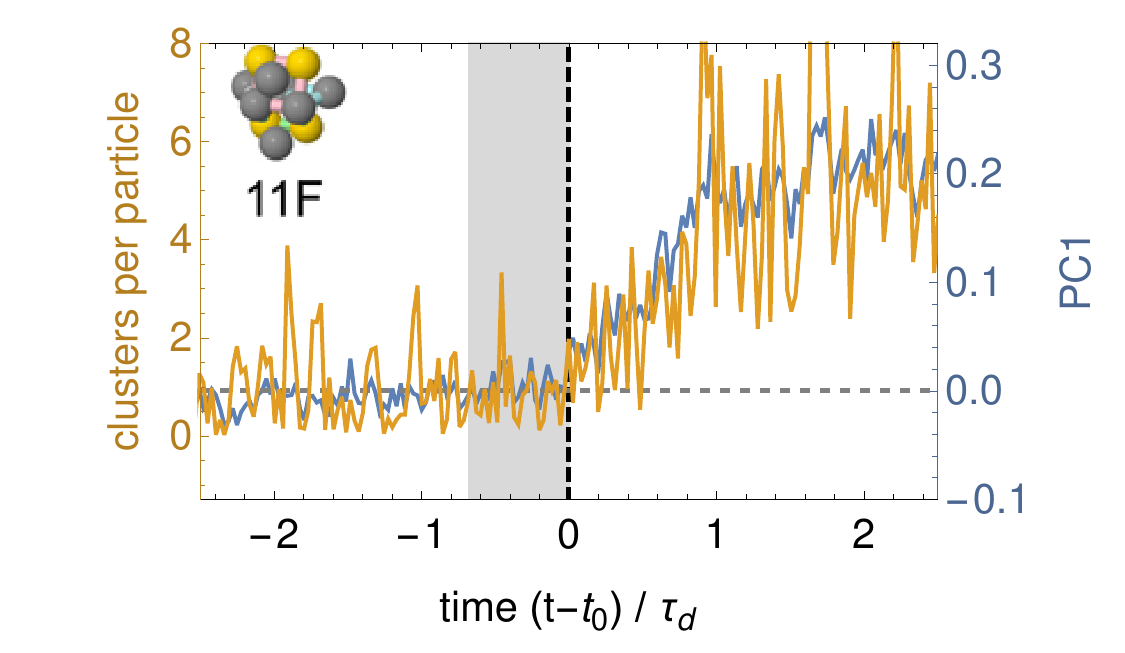} \\
     g) & \hspace{0.5cm} & h)  \\[-0.64cm]
     \includegraphics[width=\figwidthC]{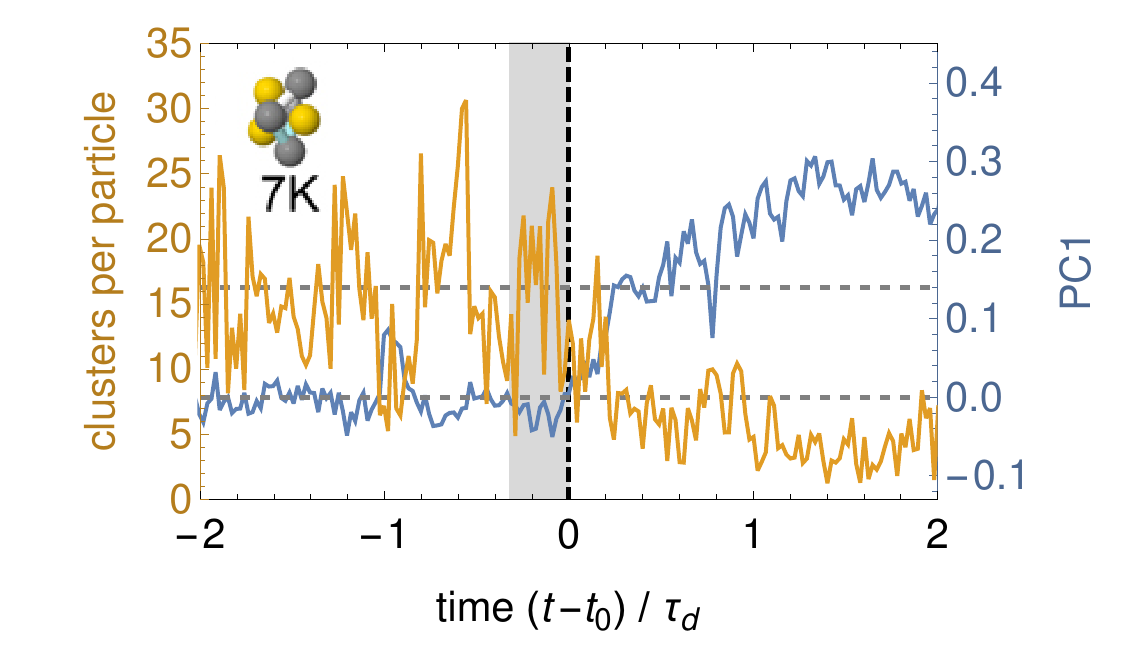} & & \includegraphics[width=\figwidthC]{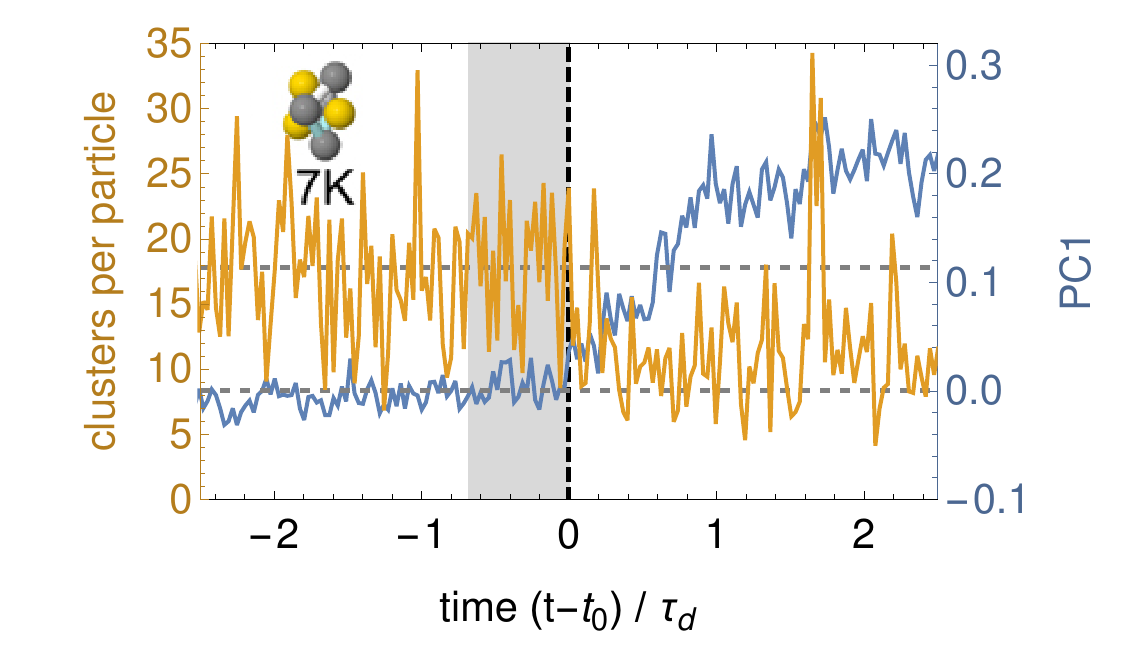} \\
     i) & \hspace{0.5cm} & j)  \\[-0.64cm]
     \includegraphics[width=\figwidthC]{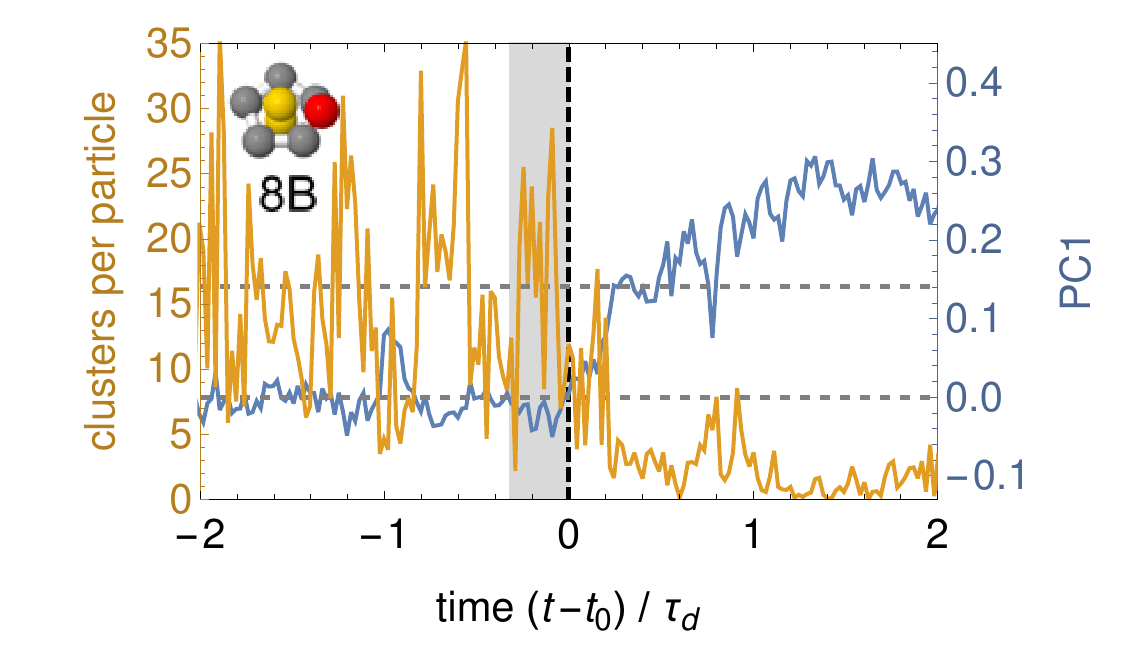} & & \includegraphics[width=\figwidthC]{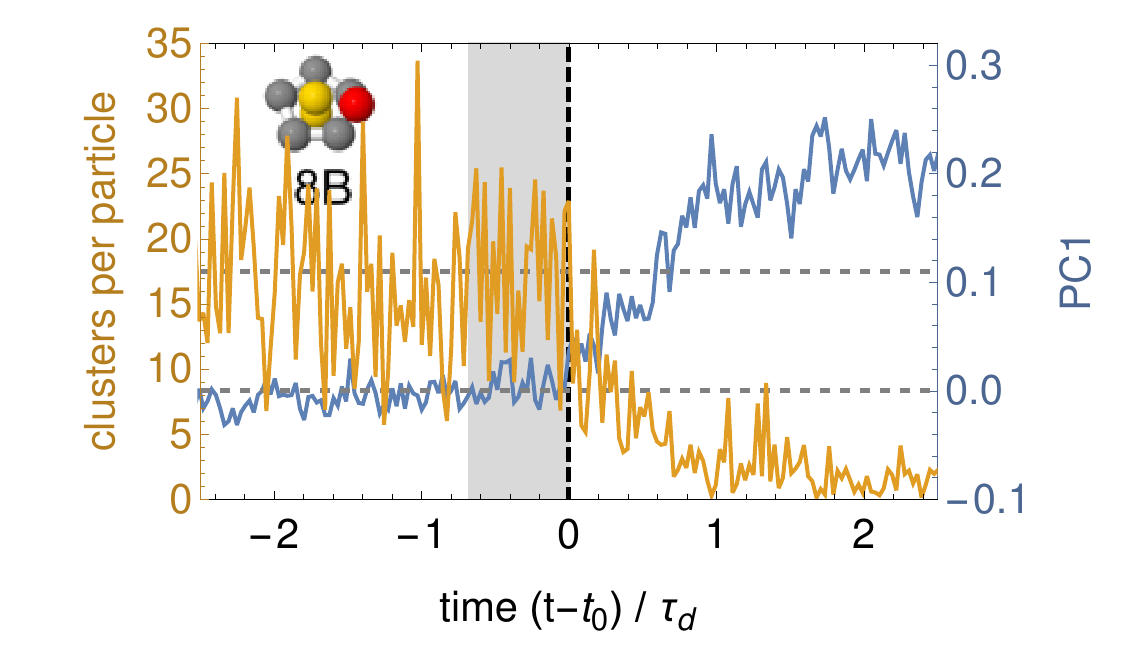} 
\end{tabular}
    \caption[width=1\linewidth]{\label{fig:maineventstccmore} 
    For the same events as in Figs. 7 and 8 of the main paper, i.e. hard spheres (left) and soft hard-core Yukawa particles (right), the average number of clusters per particle (left axis, yellow) for a couple of TCC clusters together with the average value of PC1 (right axis, blue). The horizontal dashed lines indicate the reference values of the number of clusters per particle and PC1 in the fluid. In a-f) the left axis is scaled in such a way that these lines lie on top of each other. In g-j) this was not possible without inverting one of the axes.
    }
\end{figure*}

Figure \ref{fig:maineventstccmore} shows for the same events as in the main paper the average number of a certain TCC cluster a particle is involved in. It is an addition to Fig. 8 of the main paper, which already gives these results for the 6A, 8A, FCC, and 7A clusters. Again, we see that there is no significant change prior to the start of nucleation. The BCC\_9, 9K, and 11F clusters all increase as soon as nucleation starts, while the 7K and 8B clusters both decrease as soon as nucleation starts. This is to be expected as the former group have a positive correlation with PC1 and the latter a negative. However, notice that BCC\_9 is more present in the nucleation of the FCC phase than of the BCC phase. This is not entirely surprising as bulk FCC contains more BCC\_9 clusters than bulk BCC. In contrast, the 9K and 11F clusters are found in bulk BCC and not in bulk FCC. Hence, similar to the 8A cluster, we first observe an initial increase in 9K and 11F clusters for the nucleation of the FCC phase, later this increase is followed by a decrease.



\begin{figure*}[t!]
\begin{tabular}{lll}
     a) & \hspace{0.5cm} & b)  \\[-0.6cm]
     \includegraphics[width=\figwidthC]{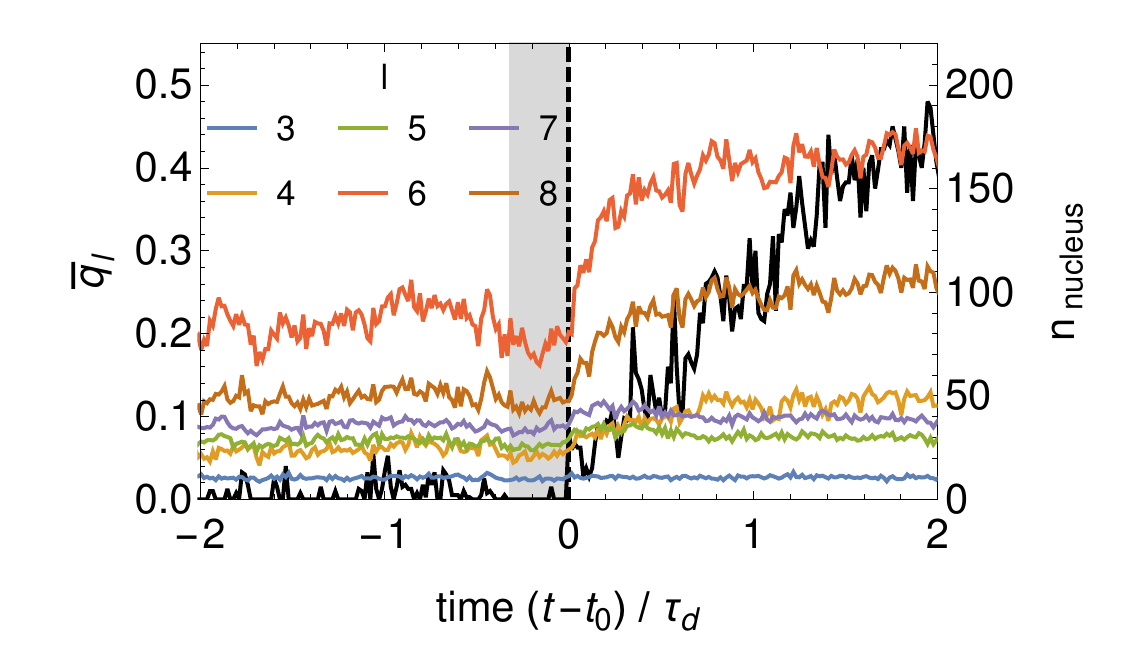} & & \includegraphics[width=\figwidthC]{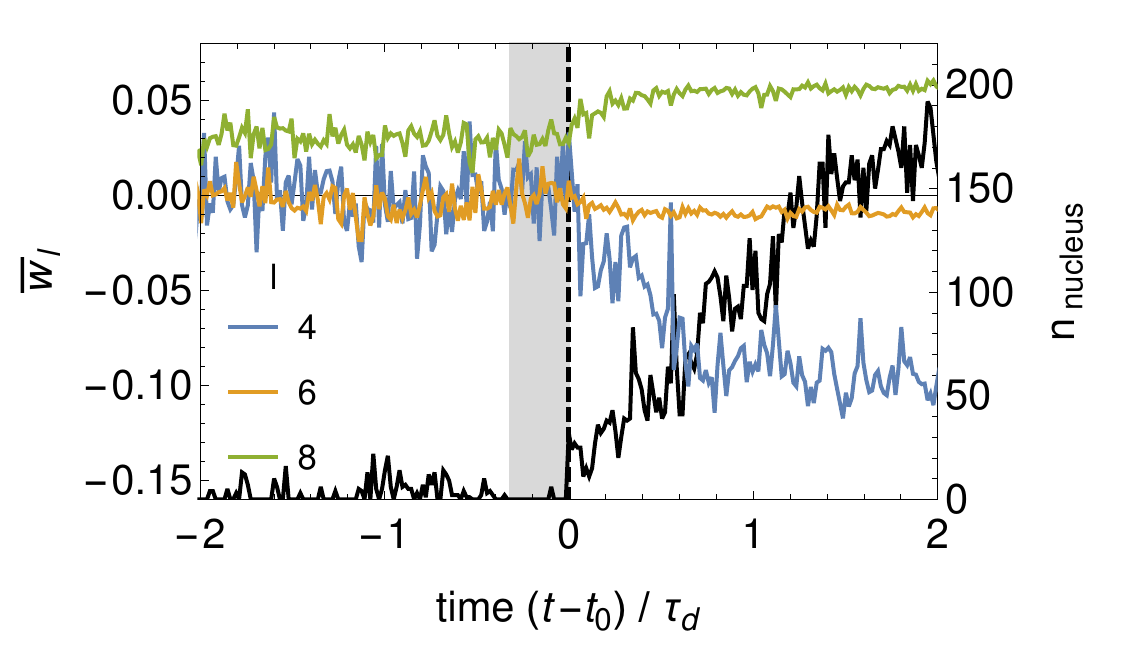} \\
     c) & \hspace{0.5cm} & d)  \\[-0.6cm]
     \includegraphics[width=\figwidthC]{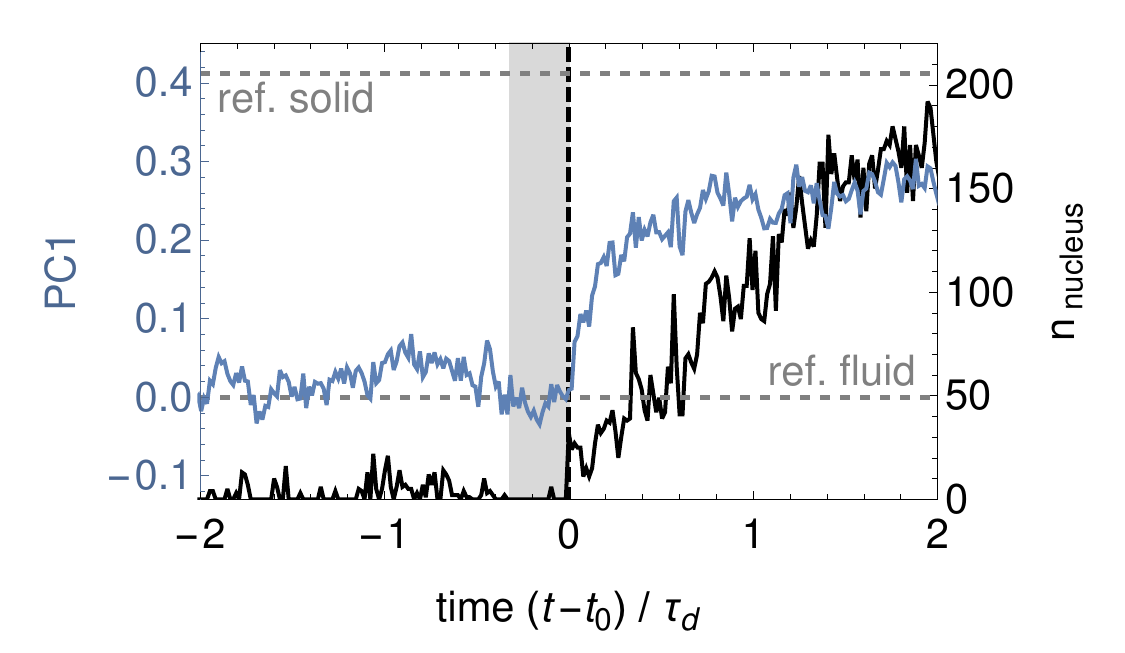} & & \includegraphics[width=\figwidthC]{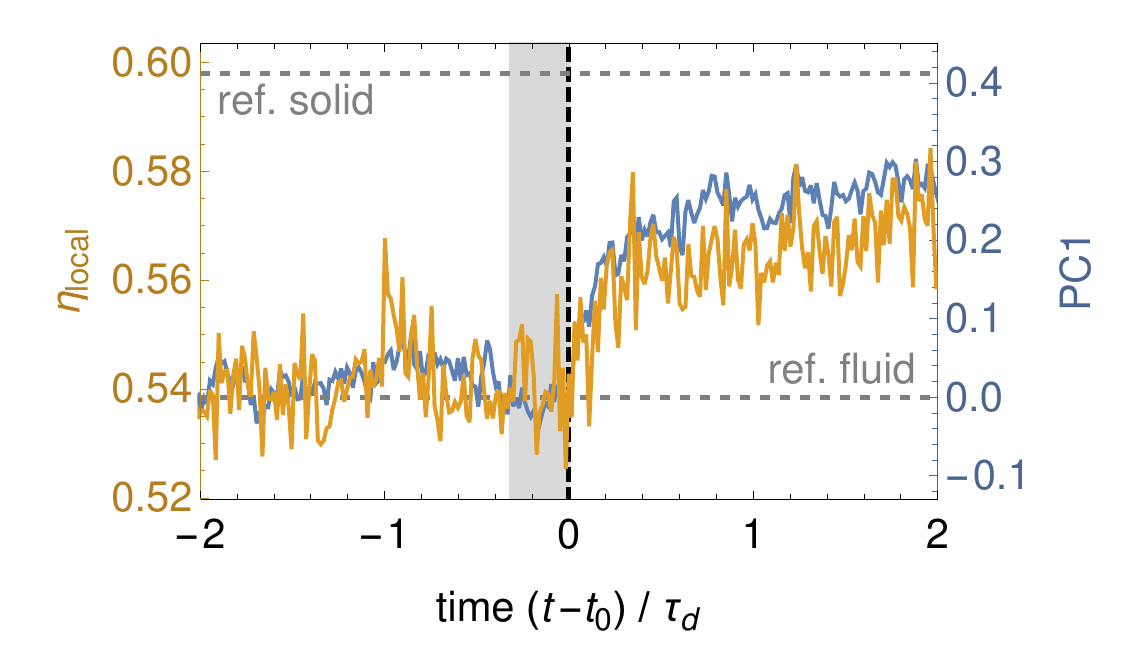} \\
     e) & \hspace{0.5cm} & f)  \\[-0.6cm]
     \includegraphics[width=\figwidthC]{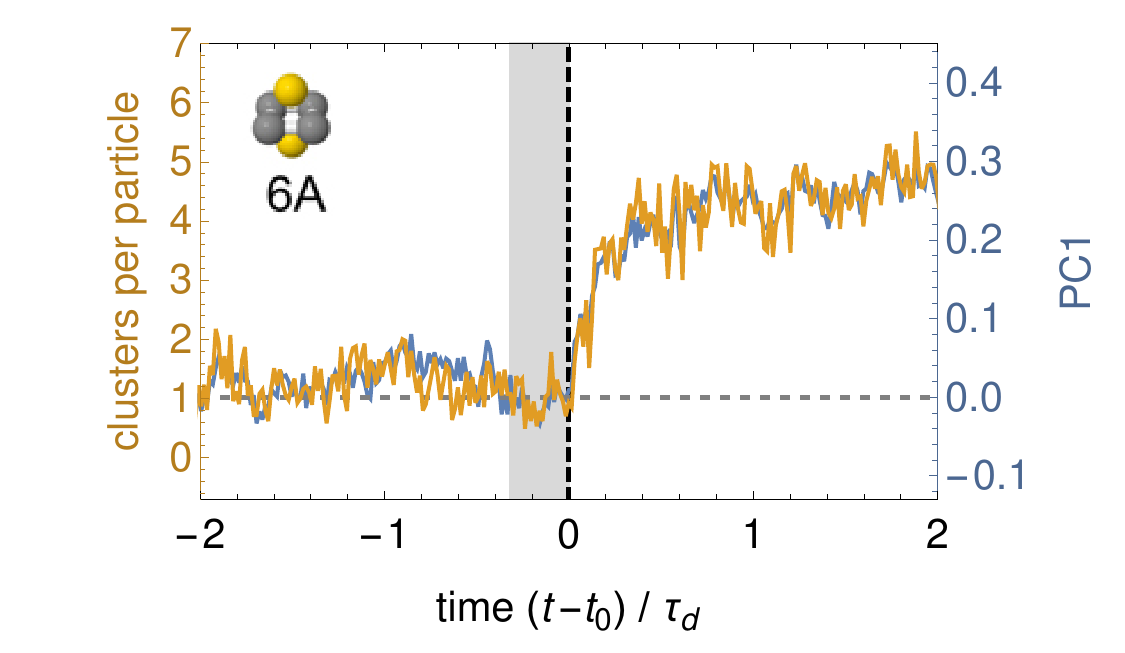} & & \includegraphics[width=\figwidthC]{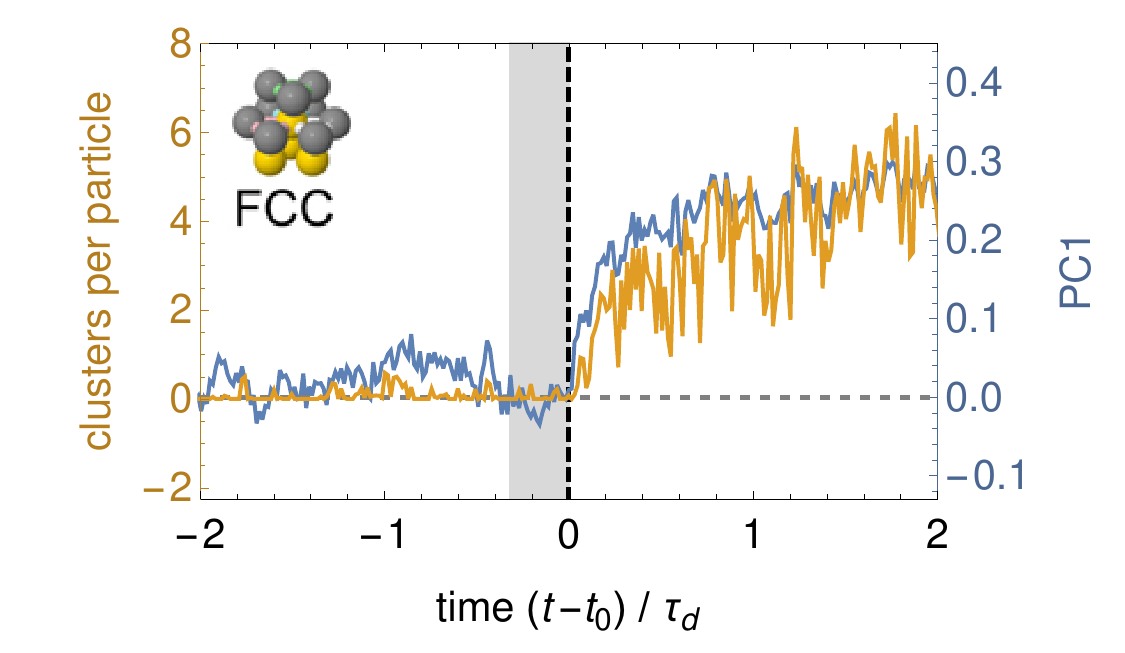} \\
     g) & \hspace{0.5cm} & h)  \\[-0.6cm]
     \includegraphics[width=\figwidthC]{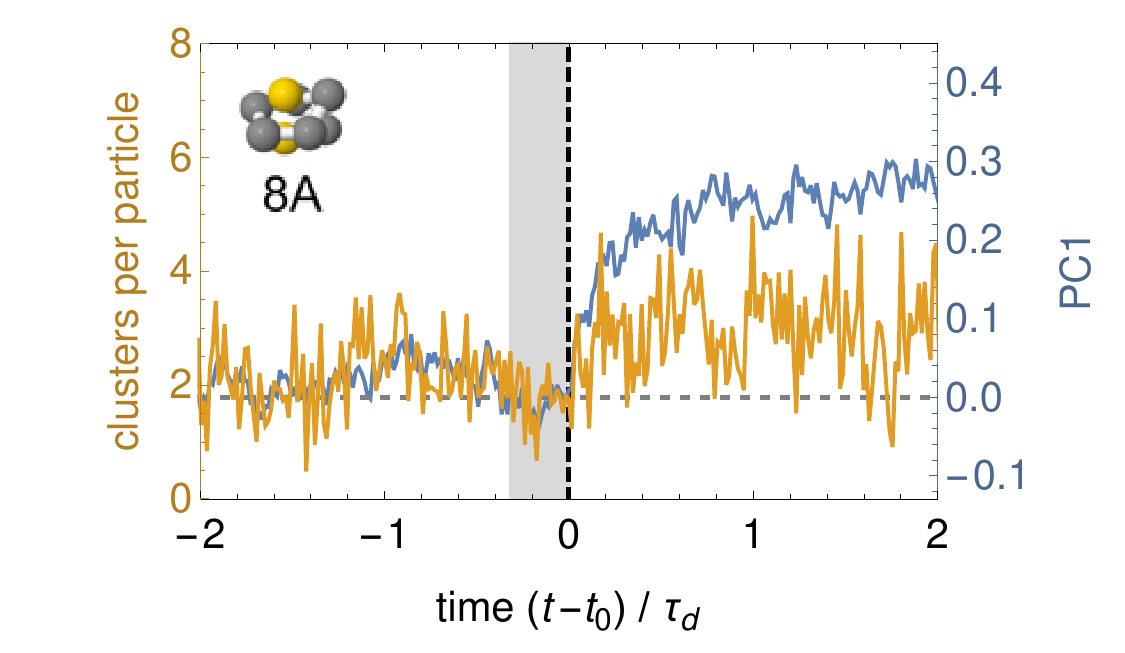} & & \includegraphics[width=\figwidthC]{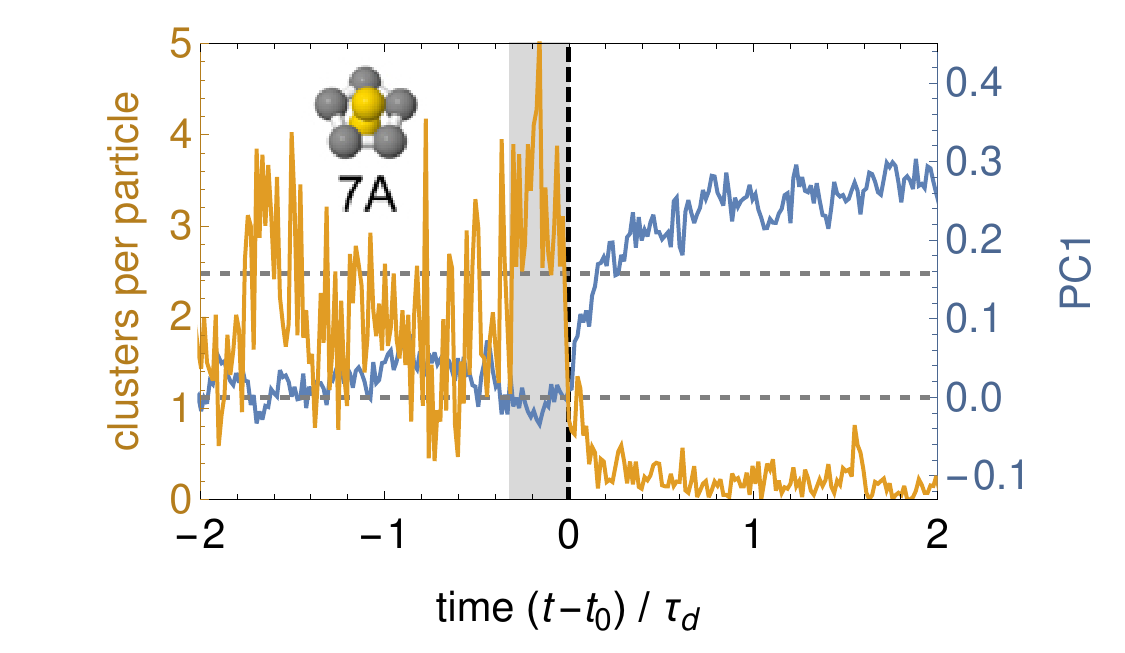}
\end{tabular}
    \caption[width=1\linewidth]{\label{fig:sieventshskmc} 
    Typical nucleation event of hard spheres ($\eta=0.5385$) obtained using KMC. In a-c) the black line (right axis) gives the size of the biggest nucleus present in the studied region, while the other lines (left axis) give the average value of a-b) the BOPs $\bar{q}_l$ and $\bar{w}_l$, and c) PC1. In d-h) blue line (right axis) gives PC1, while the yellow line (left axis) gives d) the local packing fraction and e-h) the average number of clusters per particle for a couple of TCC clusters.}
\end{figure*}

\begin{figure*}[t!]
\begin{tabular}{lll}
     a) & \hspace{0.5cm} & b)  \\[-0.6cm]
     \includegraphics[width=\figwidthC]{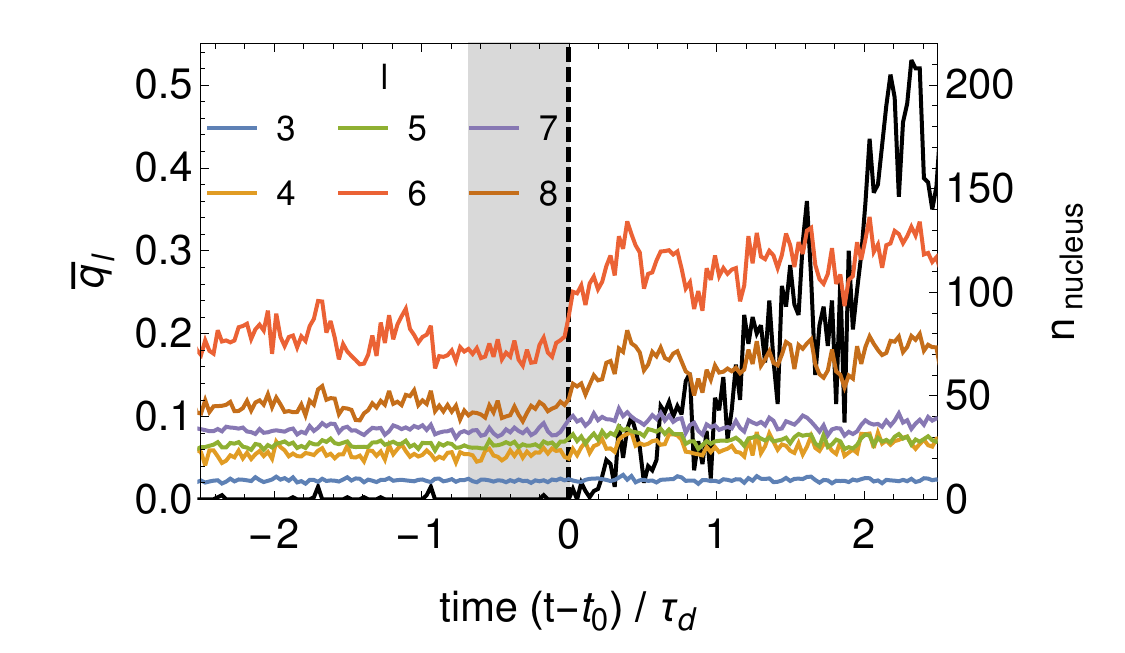} & & \includegraphics[width=\figwidthC]{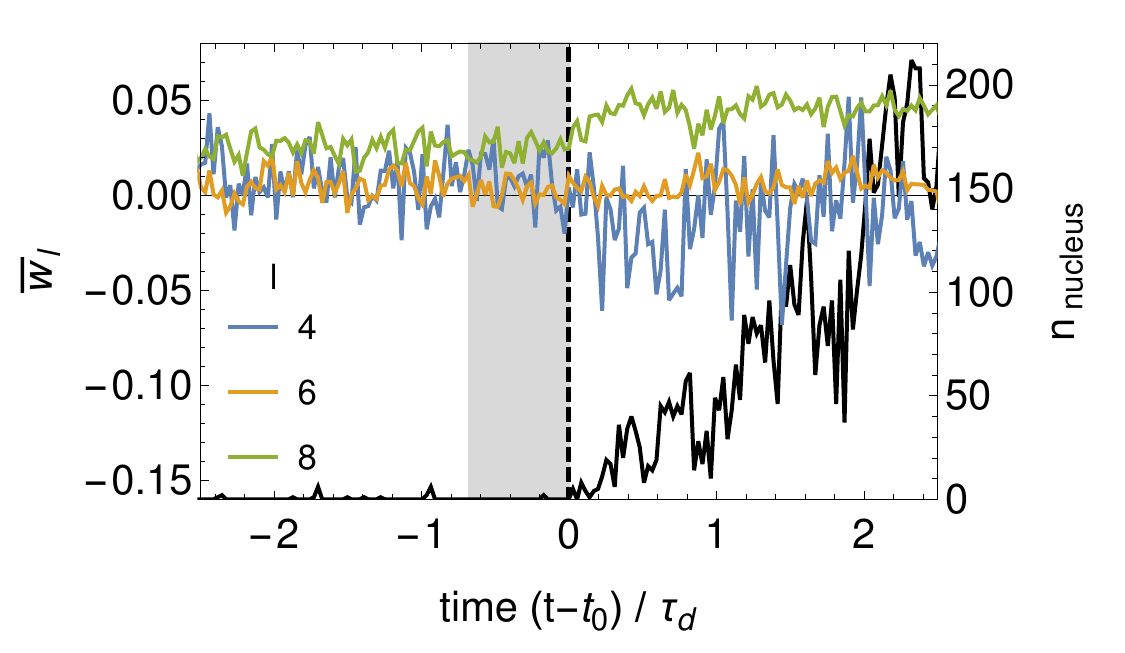} \\
     c) & \hspace{0.5cm} & d)  \\[-0.6cm]
     \includegraphics[width=\figwidthC]{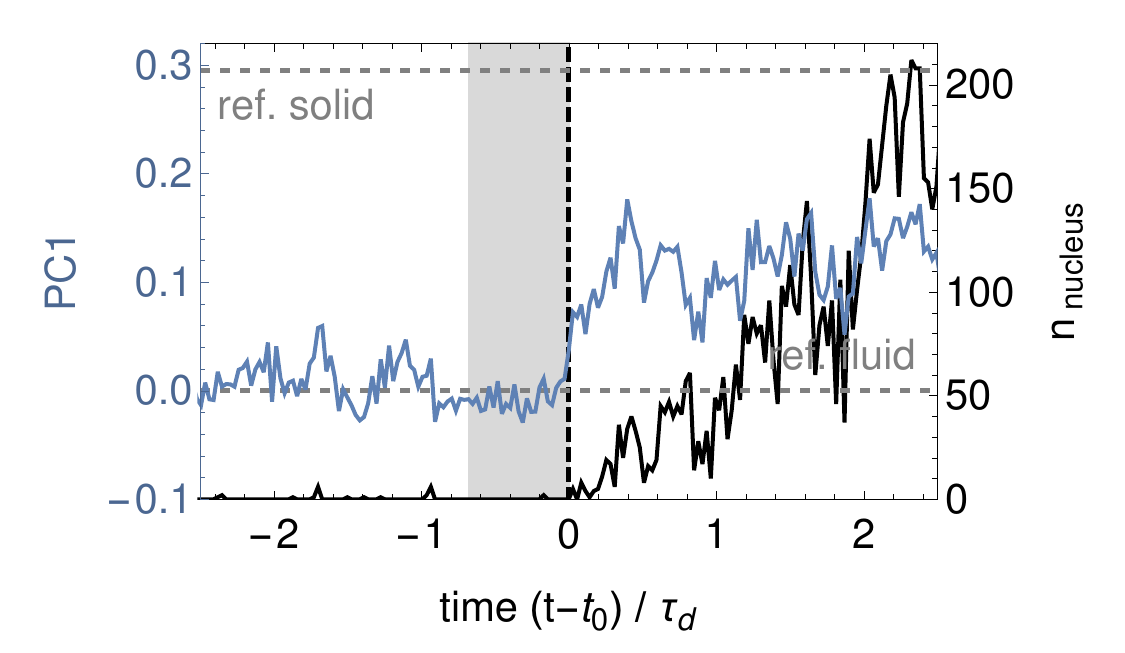} & & \includegraphics[width=\figwidthC]{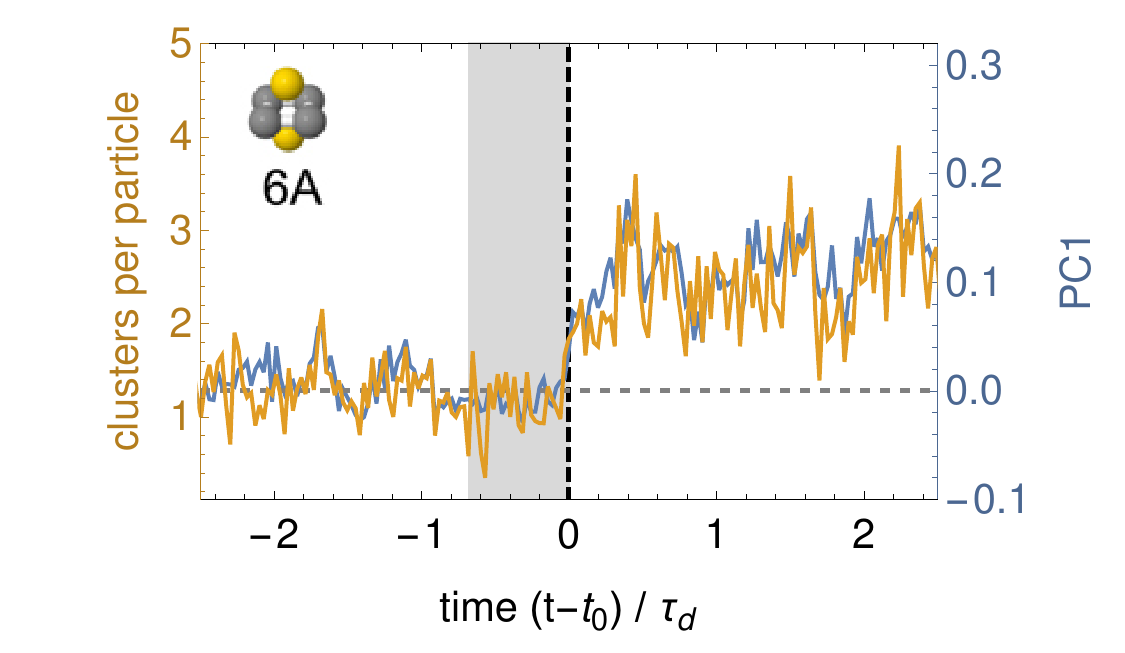} \\
     e) & \hspace{0.5cm} & f)  \\[-0.6cm]
     \includegraphics[width=\figwidthC]{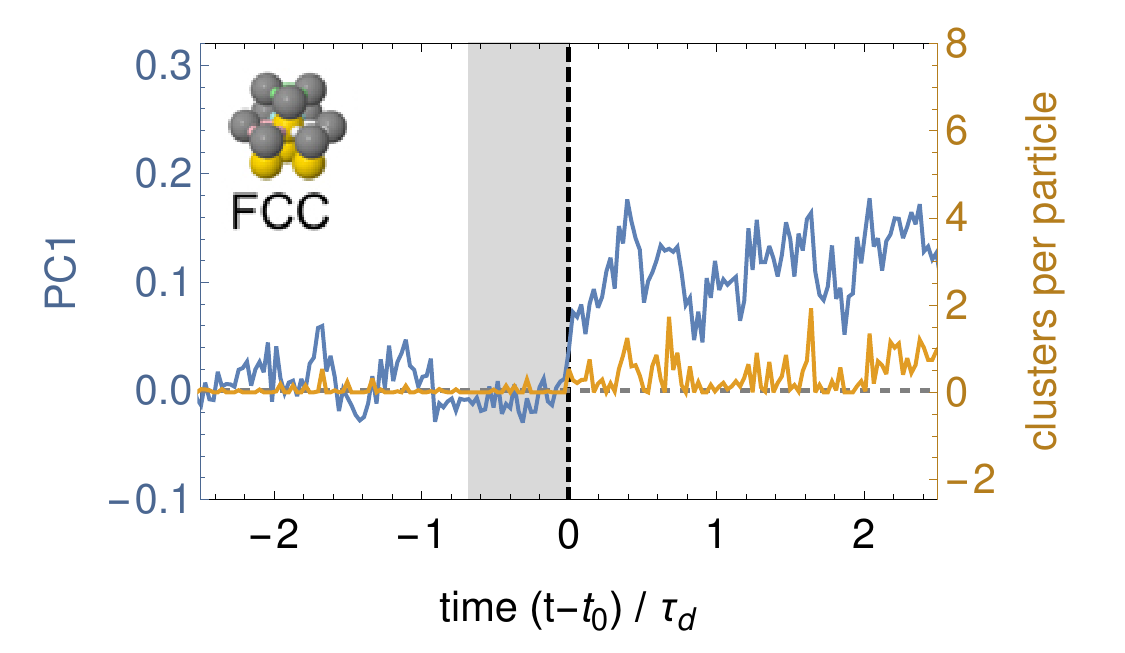} & & \includegraphics[width=\figwidthC]{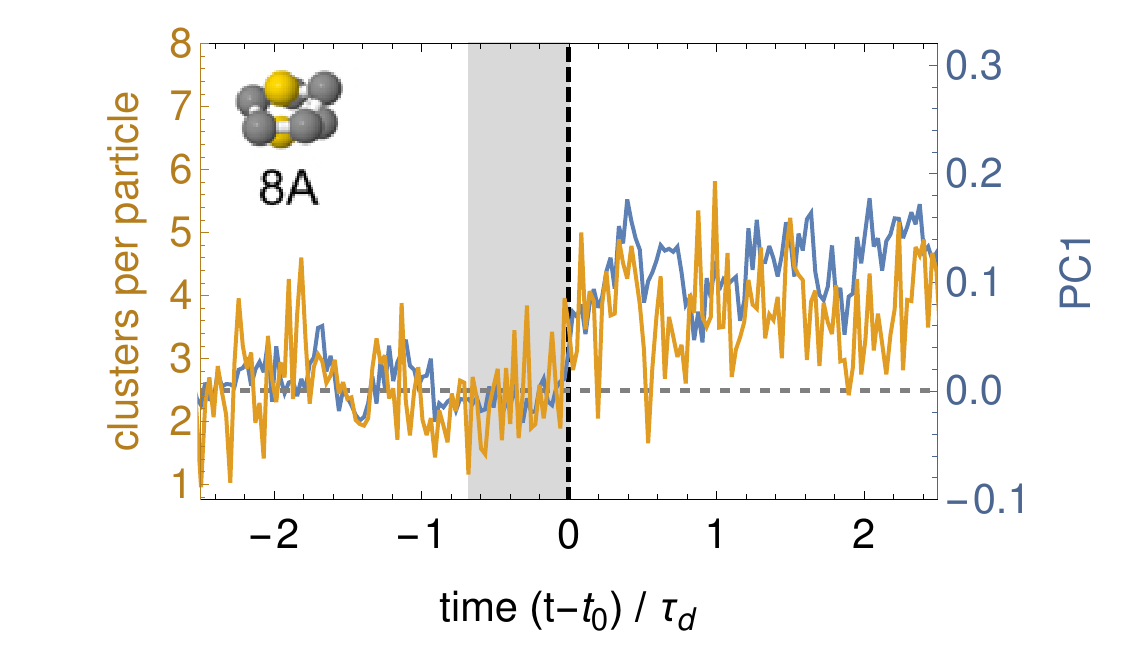} \\
     g) & \hspace{0.5cm} & h)  \\[-0.6cm]
     \includegraphics[width=\figwidthC]{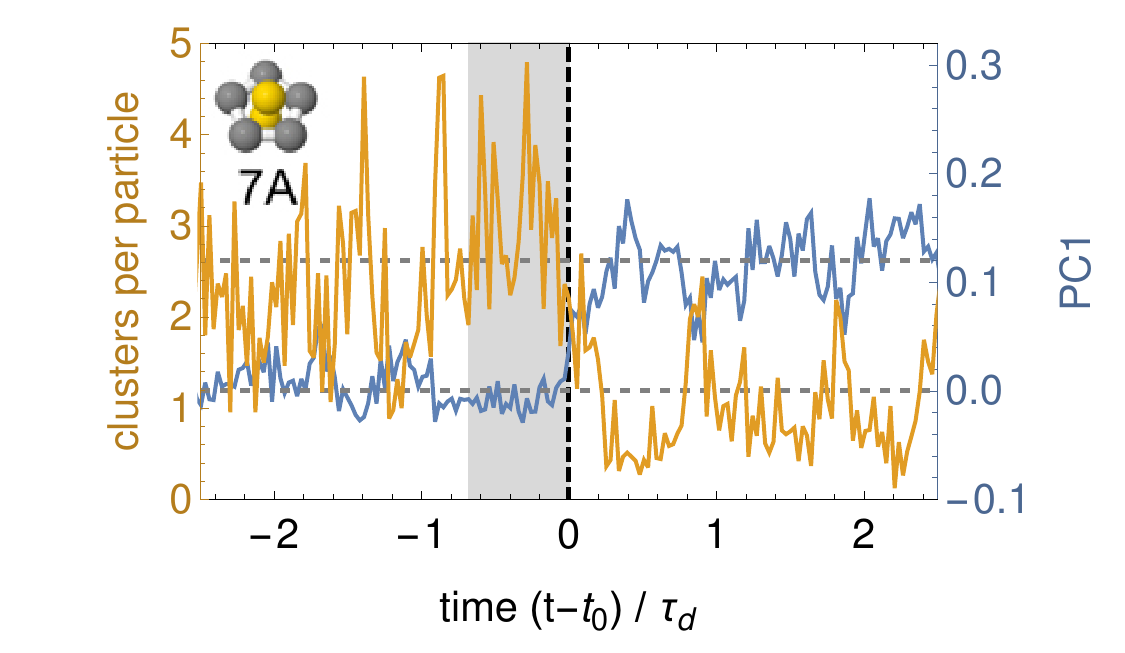} & & \includegraphics[width=\figwidthC]{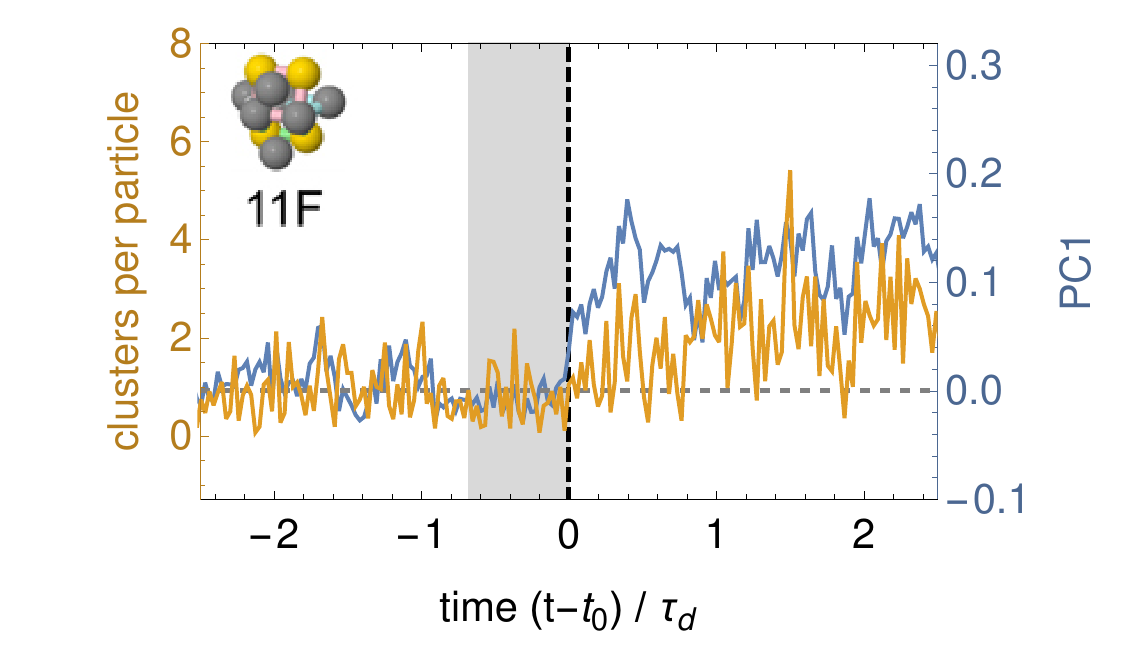}
\end{tabular}
    \caption[width=1\linewidth]{\label{fig:sieventsyukbcckmc}
    Typical nucleation event of soft hard-core Yukawa particles ($\beta\epsilon=81$, $1/\kappa\sigma=0.40$, $\eta=0.1305$) obtained using KMC. In a-c) the black line (right axis) gives the size of the biggest nucleus present in the studied region, while the other lines (left axis) give the average value of a-b) the BOPs $\bar{q}_l$ and $\bar{w}_l$, and c) PC1. In d-h) blue line (right axis) gives PC1, while the yellow line (left axis) gives the average number of clusters per particle for a couple of TCC clusters.
    }
\end{figure*}

\begin{figure*}[t!]
\begin{tabular}{lll}
     a) & \hspace{0.5cm} & b)  \\[-0.6cm]
     \includegraphics[width=\figwidthC]{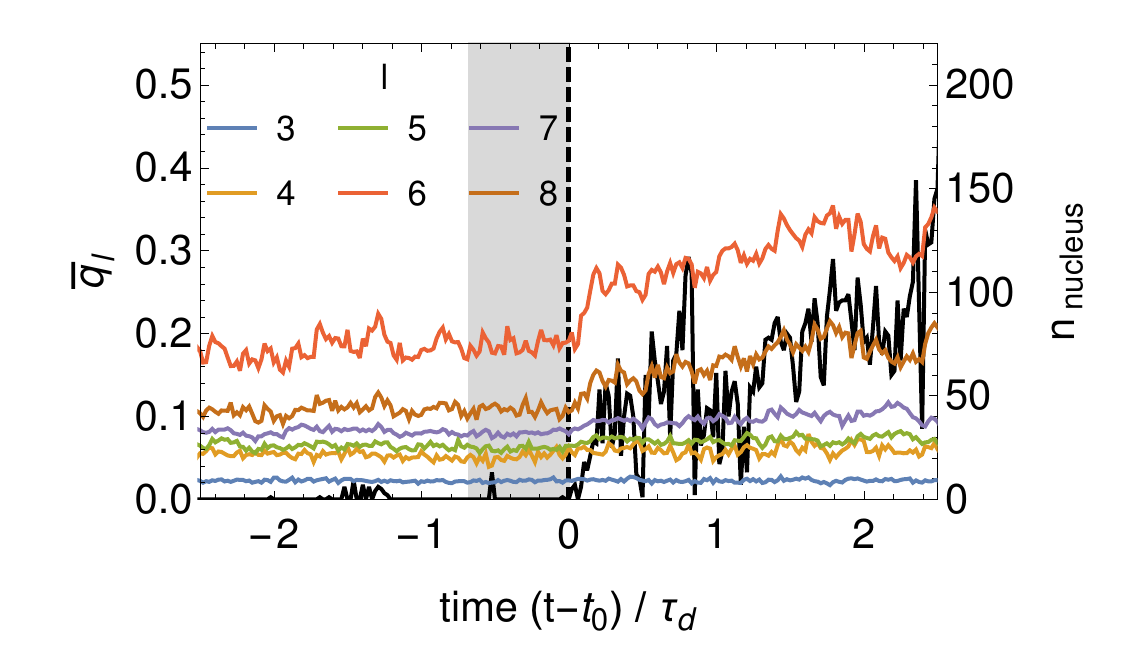} & & \includegraphics[width=\figwidthC]{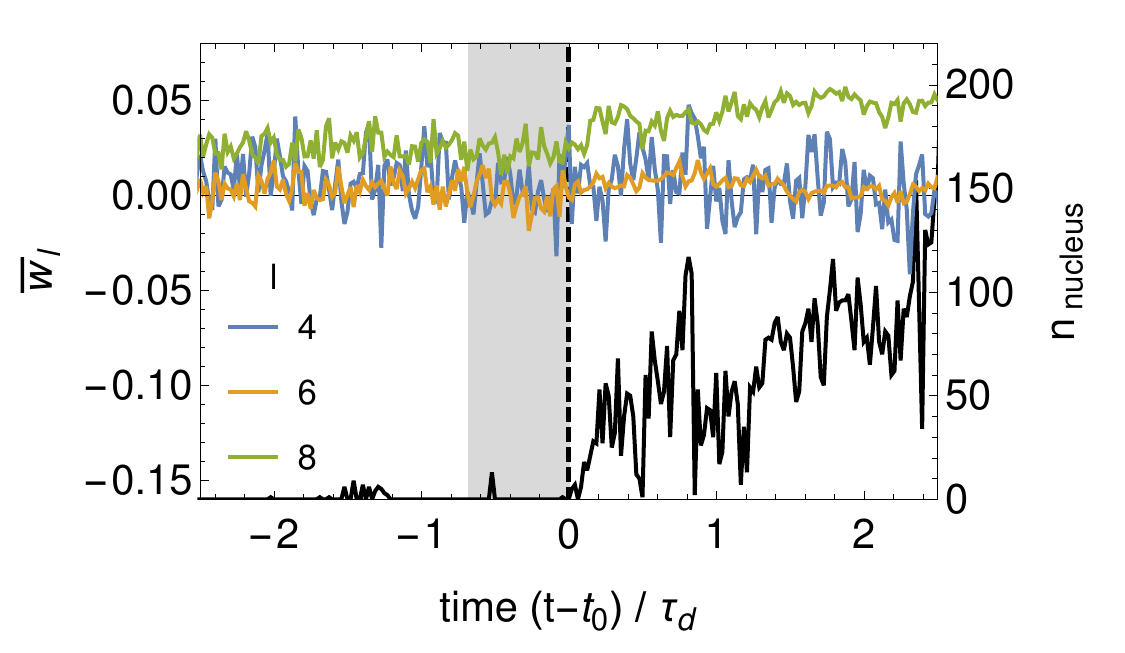} \\
     c) & \hspace{0.5cm} & d)  \\[-0.6cm]
     \includegraphics[width=\figwidthC]{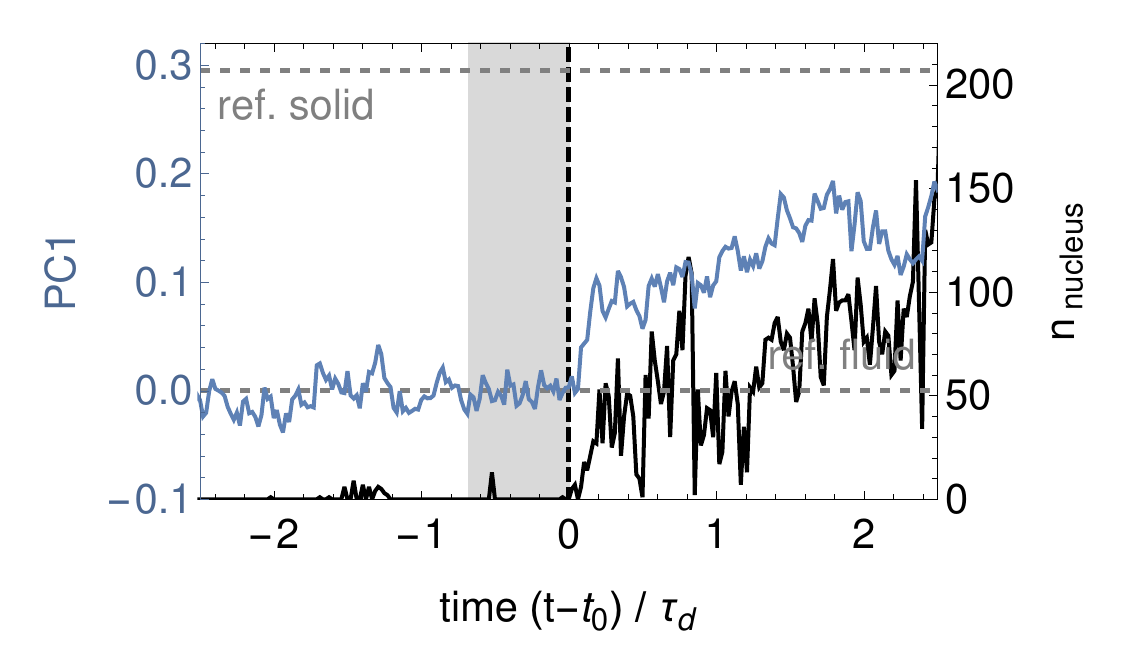} & & \includegraphics[width=\figwidthC]{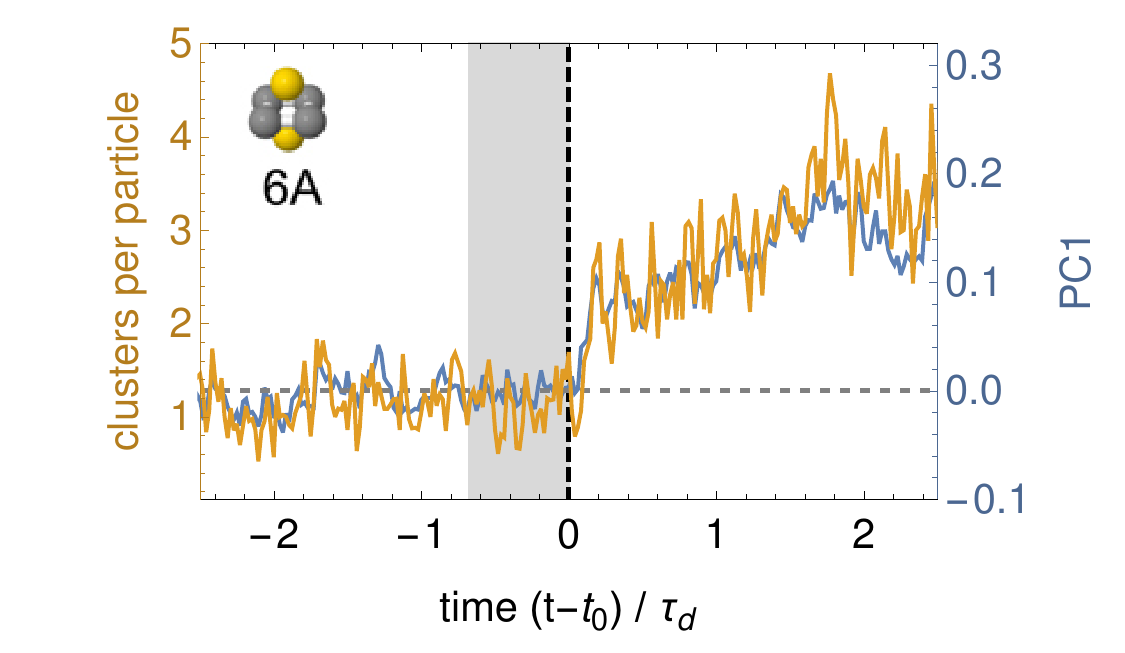} \\
     e) & \hspace{0.5cm} & f)  \\[-0.6cm]
     \includegraphics[width=\figwidthC]{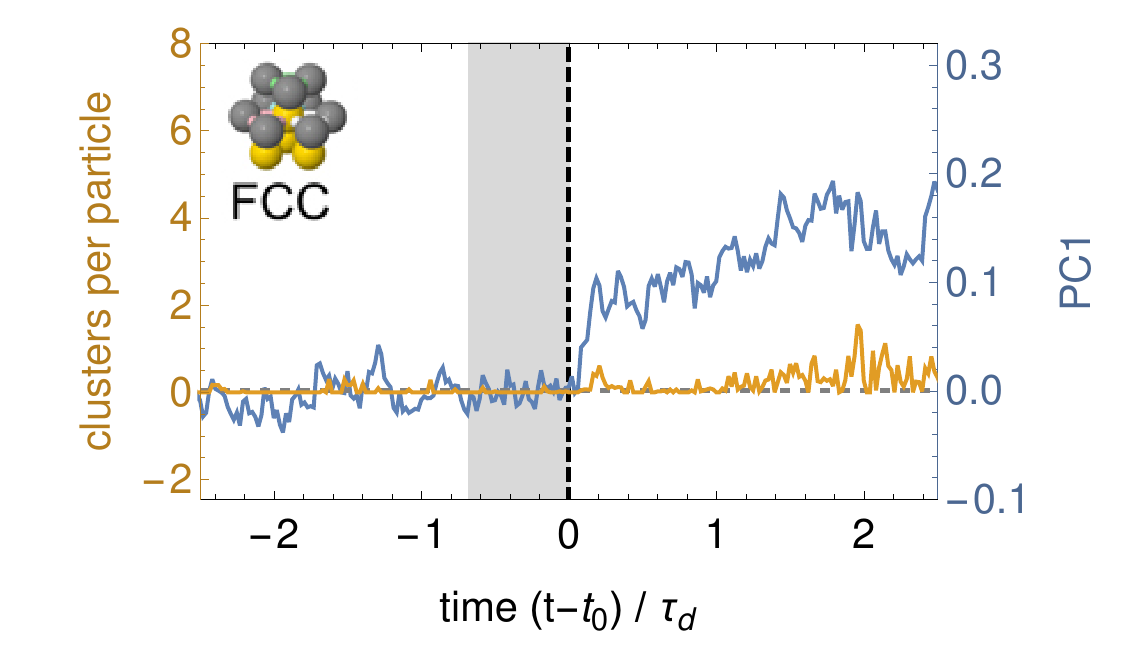} & & \includegraphics[width=\figwidthC]{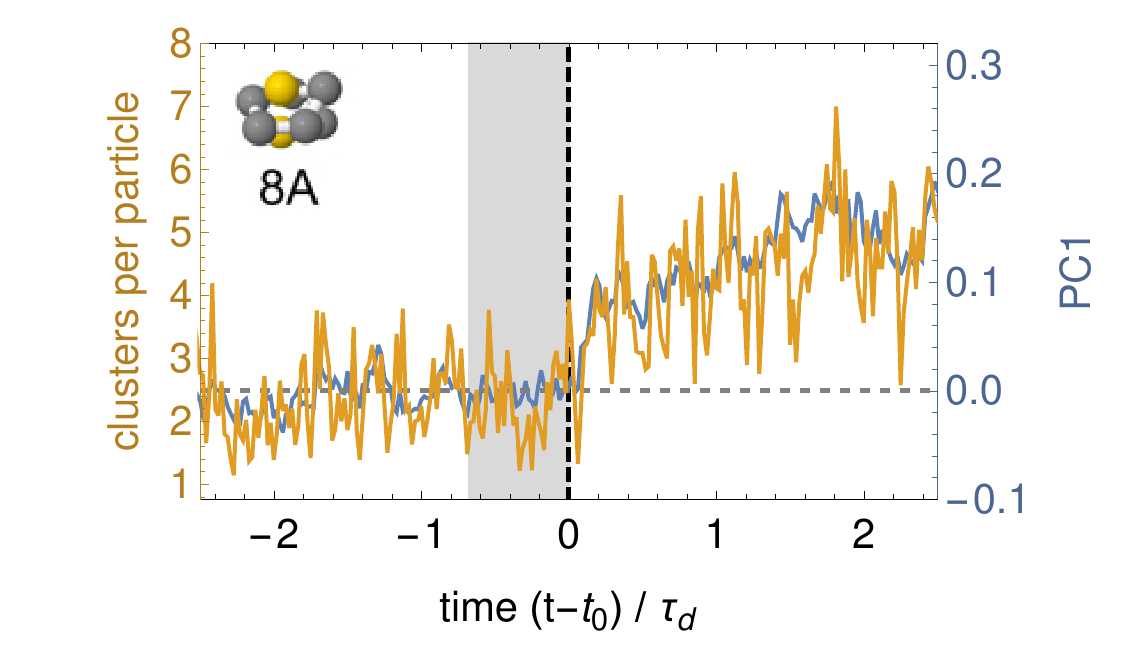} \\
     g) & \hspace{0.5cm} & h)  \\[-0.6cm]
     \includegraphics[width=\figwidthC]{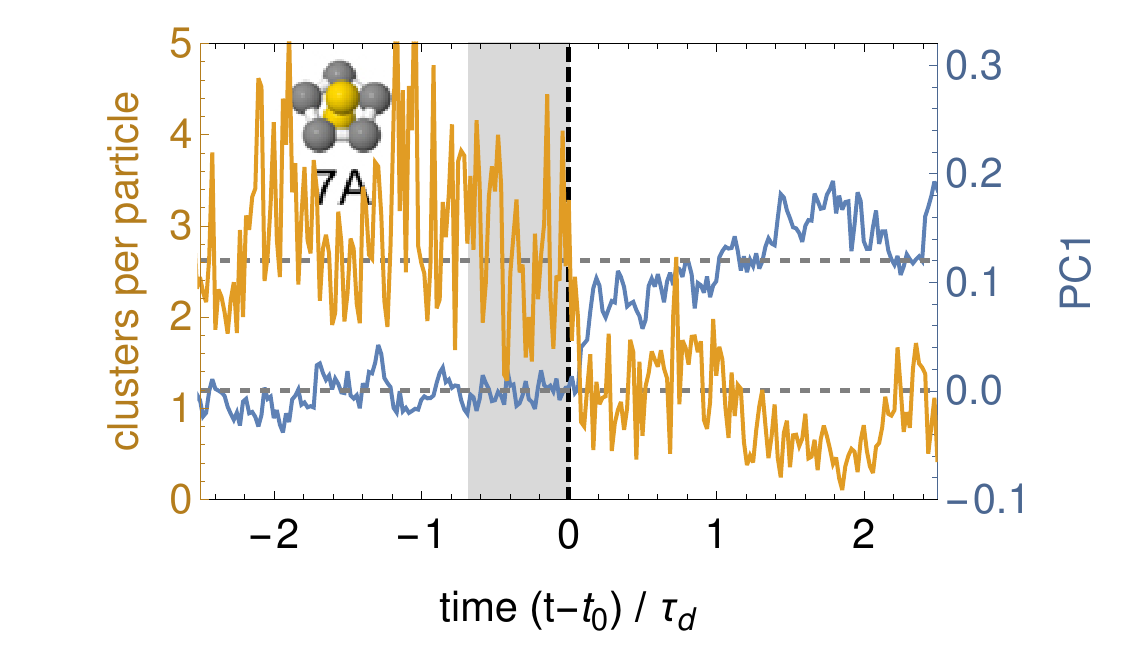} & & \includegraphics[width=\figwidthC]{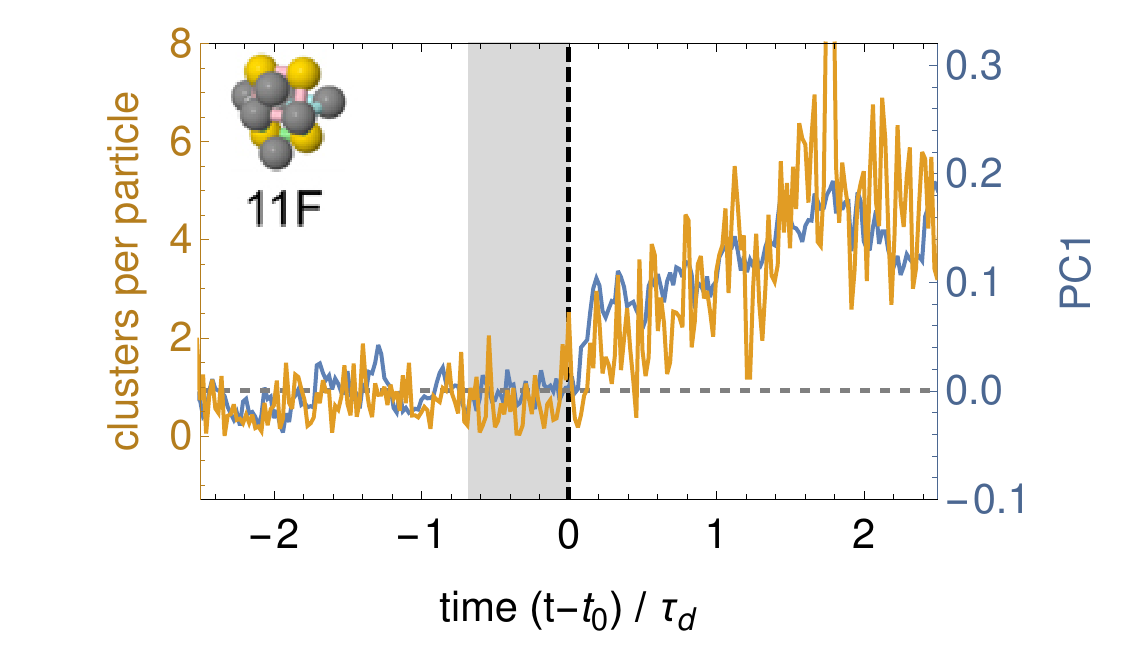}
\end{tabular}
    \caption[width=1\linewidth]{\label{fig:sieventsyukbccmd}
    Typical nucleation event of soft hard-core Yukawa particles ($\beta\epsilon=81$, $1/\kappa\sigma=0.40$, $\eta=0.1305$) obtained using MD. In a-c) the black line (right axis) gives the size of the biggest nucleus present in the studied region, while the other lines (left axis) give the average value of a-b) the BOPs $\bar{q}_l$ and $\bar{w}_l$, and c) PC1. In d-h) blue line (right axis) gives PC1, while the yellow line (left axis) gives the average number of clusters per particle for a couple of TCC clusters.
    }
\end{figure*}

The two events discussed in the main paper were obtained using MC simulations. In order to demonstrate that our observations and conclusions do not depend on the specific dynamics of the simulations method, we also show the results on three more nucleation events of the same systems but then obtained using a different simulation method. 
In Fig. \ref{fig:sieventshskmc} we show the tracking results for a typical nucleation event of the hard-spheres system obtained using kinetic MC (KMC), and in Figs. \ref{fig:sieventsyukbcckmc} and \ref{fig:sieventsyukbccmd} two of the soft spheres obtained using, respectively, KMC and molecular dynamics (MD). 
The results are very similar to those of the nucleation events obtained using MC. Again we observe no anomalous behavior in the metastable fluid prior to the start of nucleation, and all local order parameters abruptly change as soon as nucleation starts. We, thus, conclude that our results and conclusions do not depend on the specific simulation method used.


\begin{figure*}[t!]
\begin{tabular}{lll}
     a) & \hspace{0.5cm} & b)  \\[-0.6cm]
     \includegraphics[width=\figwidthC]{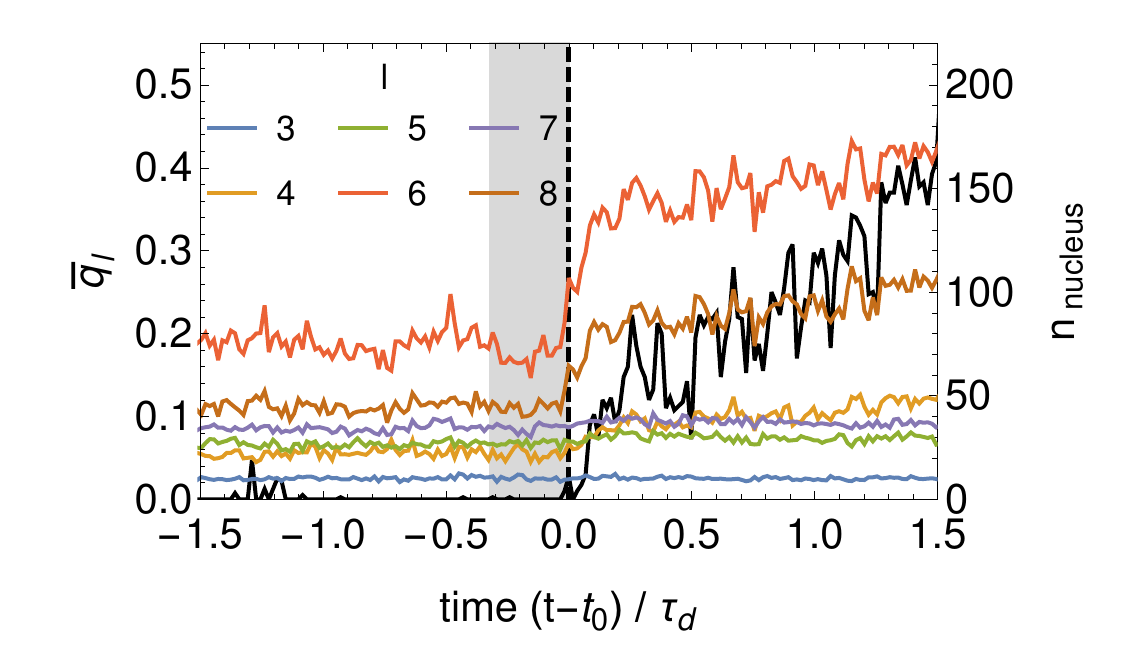} & & \includegraphics[width=\figwidthC]{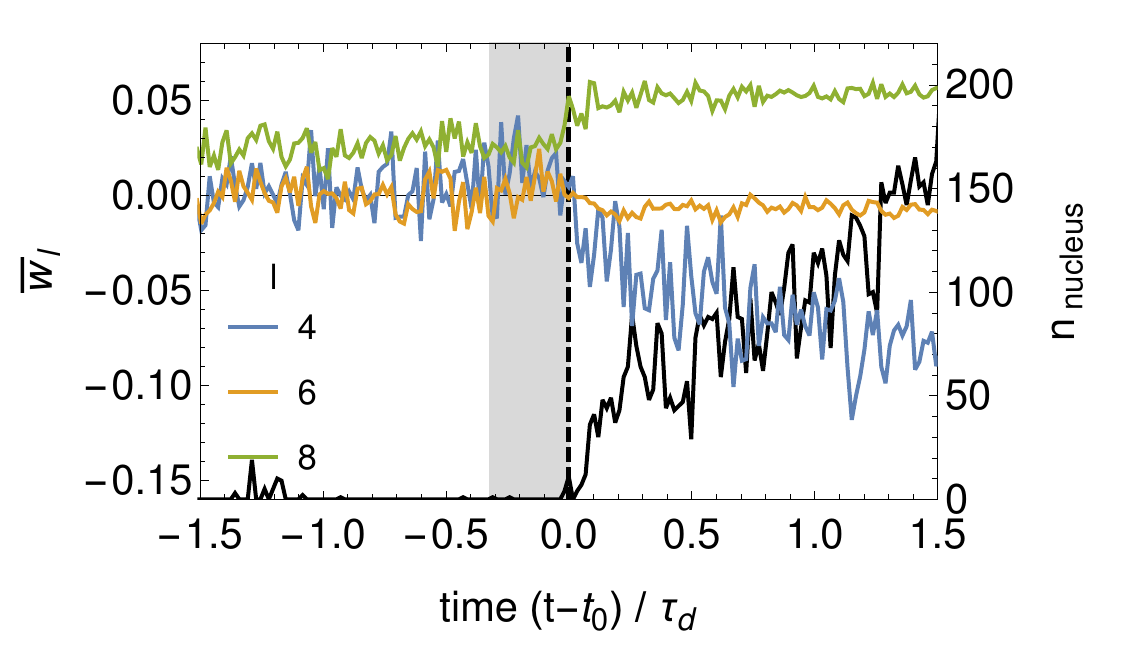} \\
     c) & \hspace{0.5cm} & d)  \\[-0.6cm]
     \includegraphics[width=\figwidthC]{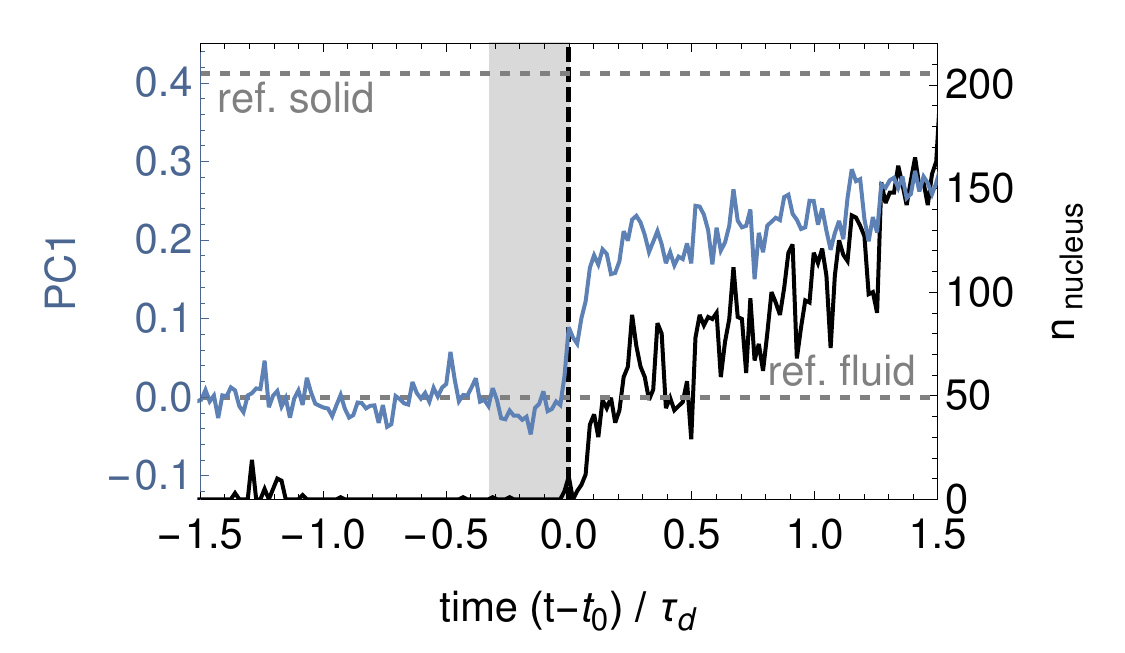} & & \includegraphics[width=\figwidthC]{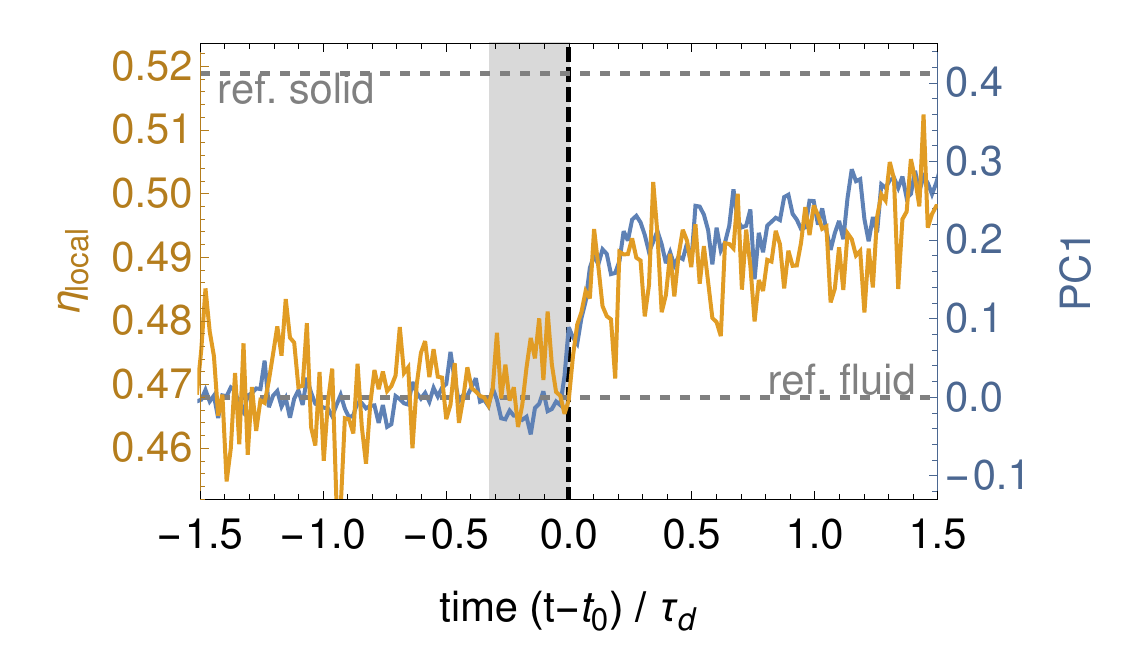} \\
     e) & \hspace{0.5cm} & f)  \\[-0.6cm]
     \includegraphics[width=\figwidthC]{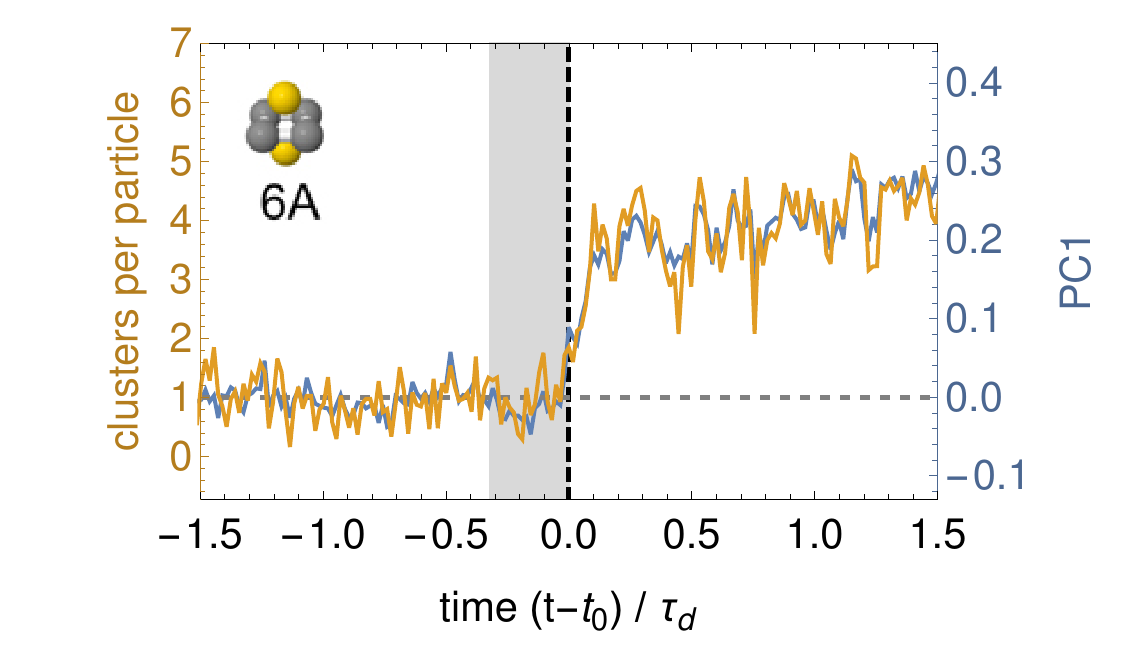} & & \includegraphics[width=\figwidthC]{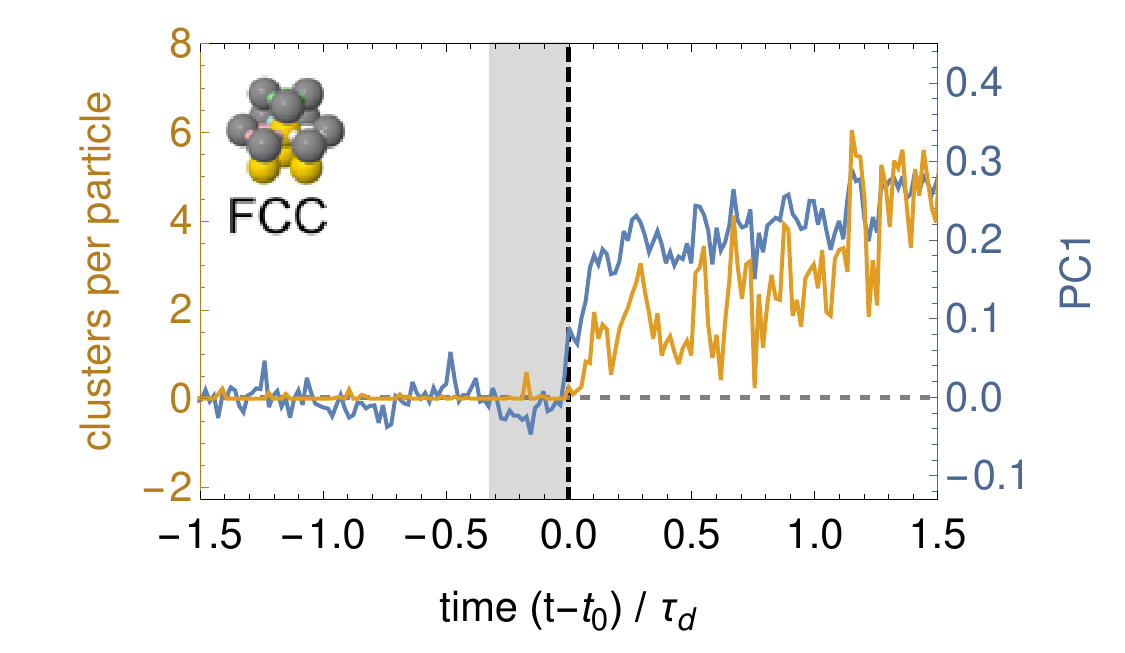} \\
     g) & \hspace{0.5cm} & h)  \\[-0.6cm]
     \includegraphics[width=\figwidthC]{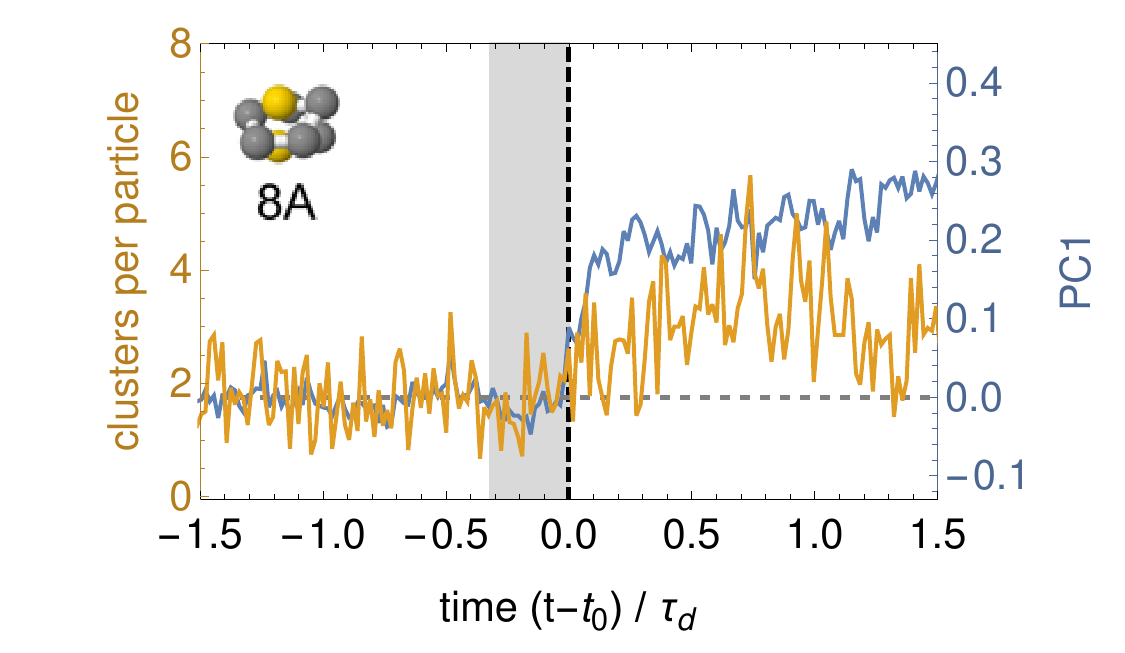} & & \includegraphics[width=\figwidthC]{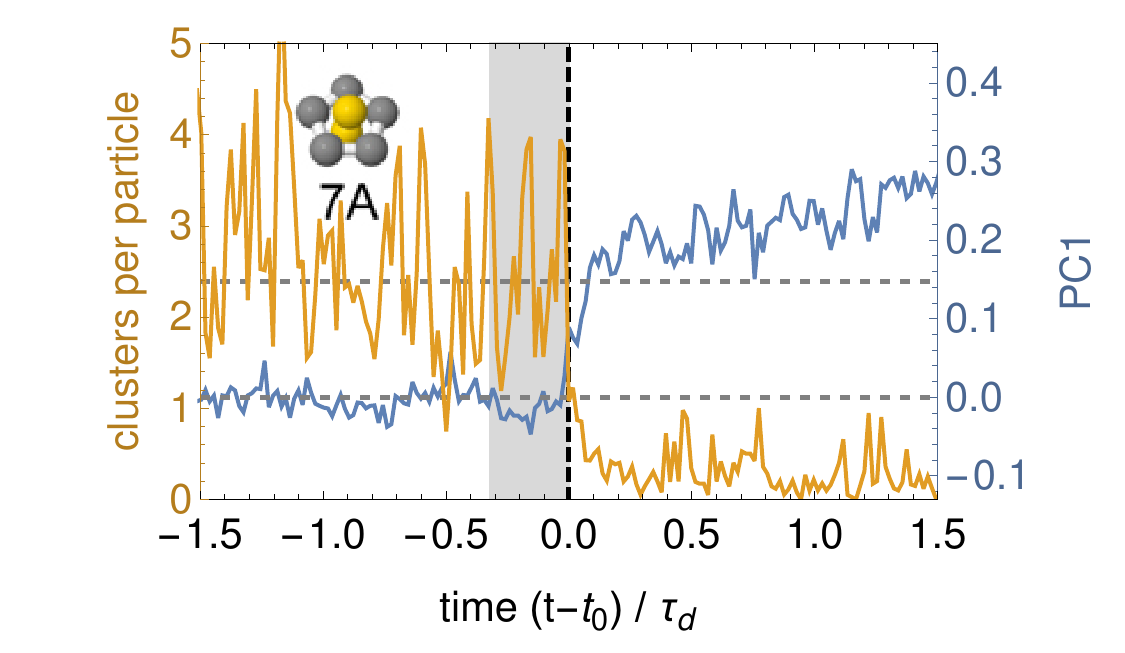}
\end{tabular}
    \caption[width=1\linewidth]{\label{fig:sieventsyukfcce81k100} 
    Typical nucleation event of essentially-hard hard-core Yukawa particles ($\beta\epsilon=81$, $1/\kappa\sigma=0.01$, $\eta=0.4681$) obtained using MC. In a-c) the black line (right axis) gives the size of the biggest nucleus present in the studied region, while the other lines (left axis) give the average value of a-b) the BOPs $\bar{q}_l$ and $\bar{w}_l$, and c) PC1. In d-h) blue line (right axis) gives PC1, while the yellow line (left axis) gives d) the local packing fraction and e-h) the average number of clusters per particle for a couple of TCC clusters.}
\end{figure*}

\begin{figure*}[t!]
\begin{tabular}{lll}
     a) & \hspace{0.5cm} & b)  \\[-0.6cm]
     \includegraphics[width=\figwidthC]{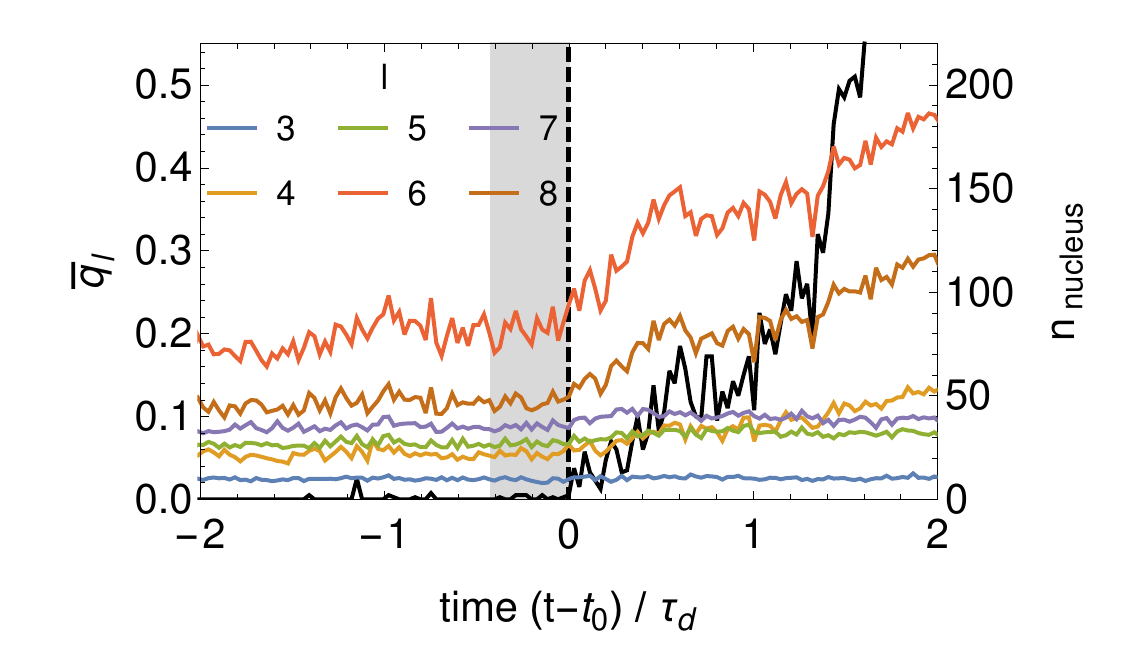} & & \includegraphics[width=\figwidthC]{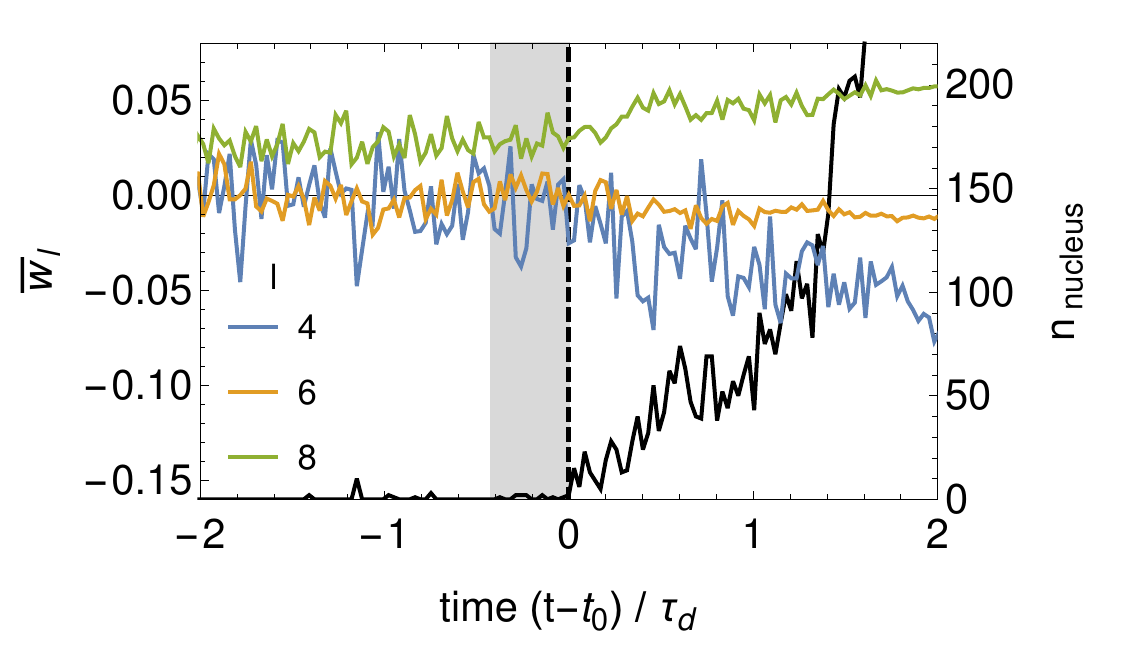} \\
     c) & \hspace{0.5cm} & d)  \\[-0.6cm]
     \includegraphics[width=\figwidthC]{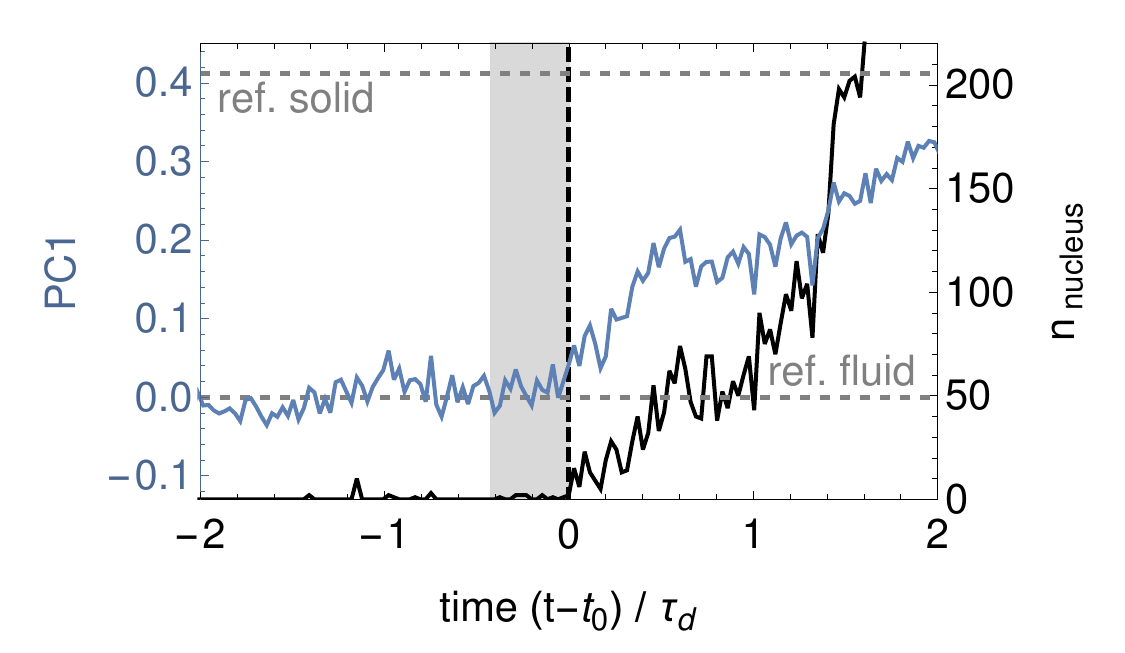} & & \includegraphics[width=\figwidthC]{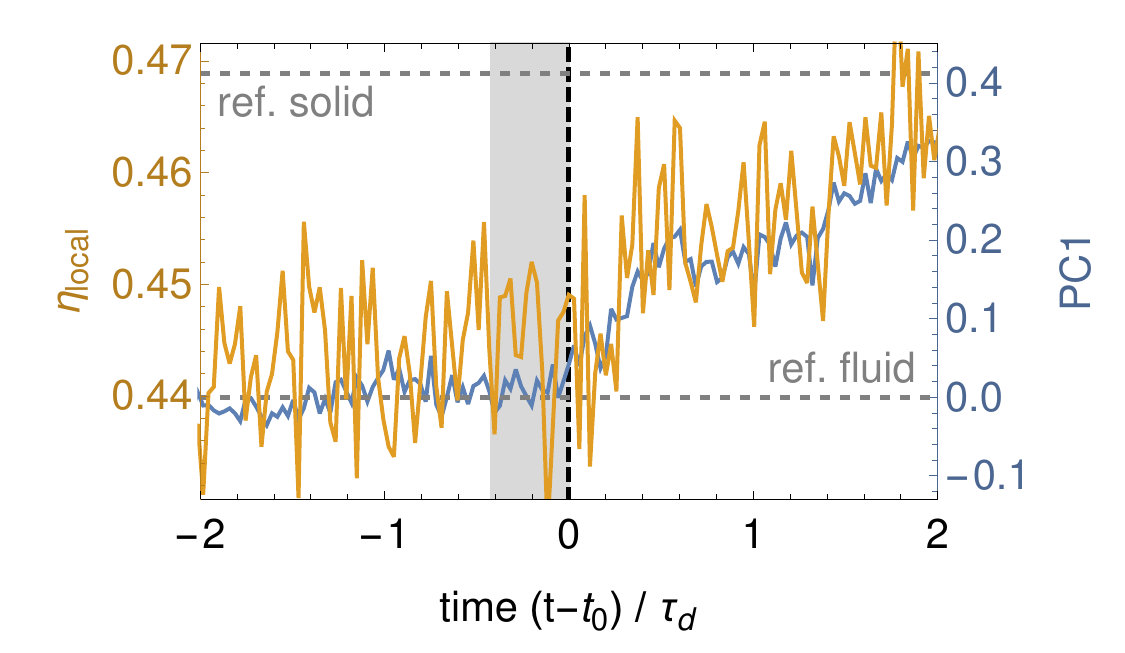} \\
     e) & \hspace{0.5cm} & f)  \\[-0.6cm]
     \includegraphics[width=\figwidthC]{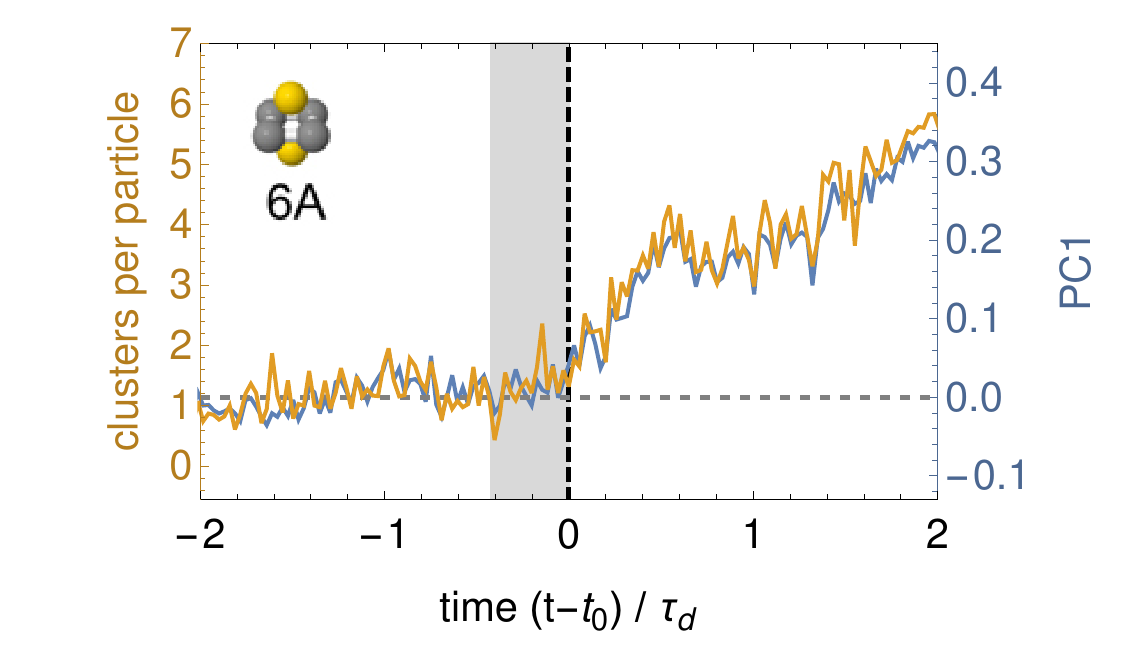} & & \includegraphics[width=\figwidthC]{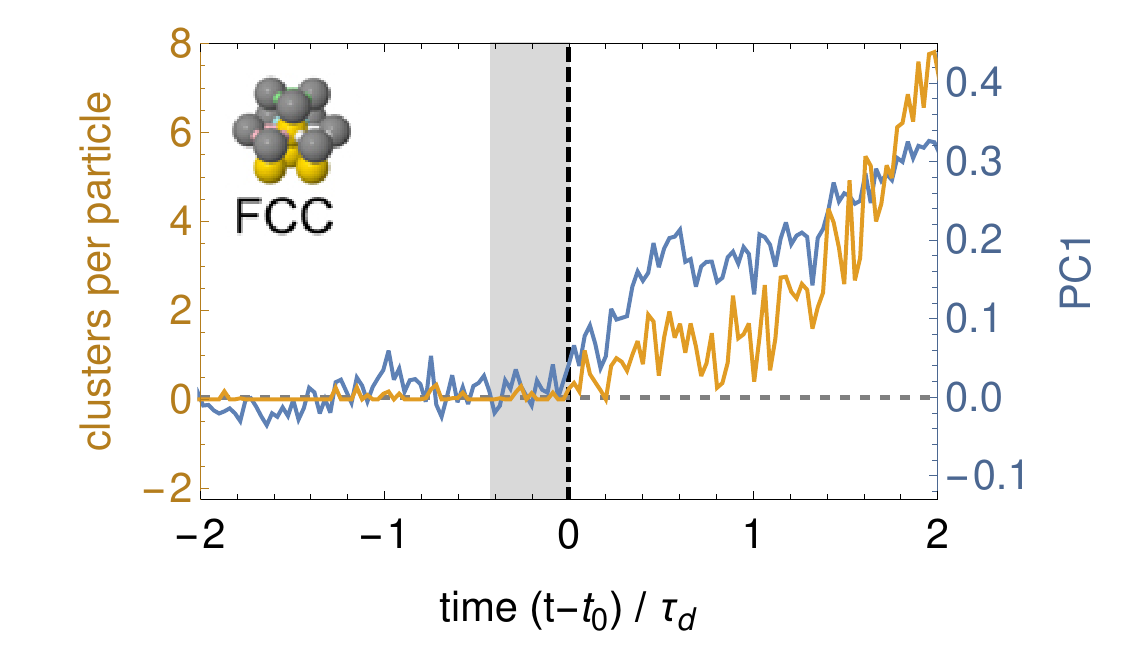} \\
     g) & \hspace{0.5cm} & h)  \\[-0.6cm]
     \includegraphics[width=\figwidthC]{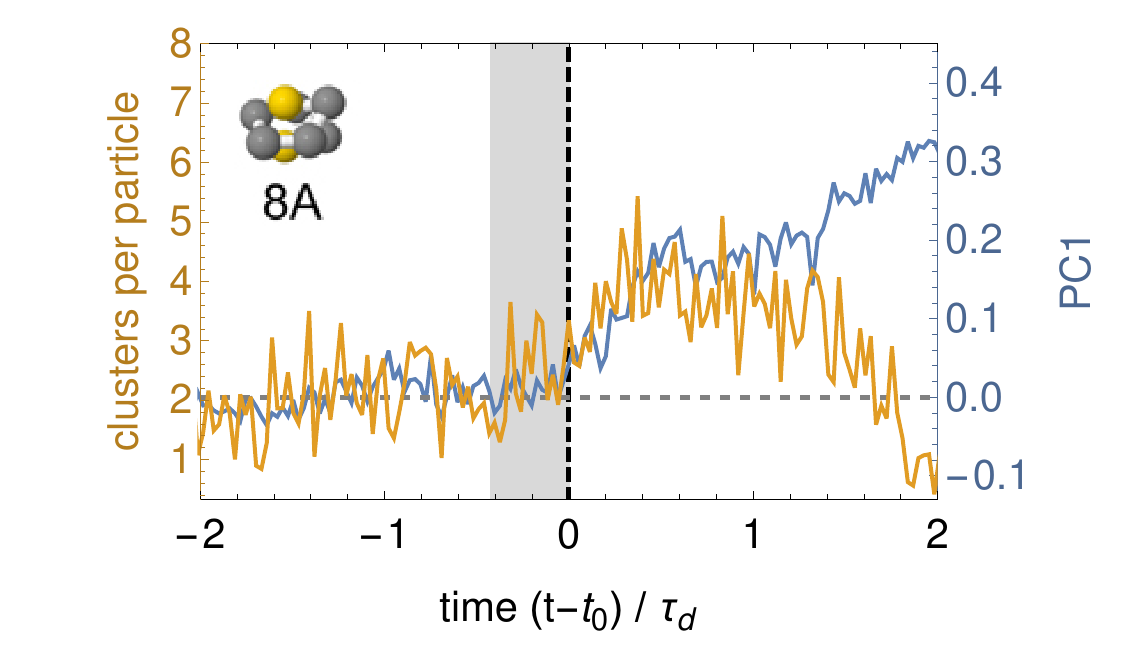} & & \includegraphics[width=\figwidthC]{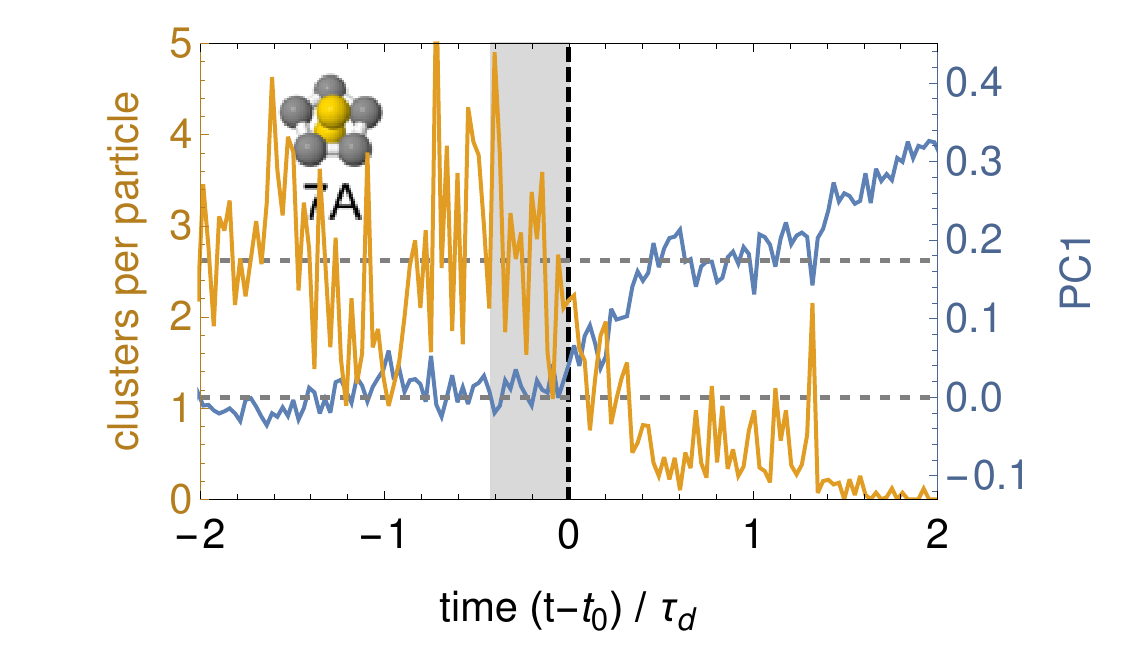}
\end{tabular}
    \caption[width=1\linewidth]{\label{fig:sieventsyukfcc} 
    Typical nucleation event of nearly-hard hard-core Yukawa particles ($\beta\epsilon=8$, $1/\kappa\sigma=0.04$, $\eta=0.4400$) obtained using MC. In a-c) the black line (right axis) gives the size of the biggest nucleus present in the studied region, while the other lines (left axis) give the average value of a-b) the BOPs $\bar{q}_l$ and $\bar{w}_l$, and c) PC1. In d-h) blue line (right axis) gives PC1, while the yellow line (left axis) gives d) the local packing fraction and e-h) the average number of clusters per particle for a couple of TCC clusters.}
\end{figure*}

Lastly, we show a typical nucleation event of a system of essentially-hard spheres ($\beta\epsilon=81$, $1/\kappa\sigma=0.01$), see Fig. \ref{fig:sieventsyukfcce81k100}, and of a system of nearly-hard spheres ($\beta\epsilon=8$, $1/\kappa\sigma=0.04$), see Fig. \ref{fig:sieventsyukfcc}. Again we observe no anomalous behavior in the metastable fluid prior to the start of nucleation, and all local order parameters abruptly change as soon as nucleation starts. 


\bibliography{paper}